\documentstyle[psfig]{l-aa}

\input psfig.sty

\newcommand{\gsim}{\raisebox{-3.8pt}{$\;\stackrel{\textstyle >}{\sim}\;$}}
\newcommand{\lsim}{\raisebox{-3.8pt}{$\;\stackrel{\textstyle <}{\sim}\;$}}

\newcommand{\Msol}{$M_{\odot}$}
\newcommand{\Zsol}{$Z_{\odot}$}
\newcommand{\MHe}{$M_{He}$}
\newcommand{\MCO}{$M_{CO}$}
\newcommand{\Mfin}{$M_{fin}$}

\newcommand{\etal}{\mbox{{\rm et~al.\ }}}
\newcommand{\Hydrogen}{\mbox{${\rm {\rm ^{1}H}}$}}
\newcommand{\Hetre}{\mbox{${\rm {\rm ^{3}He}}$}}
\newcommand{\Helium}{\mbox{${\rm {\rm ^{4}He}}$}}
\newcommand{\Carbon}{\mbox{${\rm {\rm ^{12}C}}$}}
\newcommand{\Ctredici}{\mbox{${\rm {\rm ^{13}C}}$}}
\newcommand{\Nitrogen}{\mbox{${\rm {\rm ^{14}N}}$}}
\newcommand{\Nquindici}{\mbox{${\rm {\rm ^{15}N}}$}}
\newcommand{\Oxygen}{\mbox{${\rm {\rm ^{16}O}}$}}
\newcommand{\Odiciassette}{\mbox{${\rm {\rm ^{17}O}}$}}
\newcommand{\Odiciotto}{\mbox{${\rm {\rm ^{18}O}}$}}
\newcommand{\Neon}{\mbox{${\rm {\rm ^{20}Ne}}$}}
\newcommand{\Neventidue}{\mbox{${\rm {\rm ^{22}Ne}}$}}
\newcommand{\Mgventicinque}{\mbox{${\rm {\rm ^{25}Mg}}$}}
\newcommand{\Magnesium}{\mbox{${\rm {\rm ^{24}Mg}}$}}
\newcommand{\Silicon}{\mbox{${\rm {\rm ^{28}Si}}$}}
\newcommand{\Sulfur}{\mbox{${\rm {\rm ^{32}S}}$}}
\newcommand{\Calcium}{\mbox{${\rm {\rm ^{40}Ca}}$}}
\newcommand{\Fe}{\mbox{${\rm {\rm ^{56}Fe}}$}}
\newcommand{\Nickel}{\mbox{${\rm {\rm ^{56}Ni}}$}}

\def\smallskip{\vskip 6pt}

\begin{document}

\thesaurus{       }

\title {Galactic chemical enrichment with new metallicity 
dependent stellar yields } 

\author{L.\ Portinari$^{1}$, C.\ Chiosi$^{2,1}$, 
and A.\ Bressan$^{3}$  }

\institute{
$^1$ Department of Astronomy, University of Padova,
Vicolo dell'Osservatorio 5, 35122 Padova, Italy \\         
$^2$ European Southern Observatory, K-Schwarzschild-strasse 2, D-85748,
Garching bei M\"unchen, Germany\\ 
$^3$ Astronomical Observatory of Padova, Vicolo dell'Osservatorio 5, 
35122 Padova, Italy \\
e-mail: portinari, chiosi, bressan@pd.astro.it}

\offprints{L.\ Portinari}
\date{Received  December 1997; accepted }

\maketitle
\markboth{Portinari \etal: Chemical evolution
with metallicity
dependent yields}{}


\begin{abstract}

New detailed stellar yields of several elemental species are derived for
massive stars in a wide range of masses (from 6 to 120~\Msol) and metallicities
($Z =$0.0004, 0.004, 0.008, 0.02, 0.05). Our calculations are based on the
Padova evolutionary tracks and take into account 
recent results on stellar
evolution, such as overshooting and quiescent mass-loss, paying major
attention to the effects of the initial chemical composition of the star.
We finally include modern results on explosive nucleosynthesis in SN\ae\
by Woosley \& Weaver 1995.
The issue of the chemical yields of Very Massive Objects (from 120 to
1000~\Msol) is also addressed. 

Our grid of stellar yields for massive stars is complementary to the results
by Marigo \etal (1996, 1997) on the evolution and nucleosynthesis of low and
intermediate mass stars, also based on the Padova evolutionary tracks.
Altogether, they represent a complete set of stellar yields of unprecedented
homogeneity and self-consistency.

Our new stellar yields are inserted in a code for the chemical evolution of
the Galactic disc with infall of primordial gas, according to the formulation
first suggested by Talbot \& Arnett (1971, 1973, 1975) and Chiosi (1980).
As a first application, the code is used to develop a model of the
chemical evolution of the Solar Vicinity, with a detailed comparison to
the available observational constraints.


\end{abstract}


\newpage
\section{Introduction}

In modelling the chemical evolution of our own Galaxy as well as of external
galaxies, we deal with three basic 
ingredients: (1) the Star
Formation Rate (SFR), (2) the Initial Mass Function (IMF) and (3) the stellar
ejecta of different elements. Indeed, the chemical
enrichment of galaxies is basically due to stars:
while primordial nucleosynthesis produced mainly light
elements, say up to $^{7}Li$ (Adouze 1986, Walker \etal 1991), heavier species
(metals) are due to stellar nucleosynthesis, either hydrostatic or explosive 
as in
the case of supernov\ae\ (SN\ae). Therefore, a chemical model should estimate
accurately the role of stars of different masses in enriching the
interstellar medium (ISM) with various elemental species. To this aim, one
needs to consider the main nucleosynthetic processes taking place in stellar
interiors, combined with a modelling of stellar evolution. 

New extended sets of stellar tracks for fine grids of masses and metallicities
have recently become available from various research groups. The body of
these models shows that the initial chemical composition of a star bears on the
details of its evolution.
Therefore, we expect
that stellar ejecta depend not only on the mass of a star, but also on its
metallicity. In spite of this evidence, most chemical evolution models 
rest upon grids of stellar yields derived for stars with solar
composition, and only few examples of metallicity-dependent yields can be found
in literature (Maeder 1992, hereinafter M92; Maeder 1993; Woosley \& Weaver
1995, hereinafter WW95). 
   
In order to improve upon this, we derive here a wide and up-to-date set of
ejecta in which the effect of the initial chemical composition is taken into
account. We adopt the stellar models from the Padua library (Bressan \etal
1993; Fagotto \etal 1994a,b), which span the mass range from 0.6 to 120~\Msol.
Full sets are available for the metallicities $Z$=0.0004, $Z$=0.004, $Z$=0.008,
$Z$=\Zsol=0.02 and $Z$=0.05, and we derive five corresponding sets of stellar
ejecta. Thus we get coherent prescriptions for stellar yields deduced from a
unique and homogeneous grid of stellar models covering all mass ranges, instead
of patching together data from different sources. In addition, in the range of
massive stars we take the effects of quiescent mass-loss starting from the ZAMS
into account. We also point out that, unlike most stellar yields adopted in
chemical models up to now, here we refer to evolutionary tracks calculated with
convective overshooting.

As a first application of our grid of stellar yields, we develop a chemical
evolution model for the Solar Neighbourhood, which is the first template
environment where nucleosynthetic yields and chemical models are to be
calibrated and tested.

This paper is organized as follows. In Sec.~2 we briefly summarize the 
various groups in which stars of different mass can be classified
according to the dominant physical process and type of evolution. 
In Sec.~3 we deal with the role of mass
loss in the evolution and nucleosynthesis of massive stars, and calculate
the ejecta of the stellar winds predicted by our stellar models.
In Sec.~4 we discuss the effects of mass loss on the outcoming SN\ae\
and outline a recipe to calculate the explosive ejecta of SN\ae\ originating
from our pre-SN models. Sec.~5 summarizes and comments upon the total ejecta
of mass losing massive stars with different initial metallicity.
Sec.~6 describes the adopted prescriptions for the yields of low and
intermediate mass stars. By means of simple semi-analytical prescriptions,
in Sec.~7 we extend our grid of ejecta to Very Massive Objects up to
1000~\Msol\ (if they can form now or have existed in the past), 
since in some particular astrophysical problems it can be useful
to consider the possibility of chemical enrichment by these stars. 
Sec.~8 introduces the mathematical
formulation of our chemical model. In Sec.~9 we apply our model
to the Solar Neighbourhood, discuss the space of parameters with respect
to the available observational constraint and comment upon the 
chemical abundances and abundance ratios our model predicts. Finally,
in Sec. 10 we draw some concluding remarks.


\section{Definition of the mass intervals}

For  the sake of better understanding, we summarize
here  how stars are classified according to the dominant physical 
processes governing their evolution and fate. This is
especially needed since we adopt models with convective overshoot
for which the mass grouping is different as compared to standard models
(cf. Chiosi 1986; Woosley 1986; Chiosi \etal 1992; and references therein).

\begin{description}

\item[{\sl Low mass stars}:] They develop degenerate He-cores, undergo core
He-flash, proceed to the formation of a degenerate CO-core, and after
the AGB phase terminated by the  loss of the envelope by stellar wind,
 become white
dwarfs. In models with overshoot the upper mass limit ($M_{HeF}$) 
for this to occur is about $1.6\div 1.7$~\Msol, 
depending on the chemical composition. 

\item[{\sl Intermediate mass stars}:] They avoid core He-flash (helium is
ignited in a non degenerate core), proceed to the formation of a
degenerate CO-core, undergo the TP-AGB phase, and losing mass by stellar
wind end up as white dwarfs. In models with convective overshoot the upper
mass limit ($M_{up}$) of
this group is about 5~\Msol\ depending on the chemical composition.

\item[{\sl Quasi-massive stars}:] In models with convective overshoot, stars
 in the range 6 to 8~\Msol\ undergo core C-burning in non degenerate
 conditions, but develop highly degenerate O-Ne-Mg-cores. They become
dynamically unstable to  {\it e-capture} instability and explode as supernovae.

\item[{\sl Massive stars}:] In the mass range 9 to 120~\Msol\ there is the 
simultaneous occurrence of two phenomena: the dominant mass loss by
stellar wind during the whole evolutionary history, and the completion of
the nuclear sequence down to the formation of an iron core in presence
of strong neutrino cooling. The supernova explosion
is triggered by  core collapse induced by {\it e-captures} on heavy nuclei,
{\it photo-dissociation}  of Fe into $\alpha$-particles, and rapid
neutronization of the
collapsing material. This is the range of classical Type II SN\ae. 
Either a neutron star of about 1.4 \Msol\ or a black hole of larger
mass is left over, depending on the efficiency of neutrino cooling 
during the  previous stages.

\item[{\sl Very massive stars}:] Starting from the very initial stages,
objects more massive than, say, 120~\Msol\ are strongly pulsationally unstable,
suffer from violent mass loss, and undergo {\it pair-creation} instability
during core O-burning. The final outcome is regulated by the mass of the 
CO-core: at decreasing core mass, these stars can either collapse later to
a black hole, or suffer complete thermonuclear explosion, or recover the
behaviour of the massive stars above.

\end{description}


\section{Ejecta from  the stellar wind of massive stars}

Massive stars contribute to the chemical enrichment of the ISM mainly by means
of their final explosion in supernova (SN). Anyway, mass loss is also known to
play a major role in the evolution of massive stars and stellar
winds can have a non--negligible influence on stellar ejecta (Chiosi \& Maeder
1986, hereinafter CM86; Chiosi \& Caimmi 1979; Chiosi 1979; Maeder 1981, 1983,
1992, 1993). Mass loss has got two main effects on final ejecta: 

(1) stellar layers peeled off by stellar winds directly mix with and enrich the
surrounding ISM, and evade further burning stages (direct effect); 

(2) mass loss affects the final mass \Mfin\ of the star and the final masses
of its He-- and CO--core (\MHe\ and \MCO, respectively), thus affecting
the resulting SN (indirect effect).

\begin{figure}[t]
\psfig{file=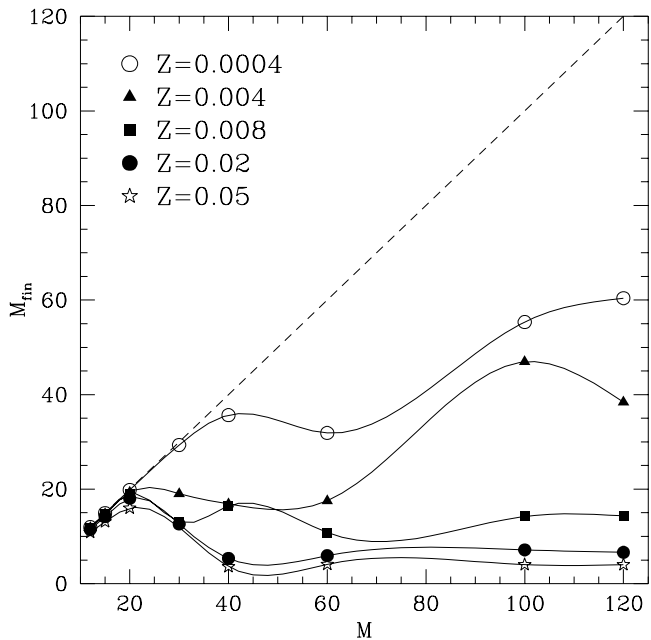,width=9truecm}
\caption{Final mass vs.\ initial mass relation (in \Msol) for massive stars of
five metallicities (legend on top left). The dashed line corresponds to the
case of constant mass}
\label{Mfinfig}
\end{figure}

In our adopted models, stars more massive than  $M >$ 12~\Msol\ are
affected by stellar winds during their core H- and He-burning
phases. All remaining evolutionary stages are so short-lived that mass
loss has in practice no effect at all and the mass of star is frozen
to the value with which it will undergo the final supernova
explosion. This value of the mass is thereinafter referred to as the
final mass \Mfin. The mass loss
rates are included according to de~Jager \etal (1988) from the ZAMS up to the
de~Jager limit; beyond such limit the star becomes a LBV candidate and its mass
loss rate is increased up to 10$^{-3}$~\Msol/yr. When the surface hydrogen
abundance falls below 0.3, the star is considered to enter its Wolf--Rayet
stage and from then on the assumed mass loss rate follows Langer (1989). Since
metallicity affects the bound--free and line opacity in the surface layers of
massive stars, the efficiency of radiation pressure driven winds is expected to
depend on metallicity according to $\dot{M} \propto Z^{\alpha}$, $\alpha \sim$
0.5 $\div$ 1.0  
(Kudritzki \etal 1989; Kudritzki 1997; Leitherer \& Langer 1991;
Lamers \& Cassinelli 1996). Evidence for the dependence of mass loss rates on
metallicity comes from statistics of blue and red supergiants and of WR stars
in our own Galaxy, in the Magellanic Clouds and in other galaxies of the Local
Group (Maeder \& Conti 1994). In our models,
mass loss rates are scaled with metallicity  according to
Kudritzki \etal (1989, $\dot{M} \propto Z^{0.5}$), therefore wind ejecta and
the structure of SN precursors sensitively depend on the initial metallicity
of the star, as shown by M92. For more details on the evolution of massive
stars in presence of mass loss the reader is referred to CM86, Chiosi
et al. (1992),  Chiosi (1997). 

\begin{table}[t]
\caption{Initial masses and corresponding final masses}
\label{Mfintab}
\begin{small}
\begin{center}
\begin{tabular}{|c|c c c c c|}
\hline
 M  &	       &	 &  \Mfin\  & 	     &	     \\
\hline
    & Z=0.0004 & Z=0.004 & Z=0.008 & Z=0.02  &  Z=0.05\\
\hline
 12 &   11.97  &  11.84  &  11.77  &  11.46  &  10.85 \\
 15 &   14.94  &  14.76  &  14.63  &  14.21  &  13.11 \\
 20 &   19.83  &  19.30  &  19.03  &  18.06  &  16.00 \\
 30 &   29.35  &  19.05  &  13.03  &  12.63  &   ---  \\
 40 &   35.65  &  16.86  &  16.44  &   5.35  &   3.62 \\
 60 &   31.88  &  17.55  &  10.81  &   5.93  &   4.09 \\
100 &   55.35  &  46.93  &  14.23  &   7.16  &   4.03 \\
120 &   60.41  &  38.43  &  14.35  &   6.63  &   4.04 \\
\hline
\end{tabular}
\end{center}
\end{small}
\end{table}

In Fig.~\ref{Mfinfig} we plot the final mass \Mfin\ versus the initial mass
$M$ for our five sets of models between 12 and 120~\Msol; the corresponding
values are listed in Tab.~\ref{Mfintab}. As expected, higher metallicities
correspond to lower final masses, due to larger mass loss rates. In particular,
the most massive stars ($M >$ 40~\Msol) in the higher metallicity sets end up
with quite low final masses, typically 4 $\div$ 6~\Msol. These stars go through
a WR phase, during which mass loss rates are high and strongly dependent on the
stellar mass ($\dot{M} \propto M^{2.5}$, Langer 1989); in this stage the
stellar mass is decreasing rapidly, and the mass loss rate is also
correspondingly decreasing, until all these stars reach very similar final
masses. The final mass is very low especially in the case of high metallicity,
because a more efficient wind during the Main Sequence phase allows the WR
stage to be reached sooner and the intense WR mass loss to be active for a
longer time. Indeed, to reach the final convergence mass 
the star needs to lose its H--rich envelope soon enough 
(before the helium abundance
in the core gets to $Y \sim$ 0.5, according to Woosley \etal 1995). 

The ejecta due to stellar winds are calculated as 

\begin{equation}
 E_{iM}^{wind} = \int_{0}^{\tau_{M}}\dot{M}(M,t)\,X_{i}^{S}(t)\,dt 
\nonumber
\end{equation}

\noindent
where $E_{iM}^{wind}$ is the amount (in~\Msol) of species $i$ expelled in the
wind by a star of mass $M$, $\dot{M}$ is the mass loss rate and $X_{i}^{S}$ is
the surface abundance of species $i$. The integration runs over the whole
lifetime $\tau_{M}$ of a star of mass $M$. In practice, 
$\tau_{M}$ is the time when carbon
ignites in the core, i.e.\ when the track ends. Wind ejecta are calculated for
H, $^{3}$He, $^{4}$He, $^{12}$C, $^{13}$C, $^{14}$N, $^{15}$N, $^{16}$O,
$^{17}$O, $^{18}$O, $^{20}$Ne, $^{22}$Ne and are listed in
Tab.~2. 

\begin{table*}[t]
\begin{minipage}{18truecm}
\caption{Ejecta (in \Msol) of the stellar wind for our model massive stars}
\scriptsize
\begin{tabular}{|r|c|c|c|c|c|c|c|c|c|c|c|c|}
\multicolumn{13}{c}{{\bf $Z$=0.0004}}\\
\hline
 $M$ & \Hydrogen &   \Hetre & \Helium & \Carbon  & \Ctredici & \Nitrogen & \Nquindici & \Oxygen & \Odiciassette & \Odiciotto &
  \Neon  & \Neventidue \\
\hline
  12 &   0.024   & 4.23E-07 &  0.007  & 2.13E-06 &  1.09E-07 & 1.93E-06  &  1.58E-09  & 6.47E-06 & 6.24E-09     & 1.22E-08   &
9.55E-07 & 1.28E-07    \\
  15 &   0.049   & 5.18E-07 &  0.015  & 5.10E-06 &  2.21E-07 & 2.84E-06  &  3.81E-09  & 1.32E-05 & 1.09E-08     & 2.71E-08   &
1.93E-06 & 2.59E-07    \\
  20 &   0.128   & 1.52E-07 &  0.038  & 1.63E-05 &  2.00E-07 & 4.28E-06  &  1.62E-08  & 3.50E-05 & 1.39E-08     & 8.03E-08   &
5.07E-06 & 6.80E-07    \\
  30 &   0.500   & 6.70E-07 &  0.150  & 6.43E-05 &  7.74E-07 & 1.62E-05  &  5.94E-08  & 1.37E-04 & 8.84E-08     & 2.99E-07   &
1.99E-05 & 2.67E-06    \\
  40 &   3.34    & 6.62E-06 &  1.00   & 4.28E-04 &  5.45E-06 & 1.11E-04  &  3.21E-07  & 9.17E-04 & 9.67E-07     & 1.92E-06   &
1.33E-04 & 1.78E-05    \\
  60 &   17.08   & 1.09E-05 &  11.03  & 1.01E-03 &  6.42E-05 & 5.60E-03  &  8.11E-07  & 2.64E-03 & 2.53E-05     & 3.31E-06   &
8.59E-04 & 1.15E-04    \\
 100 &   27.44   & 5.20E-05 &  17.19  & 1.73E-03 &  2.05E-04 & 8.61E-03  &  1.14E-06  & 4.26E-03 & 1.23E-05     & 7.59E-06   &
1.36E-03 & 1.83E-04    \\
 120 &   32.52   & 5.05E-05 &  27.06  & 1.65E-03 &  1.43E-04 & 1.38E-02  &  1.77E-06  & 4.03E-03 & 5.15E-05     & 8.80E-06   &
1.82E-03 & 2.45E-04    \\
\hline
\multicolumn{13}{c}{}\\
\multicolumn{13}{c}{{\bf $Z$=0.004}}\\
\hline
 $M$ & \Hydrogen & \Hetre   & \Helium & \Carbon  & \Ctredici & \Nitrogen & \Nquindici & \Oxygen & \Odiciassette & \Odiciotto &
  \Neon  & \Neventidue \\
\hline
  12 &   0.115   & 2.43E-06 &  0.042  & 1.08E-04 &  5.26E-06 & 1.26E-04  & 8.37E-08   & 2.90E-04 & 6.35E-07     & 6.09E-07   &
4.80E-05 & 6.44E-06    \\
  15 &   0.173   & 2.36E-06 &  0.064  & 1.78E-04 &  6.20E-06 & 1.75E-04  & 1.82E-07   & 4.42E-04 & 6.42E-07     & 1.01E-06   &
7.27E-05 & 9.76E-06    \\
  20 &   0.509   & 6.76E-06 &  0.185  & 5.69E-04 &  1.82E-05 & 4.18E-04  & 4.95E-07   & 1.34E-03 & 1.58E-06     & 3.02E-06   &
2.13E-04 & 2.86E-05    \\
  30 &   7.68    & 8.21E-05 &  3.23   & 7.91E-03 &  2.85E-04 & 9.52E-03  & 4.33E-06   & 1.90E-02 & 8.61E-05     & 3.44E-05   &
3.35E-03 & 4.49E-04    \\
  40 &   14.39   & 7.77E-05 &  8.66   & 1.03E-02 &  5.55E-04 & 3.80E-02  & 8.09E-06   & 2.82E-02 & 3.57E-04     & 3.31E-05   &
7.07E-03 & 9.50E-04    \\
  60 &   19.54   & 1.73E-04 &  15.95  & 4.33     &  1.28E-03 & 5.77E-02  & 1.53E-05   & 2.51     & 2.05E-04     & 8.42E-05   &
1.30E-02 & 1.74E-03    \\
 100 &   28.02   & 1.26E-04 &  24.84  & 1.78E-02 &  8.75E-04 & 0.118     & 1.50E-05   & 3.83E-02 & 3.97E-04     & 6.04E-05   &
1.62E-02 & 2.18E-03    \\
 120 &   31.93   & 1.20E-04 &  49.09  & 0.127    &  1.55E-03 & 0.214     & 2.01E-05   & 0.147    & 3.03E-04     & 6.12E-05   &
2.49E-02 & 3.62E-03    \\
\hline
\multicolumn{13}{c}{}\\
\multicolumn{13}{c}{{\bf $Z$=0.008}}\\
\hline
 $M$ & \Hydrogen & \Hetre   & \Helium & \Carbon  & \Ctredici & \Nitrogen & \Nquindici & \Oxygen & \Odiciassette & \Odiciotto &
  \Neon  & \Neventidue \\
\hline
  12 &   0.168   & 5.94E-06 &  0.065  & 3.36E-04 &  1.61E-05 & 3.54E-04  & 2.82E-07   & 8.78E-04 & 5.09E-07     & 1.88E-06   &
1.43E-04 & 1.93E-05    \\
  15 &   0.262   & 8.42E-06 &  0.103  & 5.74E-04 &  2.01E-05 & 5.02E-04  & 4.87E-07   & 1.39E-03 & 7.44E-07     & 2.98E-06   &
2.26E-04 & 3.03E-05    \\
  20 &   0.693   & 1.53E-05 &  0.272  & 1.61E-03 &  4.66E-05 & 1.17E-03  & 1.44E-06   & 3.71E-03 & 8.32E-06     & 8.15E-06   &
5.94E-04 & 7.98E-05    \\
  30 &   10.81   & 1.14E-04 &  6.02   & 1.81E-02 &  7.55E-04 & 4.52E-02  & 1.27E-05   & 4.96E-02 & 4.19E-04     & 5.51E-05   &
1.78E-03 & 1.39E-03    \\
  40 &   13.92   & 2.19E-04 &  9.46   & 2.41E-02 &  9.89E-04 & 7.23E-02  & 2.63E-05   & 5.95E-02 & 4.83E-04     & 1.19E-04   &
1.44E-02 & 1.93E-03    \\
  60 &   18.80   & 2.67E-04 &  21.53  &   7.02   &  1.97E-03 & 0.133     & 2.96E-05   & 1.50     & 6.99E-04     & 1.37E-04   &
3.01E-02 & 4.04E-03    \\
 100 &   25.91   &    ---   &  38.38  &   15.5   &  3.79E-03 & 0.267     &   ---      & 5.28     &    ---       &    ---     &
---      &    ---      \\
 120 &   30.12   & 3.51E-04 &  49.28  &   19.2   &  4.79E-03 & 0.349     & 5.54E-05   & 6.19     & 8.41E-04     & 1.93E-04   &
6.51E-02 & 0.295       \\
\hline
\multicolumn{13}{c}{}\\
\multicolumn{13}{c}{{\bf $Z$=\Zsol=0.02}}\\
\hline
 $M$ & \Hydrogen &  \Hetre  & \Helium & \Carbon  & \Ctredici & \Nitrogen & \Nquindici & \Oxygen  & \Odiciassette & \Odiciotto &
  \Neon  & \Neventidue \\
\hline
  12 &   0.365   & 2.00E-05 &  0.168  & 2.01E-03 &  9.45E-05 & 1.89E-03  & 1.84E-06   & 5.17E-03 & 3.30E-05      & 1.14E-05   &
8.32E-04 & 1.12E-04    \\
  15 &   0.532   & 2.93E-05 &  0.246  & 3.11E-03 &  1.15E-04 & 2.55E-03  & 2.93E-06   & 7.56E-03 & 3.29E-05      & 1.69E-05   &
1.21E-03 & 1.63E-04    \\
  20 &   1.29    & 7.19E-05 &  0.609  & 7.89E-03 &  2.11E-04 & 6.06E-03  & 7.51E-06   & 1.84E-02 & 6.68E-05      & 4.09E-05   &
2.96E-03 & 3.98E-04    \\
  30 &   10.31   & 3.96E-04 &   6.72  & 5.37E-02 &  1.83E-03 & 0.103     & 4.07E-05   & 0.131    & 1.41E-03      & 2.31E-04   &
2.65E-02 & 3.56E-03    \\
  40 &   12.62   & 4.72E-04 &  18.81  &   2.39   &  3.50E-03 & 0.211     & 5.37E-05   & 0.280    & 3.35E-03      & 7.00E-03   &
5.29E-02 & 0.264       \\
  60 &   16.42   & 5.27E-04 &  31.95  &   4.28   &  5.30E-03 & 0.343     & 7.79E-05   & 0.440    & 3.82E-03      & 2.89E-04   &
8.26E-02 & 0.318       \\
 100 &   23.84   & 5.51E-04 &  55.35  &   10.7   &  9.68E-03 & 0.682     & 9.96E-05   & 1.14     & 5.56E-03      & 3.04E-04   &
0.142    & 0.596       \\
 120 &   28.57   & 1.37E-03 &  73.88  &   7.93   &  1.15E-02 & 1.01      & 1.92E-04   & 0.781    & 2.15E-03      & 7.53E-04   &
0.173    & 0.525       \\
\hline
\multicolumn{13}{c}{}\\
\multicolumn{13}{c}{{\bf $Z$=0.05}}\\
\hline
 $M$ & \Hydrogen &  \Hetre  & \Helium & \Carbon  & \Ctredici & \Nitrogen & \Nquindici & \Oxygen  & \Odiciassette & \Odiciotto &
  \Neon  & \Neventidue \\
\hline
  12 &   0.668   & 9.06E-05 &  0.426  & 1.13E-02 &  4.86E-04 & 7.72E-03  & 9.45E-06   & 2.90E-02 & 2.45E-04      & 6.08E-05   &
4.40E-03 & 5.90E-04    \\
  15 &   1.09    & 1.53E-04 &  0.712  & 1.89E-02 &  7.73E-04 & 1.29E-02  & 1.62E-05   & 4.70E-02 & 3.88E-04      & 9.92E-05   &
7.24E-03 & 9.72E-04    \\
  20 &   2.26    & 3.45E-04 &  1.53   & 4.04E-02 &  1.19E-03 & 2.88E-02  & 3.72E-05   & 9.74E-02 & 7.06E-04      & 2.07E-04   &
1.53E-02 & 2.05E-03    \\
  40 &   10.05   & 1.17E-03 &  22.56  &   1.82   &  7.50E-03 & 0.484     & 1.50E-04   & 0.507    & 1.39E-02      & 7.61E-03   &
0.139    & 0.608       \\
  60 &   13.95   & 6.47E-04 &  35.83  &   3.07   &  1.12E-02 & 1.06      & 1.25E-04   & 0.711    & 3.38E-02      & 3.67E-04   &
0.213    & 0.523       \\
 100 &   26.92   & 8.60E-04 &  60.69  &   3.22   &  1.92E-02 & 1.79      & 1.84E-04   & 1.14     & 5.51E-02      & 3.23E-02   &
0.366    & 1.23        \\
 120 &   38.44   & 1.15E-03 &  68.13  &   3.28   &  2.59E-02 & 2.11      & 2.26E-04   & 1.62     & 6.30E-02      & 0.177      &
0.443    & 1.23        \\
\hline   
\end{tabular}
\normalsize
\end{minipage}
\label{ejwindtab}
\end{table*}

\begin{table*}[t]
\begin{minipage}{18truecm}
\caption{Yields $p_{iM}^{wind}$ (see Eq.~3) of the stellar wind for our
massive stars. When
the amount of a newly synthesized/consumed element is less than 1\% of its
initial one, the corresponding yield is assumed to be not significative, i.e.\
is set equal 0}
\tiny
\begin{center}
\begin{tabular}{|r|c|r|c|r|c|c|c|r|c|r|c|c|}
\multicolumn{13}{c}{{\bf $Z$=0.0004}}\\
\hline
 $M$ & \Hydrogen & \multicolumn{1}{|c|}{\Hetre} & \Helium & \multicolumn{1}{|c|}{\Carbon} & \Ctredici & \Nitrogen & \Nquindici &
\multicolumn{1}{|c|}{\Oxygen} & \Odiciassette & \multicolumn{1}{|c|}{\Odiciotto} & \Neon & \Neventidue \\
\hline
  12 & 0 & \multicolumn{1}{|c|}{0} & 0 & \multicolumn{1}{|c|}{0} &   0   &   0   &    0    & 
\multicolumn{1}{|c|}{0} &  0  & \multicolumn{1}{|c|}{0} &  0  &  0    \\
  15 & 0 & \multicolumn{1}{|c|}{0} & 0 & \multicolumn{1}{|c|}{0} &   0   &   0   &    0    & 
\multicolumn{1}{|c|}{0} &  0  & \multicolumn{1}{|c|}{0} &  0  &  0    \\
  20 &      0    &  3.23E-10  &   0    & \multicolumn{1}{|c|}{0} &   0   &   0   &    0    & 
\multicolumn{1}{|c|}{0} &  0  & \multicolumn{1}{|c|}{0} &  0  &  0    \\
  30 &      0    & 3.26E-09   &   0    & \multicolumn{1}{|c|}{0} &  0  &   0   & -1.35E-10 & 
\multicolumn{1}{|c|}{0} & 1.13E-09 & -5.28E-10  &     0      &      0      \\
  40 &      0    & 6.99E-08   &   0    & \multicolumn{1}{|c|}{0} &  0  &   0   & -2.59E-09 & 
\multicolumn{1}{|c|}{0} & 1.50E-08 & -4.68E-09  &     0      &      0      \\
  60 & -7.60E-02 & -2.31E-07  & 7.60E-02 & -2.95E-05 &  5.15E-07  &  8.18E-05 & -3.22E-08  & 
-5.50E-05 & 3.82E-07 & -1.72E-07  &     0      &      0      \\
 100 & -6.92E-02 &  1.27E-07  & 6.92E-02 & -2.68E-05 &  1.52E-06  &  7.50E-05 & -3.21E-08  & 
-5.18E-05 & 8.52E-08 & -1.40E-07  &     0      &      0      \\
 120 &   -0.111  & -1.63E-08  &   0.111  & -3.53E-05 &  6.02E-07  &  1.03E-04 & -3.37E-08  & 
-7.14E-05 & 3.87E-07 & -1.67E-07  &     0      &      0      \\
\hline
\multicolumn{13}{c}{} \\
\multicolumn{13}{c}{{\bf $Z$=0.004}} \\
\hline
 $M$ & \Hydrogen & \multicolumn{1}{|c|}{\Hetre} & \Helium & \multicolumn{1}{|c|}{\Carbon} & \Ctredici & \Nitrogen & \Nquindici &
\multicolumn{1}{|c|}{\Oxygen} & \Odiciassette & \multicolumn{1}{|c|}{\Odiciotto} & \Neon & \Neventidue \\
\hline
  12 & -3.34E-04 &  8.70E-08  & 3.34E-04 & -3.95E-06 &  2.83E-07  & 7.29E-06  & -5.79E-09  & 
-3.45E-06 & 4.20E-08 & -1.26E-08  &     0      &      0      \\
  15 & -4.91E-04 &  1.76E-08  & 4.91E-04 & -3.80E-06 &  2.26E-07  & 7.73E-06  & -3.38E-09  & 
-4.08E-06 & 2.95E-08 & -9.34E-09  &     0      &      0      \\
  20 & -8.96E-04 &  3.13E-08  & 8.96E-04 & -5.98E-06 &  4.99E-07  & 1.22E-05  & -9.28E-09  & 
-6.65E-06 & 4.98E-08 & -1.79E-08  &     0      &      0      \\
  30 & -2.00E-02 & -4.77E-07  & 2.00E-02 & -9.74E-05 &  5.18E-06  & 2.27E-04  & -2.12E-07  & 
-1.37E-04 & 2.56E-06 & -6.22E-07  &     0      &      0      \\
  40 & -7.76E-02 & -3.15E-06  & 7.76E-02 & -3.14E-04 &  7.01E-06  & 8.06E-04  & -3.62E-07  & 
-5.17E-04 & 8.43E-06 & -1.97E-06  &     0      &      0      \\
  60 &  -2.09    & -3.34E-06  & 9.60E-02 &  7.14E-02 &  1.29E-05  & 7.85E-04  & -4.36E-07  & 
 4.04E-02 & 2.82E-06 & -2.02E-06  &     0      &      0      \\
 100 &   -0.121  & -3.41E-06  &    1.21  & -3.46E-04 &  2.47E-06  & 1.05E-03  & -3.68E-07  &
-7.39E-04 & 3.52E-06 & -1.97E-06  &     0      &      0      \\
 120 &   -0.248  & -4.98E-06  &    2.46  &  3.83E-04 &  4.90E-06  & 1.61E-03  & -4.96E-07  & 
-2.14E-04 & 1.95E-06 & -2.78E-06  &     0      & 2.25E-06    \\
\hline
\multicolumn{13}{c}{} \\
\multicolumn{13}{c}{{\bf $Z$=0.008}} \\
\hline
 $M$ & \Hydrogen & \multicolumn{1}{|c|}{\Hetre} & \Helium & \multicolumn{1}{|c|}{\Carbon} & \Ctredici & \Nitrogen & \Nquindici &
\multicolumn{1}{|c|}{\Oxygen} & \Odiciassette & \multicolumn{1}{|c|}{\Odiciotto} & \Neon & \Neventidue \\
\hline
  12 & -5.11E-04 &  1.51E-07  & 5.11E-04 & -1.06E-05 &  8.77E-07  & 1.98E-05  & -1.46E-08  & 
-9.49E-06 & 9.52E-09 & -3.27E-08  &     0      &     0       \\
  15 & -7.33E-04 &  1.28E-07  & 7.33E-04 & -1.04E-05 &  7.60E-07  & 2.13E-05  & -1.56E-08  & 
-1.14E-05 & 8.24E-09 & -3.97E-08  &     0      &     0       \\
  20 & -1.44E-03 & -9.11E-08  & 1.44E-03 & -1.55E-05 &  1.18E-06  & 3.43E-05  & -2.30E-08  & 
-2.02E-05 & 3.34E-07 & -6.36E-08  &     0      &     0       \\
  30 & -5.92E-02 & -6.15E-06  & 5.92E-02 & -5.14E-04 &  1.18E-05  & 1.23E-03  & -6.82E-07  & 
-7.36E-04 & 1.30E-05 & -3.64E-06  &     0      &     0       \\
  40 & -8.91E-02 & -4.90E-06  & 8.92E-02 & -5.62E-04 &  1.08E-05  & 1.51E-03  & -4.93E-07  & 
-1.00E-03 & 1.11E-05 & -2.74E-06  &     0      &     0       \\
  60 & -2.95 & -9.98E-06 & 0.154 & \multicolumn{1}{|c|}{0.115} & 1.33E-05 & 1.81E-03 & -1.11E-06 & 
 2.15E-02 & 1.03E-05 & -5.66E-06  &     0      &     0       \\
 100 & -0.377 & \multicolumn{1}{|c|}{---} & 0.169 & \multicolumn{1}{|c|}{0.154} & 1.76E-05 & 2.24E-03 &  --- & 
4.92E-02 & --- & \multicolumn{1}{|c|}{---} &  --- &  --- \\
 120 & -0.402 & -1.26E-05 & 0.191 & \multicolumn{1}{|c|}{0.158} & 1.91E-05 & 2.47E-03 & -1.26E-06 & 
4.79E-02 & 5.53E-06 & -6.92E-06   &     0  & 2.38E-03    \\
\hline
\multicolumn{13}{c}{} \\
\multicolumn{13}{c}{{\bf $Z$=\Zsol=0.02}} \\
\hline
 $M$ & \Hydrogen & \multicolumn{1}{|c|}{\Hetre} & \Helium & \multicolumn{1}{|c|}{\Carbon} & \Ctredici & \Nitrogen & \Nquindici &
\multicolumn{1}{|c|}{\Oxygen} & \Odiciassette & \multicolumn{1}{|c|}{\Odiciotto} & \Neon & \Neventidue \\
\hline
  12 & -1.32E-03 & -3.29E-07  & 1.32E-03 & -5.70E-05 &  5.19E-06  & 1.01E-04  & -6.80E-08  & 
-4.86E-05 & 2.56E-06 & -1.47E-07  &     0      &     0       \\
  15 & -1.58E-03 & -3.75E-07  & 1.57E-03 & -5.42E-05 &  4.57E-06  & 1.04E-04  & -6.26E-08  & 
-5.48E-05 & 1.97E-06 & -1.54E-07  &     0      &     0       \\
  20 & -3.29E-03 & -6.74E-07  & 3.28E-03 & -8.46E-05 &  4.80E-06  & 1.83E-04  & -9.78E-08  & 
-1.05E-04 & 2.94E-06 & -3.04E-07  &     0      &     0       \\
  30 & -6.17E-02 & -1.23E-05  & 6.17E-02 & -1.07E-03 &  2.67E-05  & 2.72E-03  & -1.47E-06  & 
-1.74E-03 & 4.45E-05 & -6.31E-06  &     0      &     0       \\
  40 &   -0.291  & -2.63E-05  &   0.228  &  5.55E-02 &  3.63E-05  & 4.21E-03  & -2.88E-06  & 
-2.15E-03 & 8.01E-05 & 1.54E-04   &     0      & 6.42E-03    \\
  60 &   -0.357  & -3.09E-05  &   0.280  &  6.69E-02 &  3.50E-05  & 4.61E-03  & -3.10E-06  & 
-2.18E-03 & 5.99E-05 & -1.70E-05  &     0      & 5.12E-03    \\
 100 & -0.411 & -3.53E-05 & 0.294 & \multicolumn{1}{|c|}{0.102} & 4.18E-05 & 5.67E-03 & -3.53E-06 
& 1.58E-03 & 5.17E-05 & -1.94E-05  &     0      & 5.77E-03    \\
 120 &   -0.423  & -3.02E-05  &   0.351  &  6.14E-02 &  3.99E-05  & 7.27E-03  & -3.01E-06 
& -3.48E-03 & 1.39E-05 & -1.66E-05  &     0      & 4.18E-03    \\
\hline
\multicolumn{13}{c}{} \\
\multicolumn{13}{c}{{\bf $Z$=0.05}} \\
\hline
 $M$ & \Hydrogen & \multicolumn{1}{|c|}{\Hetre} & \Helium & \multicolumn{1}{|c|}{\Carbon} & \Ctredici & \Nitrogen & \Nquindici &
\multicolumn{1}{|c|}{\Oxygen} & \Odiciassette & \multicolumn{1}{|c|}{\Odiciotto} & \Neon & \Neventidue \\
\hline
  12 & -1.73E-03 & -3.00E-06  & 1.71E-03 & -2.46E-04 &  2.63E-05  & 3.46E-04  & -3.83E-07  &
 -1.20E-04 & 1.94E-05 & -7.43E-07  &     0      &     0       \\
  15 & -3.00E-03 & -3.67E-06  & 2.98E-03 & -2.98E-04 &  3.29E-05  & 4.67E-04  & -4.59E-07  & 
-2.02E-04 & 2.46E-05 & -1.03E-06  &     0      &     0       \\
  20 & -6.31E-03 & -4.73E-06  & 6.30E-03 & -4.49E-04 &  3.00E-05  & 8.19E-04  & -5.79E-07  & 
-4.11E-04 & 3.32E-05 & -1.75E-06  &     0      &     0       \\
  40 &   -0.293  & -7.08E-05  &  0.244   &  3.44E-02 &  5.28E-05  & 9.27E-03  & -7.35E-06  & 
-1.14E-02 & 3.37E-04 & 1.35E-04   &     0      & 1.47E-02    \\
  60 &   -0.325  & -9.17E-05  &  0.269   &  3.97E-02 &  4.83E-05  & 1.47E-02  & -9.28E-06  & 
-1.28E-02 & 5.53E-04 & -5.03E-05  &     0      & 8.25E-03    \\
 100 &   -0.305  & -9.70E-05  &  0.269   &  2.03E-02 &  5.01E-05  & 1.49E-02  & -9.86E-06  & 
-1.39E-02 & 5.41E-04 & 2.65E-04   &     0      & 1.18E-02    \\
 120 &   -0.258  & -9.67E-05  &  0.228   &  1.54E-02 &  7.28E-05  & 1.46E-02  & -9.90E-06  & 
-1.20E-02 & 5.15E-04 & 1.42E-03   &     0      & 9.74E-03    \\
\hline
\end{tabular}
\end{center}
\normalsize
\end{minipage}
\label{ywindtab}
\end{table*}

\begin{table*}[t]
\begin{minipage}{18truecm}
\caption{He--core mass (in \Msol) of model stars in four different \
evolutionary stages
(see text). Here \MHe\ is defined as the mass--point where the hydrogen content
falls to 0}
\label{MHevartab}
\begin{center}
\scriptsize
\begin{tabular}{|c|c|c|c|c|c|c|c|c|c|c|c|}
\hline
{\bf Z=0.0004} & 6 & 7 & 9 & 12 & 15 & 20 & 30 & 40 & 60 & 100 & 120\\
\hline
$M_{He}(1)$ & 0.01 & 0.01 & 0.06 & 0.71 & 1.90 & 3.97 & 7.85  & 12.62 & 
21.44 & 39.83 & 48.10\\ 
$M_{He}(2)$ & 0.82 & 1.00 & 1.40 & 2.10 & 2.95 & 4.64 & 8.51  & 12.77 & 
21.54 & 39.83 & 48.10\\
$M_{He}(3)$ & 1.69 & 2.03 & 2.80 & 4.04 & 5.36 & 7.62 & 12.36 & 17.31 & 
25.84 & 49.54 & 56.76\\
$M_{He}(4)$ & 1.69 & 2.03 & 2.80 & 4.04 & 5.36 & 7.62 & 12.33 & 16.98 & 
25.40 & 49.16 & 56.67\\
\hline
\hline
{\bf Z=0.004} & 6 & 7 & 9 & 12 & 15 & 20 & 30 & 40 & 60 & 100 & 120\\
\hline
$M_{He}(1)$ & 0.01 & 0.02 & 0.38 & 1.16 & 2.16 & 3.76 & 7.58 & 11.63 &
19.47 & 37.72 & 46.71\\
$M_{He}(2)$ & 0.90 & 1.12 & 1.62 & 2.45 & 3.39 & 5.12 & 8.70 & 12.56 &
20.39 & 37.72 & 46.71\\ 
$M_{He}(3)$ & 1.70 & 2.08 & 2.88 & 4.18 & 5.44 & 7.65 & 12.02 & 16.52 &
17.77 & 45.84 & 39.70\\ 
$M_{He}(4)$ & 1.70 & 2.08 & 2.88 & 4.18 & 5.43 & 7.64 & 11.62 & 16.48 &
17.55 & 45.47 & 38.43\\ 
\hline
\hline
{\bf Z=0.008} & 6 & 7 & 9 & 12 & 15 & 20 & 30 & 40 & 60 & 100 & 120\\
\hline
$M_{He}(1)$ & 0.01 & 0.05 & 0.35 & 1.06 & 1.98 & 3.85 & 7.77 & 11.12 &
19.58 & 38.82 & 48.26\\
$M_{He}(2)$ & 0.91 & 1.14 & 1.66& 2.54 & 3.52 & 5.22 & 8.79 & 12.93 &
20.41 & 38.82 & 49.20\\ 
$M_{He}(3)$ & 1.60 & 1.98 & 2.82 & 4.13& 5.41 & 7.58 & 11.99 & 16.26 &
10.76 & 14.03 & 14.16\\ 
$M_{He}(4)$ & 1.60 & 1.98 & 2.82 & 4.13 & 5.40 & 7.57 & 11.08 & 15.98 &
10.65 & 13.83 & 13.95\\                                   
\hline
\hline
{\bf Z=0.02} & 6 & 7 & 9 & 12 & 15 & 20 & 30 & 40 & 60 & 100 & 120\\
\hline
$M_{He}(1)$ & 0.008 & 0.06& 0.38 & 1.198 & 2.13 & 3.68 & 7.36 & 10.78
& 19.08 & 39.82 & 38.52\\
$M_{He}(2)$ & 0.94 & 1.19 & 1.75 & 2.72 & 3.74 & 5.55 & 9.26 & 12.93
& 21.19 & 39.82 & 38.52\\
$M_{He}(3)$ & 1.51 & 1.91 & 2.78 & 4.11 & 5.47 & 7.66 & 12.36 & 5.29
& 5.86 & 7.07 & 6.55\\
$M_{He}(4)$ & 1.51 & 1.91 & 2.78 & 4.10 & 5.46 & 7.64 & 12.07 & 5.20
& 5.77 & 6.96 & 6.44\\
\hline
\hline
{\bf Z=0.05} & 6 & 7 & 9 & 12 & 15 & 20 & 30 & 40 & 60 & 100 & 120\\
\hline
$M_{He}(1)$ & 0.01 & 0.08 & 0.40 & 1.26 & 2.24 & 3.90 & --- & 11.03 &
10.76 & 19.22 & 19.65\\
$M_{He}(2)$ & 1.05 & 1.34 & 2.00 & 3.10 & 4.21 & 6.21 & --- & 14.40 &
14.06 & 22.42 & 22.88\\
$M_{He}(3)$ & 1.59 & 2.03 & 2.99 & 4.51 & 6.00 & 8.47 & --- & 3.63 &
4.08 & 4.03 & 4.03\\
$M_{He}(4)$ & 1.59 & 2.03 & 2.99 & 4.48 & 5.98 & 8.46 & --- & 3.53 &
3.98 & 3.92 & 3.93\\
\hline
\end{tabular}
\normalsize
\end{center}
\end{minipage}
\end{table*}

The amount and the chemical composition of the expelled wind change with the
mass and with the metallicity of the star; in order to explain this effect
better, we consider stellar yields rather than ejecta. The stellar yield of a
species $i$, $p_{iM}$, is defined as the mass fraction of a star of mass $M$
that has been newly synthesized as species $i$ and then ejected (Tinsley 1980).
If $i$ has been converted into other species rather than produced, its yield
$p_{iM}$ is negative. On the basis of this definition, the total amount of
species $i$ ejected by a star of mass $M$ is: 

\begin{equation}
 E_{iM} = (M-M_{r})\,X_{i}^{0}\,+\,M p_{iM}
\nonumber
\end{equation}

\noindent
where $M_{r}$ is the remnant mass and $X_{i}^{0}$ is the initial abundance of
species $i$ in the star when it was born. The first term is the amount of
``original'' species $i$ eventually
expelled, while the second term is the positive or negative contribution of
newly synthesized $i$. Therefore, yields allow us to estimate the
nucleosynthetic production due to the star itself, by distinguishing the newly
synthesized fraction from the fraction that was already present in the gas
whence the star formed. The stellar yields of the wind for our set of masses and
metallicities are calculated according to: 

\begin{equation}
 M\,p_{iM}^{wind} = \int_{0}^{\tau_{M}}\dot{M}(M,t)\,[X_{i}^{S}(t)-X_{i}^{0}]\,dt 
\nonumber
\end{equation}

\noindent
The yields formalism is basically due to Tinsley (1980), but the previous
formul\ae\ are taken as revised by M92. The wind yields we obtain are listed
in Tab.~3 and can be commented on as follows (see also M92). 

\begin{description} 

\item[\bf $^{1}$H:]
Hydrogen yields are always negative and increase in modulus with
increasing mass and metallicity, since the more efficient mass loss is,
the larger amount of new helium and metals (synthesized at the expense of 
\Hydrogen) is revealed on the surface and released in the wind.

\item[\bf $^{3}$He:]
The abundance of \Hetre\ increases at the very outer edge of the H--burning
region, due to the first reactions of the p-p chain; then \Hetre\ is rapidly
turned to \Helium. Its yields are generally negative, except in some cases
of low mass loss where the \Hetre--enriched layers are lost, rather than inner
ones.

\item[\bf $^{4}$He:]
Helium yields are larger for larger masses and, at any given mass, for higher
metallicities, because strong winds can take large amounts of new \Helium\ 
away. 

\item[\bf $^{12}$C---$^{16}$O:]
Carbon yields are negative in most cases, since the CNO cycle turns \Carbon\
into \Nitrogen. Only when mass loss gets very efficient, i.e.\ for high masses
and metallicities, carbon yields can be positive because He--burning products
can be revealed on the surface (WC and WO stars). Quite the same holds for
oxygen, whose yields are seldom positive. If mass loss is extreme, anyway,
\Carbon\ and \Oxygen\
yields may even decrease because most of the mass is rapidly
lost in the wind in the form of 
\Helium.

\item[\bf $^{14}$N---$^{13}$C---$^{17}$O:] 
In our models of massive stars, nitrogen is produced only as a secondary
element, therefore its yields sensitively grow with metallicity. The same holds
for \Ctredici\ and for \Odiciassette, which are secondary products of the CNO
cycle as well. 

\item[\bf $^{15}$N:] 
\Nquindici\ is quickly destroyed in the CNO cycle, therefore its yields are
always negative (although small) and increasing in modulus with metallicity, as
the efficiency of the CNO cycle increases. 

\item[\bf $^{18}$O:]
\Odiciotto\ is destroyed in the CNO cycle; it is later produced by $\alpha$
capture on \Nitrogen\ during He--burning, but then it is turned into
\Neventidue\ by a new $\alpha$ capture, so that the abundance peak of
\Odiciotto\ is very thin (CM86). Therefore, even when mass loss is so efficient
as to reveal He--burning products on the surface, \Odiciotto\ yields usually
remain negative. 

\item[\bf $^{22}$Ne:]
\Neventidue\ yields are negligible for relatively low masses and/or
metallicities, while when mass loss gets very efficient there is a sensitive
ejection of new \Neventidue\ as a by-product of He--burning remained uncovered
on the surface of WR stars.

\item[\bf $^{20}$Ne:]
\Neon\ yields are always negligible because, although \Neon\ can be produced by
$\alpha$ capture on \Oxygen\ during He--burning, this holds for so advanced
stages that even an extreme mass loss can't peel off the layers where \Neon\
is synthesized. 

\end{description} 


\section{The supernova explosion}
\label{SNae}

To determine the contribution of quasi-massive and 
massive stars in the chemical enrichment of
the ISM we need to know the ejecta of the final SN explosion, which actually
plays the major role. Although mass loss is well known to have substantial
effects in the evolution of massive stars, SN models usually
start from constant mass pre-SN structures. Only few examples
are found in literature of stellar models with mass loss followed up to the
very final stages and the explosion (e.g.\ Woosley \etal 1993, 1995),
but no extended studies exist yet of explosive nucleosynthesis of 
mass losing massive stars
for wide sets of masses and metallicities. To get the global
nucleosynthetic production for these stars one is generally forced to 
link, somehow, pre-SN models evolved including stellar winds and SN models
based on constant mass calculations. Since our tracks stop at the beginning of
C--ignition in the core, here we need to perform such a link as well. 

Under the hypothesis that the stellar core, which drives the explosion
mechanism, evolves uncoupled from the envelope, SN models were
calculated letting He--cores, i.e.\ pure helium stars, 
evolve (Arnett 1978, 1991; Nomoto \& Hashimoto 1986; Thielemann \etal 1996;
Woosley \& Weaver 1986). These bare helium stars  are later
identified with the helium core (\MHe) of stars with initial, total mass $M$.
By means of a suitable relation between the
total mass $M$ of the star and its \MHe, the ejecta obtained for bare
He--cores were then shifted into a scale of stellar masses. 
The envelope layers, and also the wind contribution
in the case of mass losing stellar models, were
added to the ejecta of the corresponding He--core (Chiosi \& Caimmi 1979;
Chiosi 1979; Baraffe \& El Eid 1991). 
Yet, as pointed out by Maeder (1981, 1984, 1992), this method
leads to some inconsistency, because a stellar core does not necessarily
evolve like an
isolated one. Unlike a bare He--core, its size may change during stellar 
evolution: it can grow or recede in time, depending
also on dredge-up episodes and mass loss. 
Table~\ref{MHevartab} shows this effect by displaying \MHe\ values in four 
evolutionary stages of our tracks: (1) when H is exhausted in the centre, (2)
when He ignites in the core, (3) when He is exhausted in the centre and (4)
when C ignites in the core. \MHe\ gets growing in the course of evolution
for less massive and/or low metallicity stars; in stars
where mass loss is very efficient, \MHe\ falls in later stages.

Therefore, it's hard to find an univocal relation between the total initial
mass of the star and \MHe. The meaningful value of \MHe\ for the sake of 
the resulting SN
might be the one corresponding to the end of core He--burning or to
C--ignition, since following stages are so fast that mass loss cannot alter the
structure of the star any more. However, even the ``final'' He--core mass may
be inadequate to fix the resulting explosion. In particular, in 
the WR stage the strong mass loss proceeding after the He--core is uncovered can
alter the evolution of the core and the final nucleosynthesis. Woosley \etal
(1993) follow the evolution of a mass losing 60~\Msol\ star evolving through a
WR stage and ending as a 4.25~\Msol\ bare He--core at the time of explosion.
Such a structure is compared to a 4.25~\Msol\ pure He star evolved without mass
loss. The final thermal and dynamical structure of the two objects results 
very similar, 
but their chemical structure is different: although the mass of the
pre-SN He--core is the same, \MCO\ is different and this affects all subsequent
burning stages, the resulting entropy distribution and the final explosion.

\begin{figure}[t]
\psfig{file=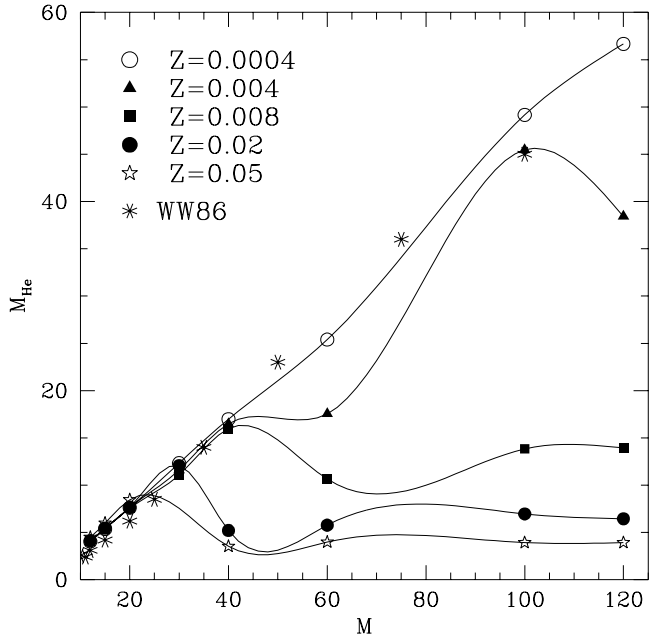,width=9truecm}
\caption{He--core mass vs.\ initial mass relation (in \Msol) for massive stars
of five metallicities. Asterisks indicate He--core masses from 
Woosley \& Weaver 1986, relevant to constant mass pre-SN models}
\label{MHefig}
\end{figure}

\begin{table}[hb]
\caption{Initial masses and corresponding He--core masses}
\label{MHetab}
\begin{small}
\begin{center}
\begin{tabular}{|c|c c c c c|}
\hline
 M  & 	       & 	 & \MHe\   &         &        \\
\hline
    & Z=0.0004 & Z=0.004 & Z=0.008 & Z=0.02  &  Z=0.05\\
\hline
  6 &    1.69  &   1.70  &   1.60  &   1.51  &   1.59 \\
  7 &    2.03  &   2.08  &   1.98  &   1.91  &   2.03 \\
  9 &    2.80  &   2.88  &   2.82  &   2.78  &   2.99 \\
 12 &    4.04  &   4.18  &   4.13  &   4.10  &   4.48 \\
 15 &    5.36  &   5.43  &   5.40  &   5.46  &   5.98 \\
 20 &    7.62  &   7.64  &   7.57  &   7.64  &   8.46 \\
 30 &   12.33  &  11.62  &  11.08  &  12.07  &   ---  \\
 40 &   16.98  &  16.48  &  15.98  &   5.20  &   3.53 \\
 60 &   25.40  &  17.55  &  10.65  &   5.77  &   3.98 \\
100 &   49.16  &  45.47  &  13.83  &   6.96  &   3.92 \\
120 &   56.67  &  38.43  &  13.95  &   6.44  &   3.93 \\
\hline
\end{tabular}
\end{center}
\end{small}
\end{table}

\begin{figure}[t]
\psfig{file=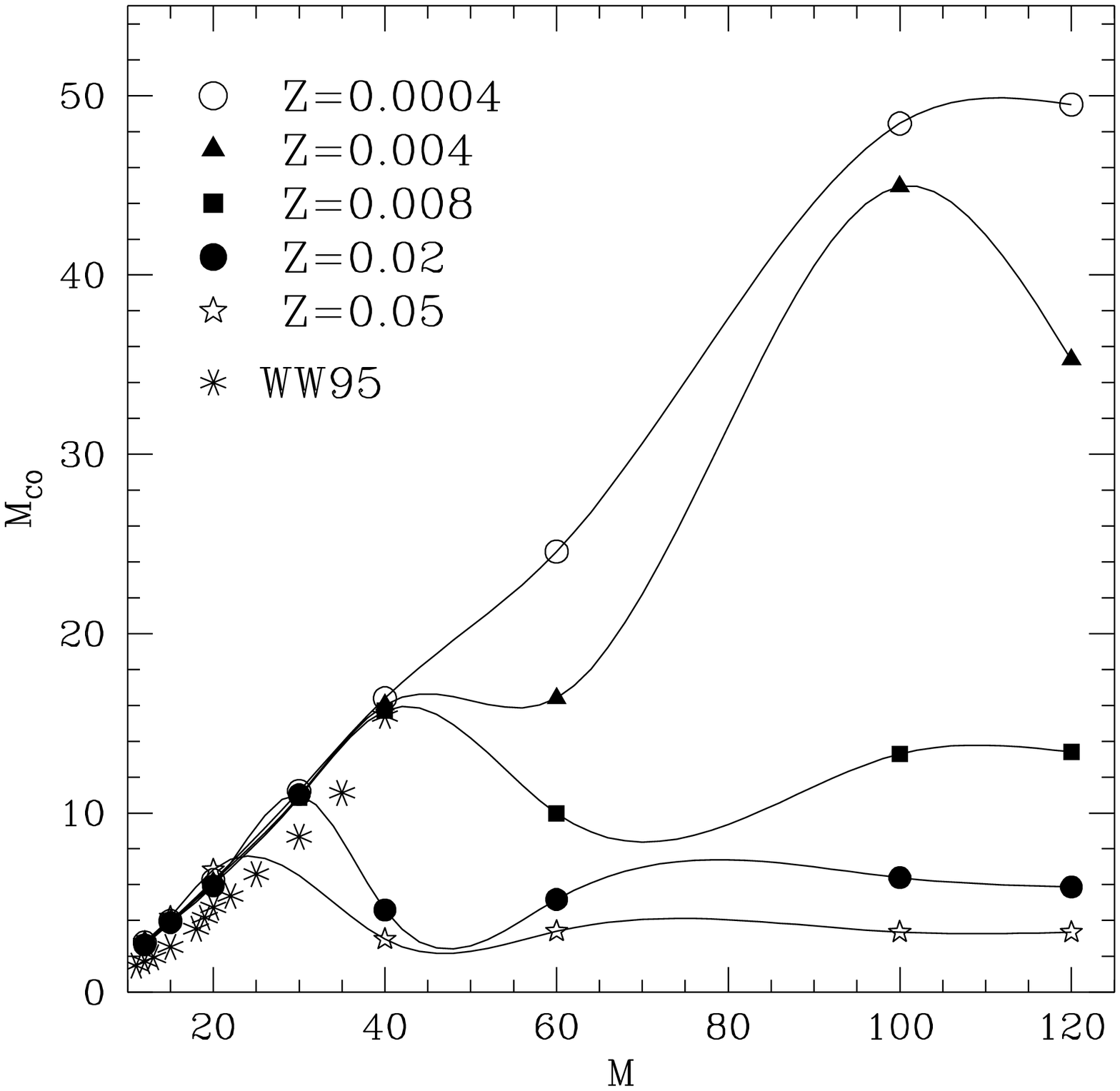,width=9truecm}
\caption{CO--core mass vs.\ initial mass relation (in \Msol) for massive stars
of five metallicities. Asterisks indicate CO--core masses from 
Woosley \& Weaver 1995,
relevant to constant mass pre-SN models}
\label{MCOfig}
\end{figure}

\begin{table}[hb]
\caption{Initial masses and corresponding CO--core masses}
\label{MCOtab}
\begin{small}
\begin{center}
\begin{tabular}{|c|c c c c c|}
\hline
 M  & 	       & 	 & \MCO\   &         &         \\
\hline
    & Z=0.0004 & Z=0.004 & Z=0.008 & Z=0.02  &  Z=0.05 \\
\hline
  6 &    1.11  &   1.11  &   1.05  &   1.01  &   1.00  \\
  7 &    1.28  &   1.20  &   1.22  &   1.10  &   1.16  \\
  9 &    1.80  &   1.83  &   1.75  &   1.68  &   1.71  \\
 12 &    2.78  &   2.82  &   2.73  &   2.64  &   2.79  \\
 15 &    3.97  &   3.96  &   3.89  &   3.89  &   4.19  \\
 20 &    6.25  &   6.15  &   6.02  &   5.94  &   6.84  \\
 30 &   11.23  &  10.95  &  10.86  &  11.02  &   ---   \\
 40 &   16.39  &  16.03  &  15.70  &   4.59  &   2.95  \\
 60 &   24.58  &  16.40  &   9.97  &   5.18  &   3.41  \\
100 &   48.45  &  44.94  &  13.29  &   6.39  &   3.36  \\
120 &   49.51  &  35.28  &  13.40  &   5.87  &   3.35  \\
\hline
\end{tabular}
\end{center}
\end{small}
\end{table}

Therefore, here we prefer to choose \MCO\ as the characterizing parameter for
the resulting SN, as suggested by M92. Indeed, C--burning and following burning
stages are so quick that mass loss hasn't got time enough to influence inner
cores and can be ignored (see also Woosley \etal 1993, 1995). \MCO\ at carbon
ignition well corresponds to \MCO\ at the time of explosion. 
Basing on \MCO\ we perform a link with SN models for all our massive stars,
i.e.\ for our models with M $\geq$ 6~\Msol. 
Even though our models include quiescent mass loss only for M $\geq$ 12~\Msol,
a link with SN models is made also for 6~\Msol~$\leq$~M~$\leq$~12~\Msol,
since in our models with overshooting $M_{\rm up} \sim$~5~\Msol\ and all higher
masses are expected to undergo SN explosion.
In general, we remind that due to overshooting any given $M$ here
corresponds to larger \MHe\ and \MCO\ with respect to ``standard'' models,
and since the final fate of a star is ultimately related to the size of its
core the significant mass ranges implying different kinds of SN are shifted
downward in mass.

\begin{table*}[ht]
\begin{minipage}{18truecm}
\caption{Total ejecta (in \Msol) for quasi-massive  stars of  6 and 7~\Msol, 
for different metallicities}
\label{ejquasimassivetab}
\tiny
\begin{center}
\begin{tabular}{|r |c |c |c |c |c |c |c |c |c |c | c| c| c|}
\multicolumn{14}{c}{{\bf $Z$=0.0004}}\\
\hline
 $M$ & \Hydrogen & \Hetre   & \Helium & \Carbon  & \Ctredici & \Nitrogen &
\Nquindici & \Oxygen  & \Odiciassette & \Odiciotto &  \Neon   & \Neventidue &
$M_{r}$ \\
\hline
  6  &    3.22   & 1.70E-04 &   1.48  & 2.42E-04 & 1.48E-05  & 5.23E-04  &
 1.82E-07  & 8.12E-04 &   3.94E-06    &  1.35E-06  & 1.93E-05 & 5.13E-06    &
 1.30   \\
  7  &    3.68   & 1.45E-04 &   2.00  & 2.67E-04 & 1.71E-05  & 7.25E-04  &
 2.02E-07  & 9.10E-04 &   4.37E-06    &  1.49E-06  & 2.34E-05 & 5.99E-06    & 
 1.30   \\
\hline
\multicolumn{14}{c}{}\\
\multicolumn{14}{c}{{\bf $Z$=0.004}}\\
\hline
 $M$ & \Hydrogen & \Hetre   & \Helium & \Carbon  & \Ctredici & \Nitrogen &
\Nquindici & \Oxygen  & \Odiciassette & \Odiciotto &  \Neon   & \Neventidue &
$M_{r}$ \\
\hline
  6  &     3.16  & 1.80E-04 &   1.52  & 2.80E-03 & 1.65E-04  & 4.72E-03  &
 2.04E-06  & 8.17E-03 &   3.99E-05    &  1.56E-05  & 1.93E-04 & 4.73E-08    &
 1.30   \\
  7  &     3.61  & 1.64E-04 &   2.07  & 3.22E-03 & 1.99E-04  & 6.42E-03  &
 2.34E-06  & 9.35E-03 &   3.93E-05    &  1.80E-05  & 2.34E-04 & 5.45E-08    &
 1.30   \\
\hline
\multicolumn{14}{c}{}\\
\multicolumn{14}{c}{{\bf $Z$=0.008}}\\
\hline
 $M$ & \Hydrogen & \Hetre   & \Helium & \Carbon  & \Ctredici & \Nitrogen &
\Nquindici & \Oxygen  & \Odiciassette & \Odiciotto &  \Neon   & \Neventidue &
$M_{r}$ \\
\hline
  6  &    3.15   & 1.97E-04 &   1.51  & 5.97E-03 & 3.30E-04  & 8.53E-03  &
 4.43E-06  & 1.69E-02 &   1.09E-04    &  3.33E-05  & 3.86E-04 & 1.03E-04    &
 1.30   \\
  7  &    3.59   & 1.80E-04 &   2.06  & 6.73E-03 & 3.98E-04  & 1.21E-02  &
 4.93E-06  & 1.91E-02 &   1.25E-04    &  3.79E-05  & 4.68E-04 & 1.20E-04    &
 1.30   \\
\hline
\multicolumn{14}{c}{}\\
\multicolumn{14}{c}{{\bf $Z$=\Zsol=0.02}}\\
\hline
 $M$ & \Hydrogen & \Hetre   & \Helium & \Carbon  & \Ctredici & \Nitrogen &
\Nquindici & \Oxygen  & \Odiciassette & \Odiciotto &  \Neon   & \Neventidue &
$M_{r}$ \\
\hline
  6  &    3.02   & 2.46E-04 &   1.58  & 1.60E-02 & 8.49E-04  & 1.84E-02  &
 1.18E-05  & 4.38E-02 &   5.24E-04    &  8.83E-05  & 9.64E-04 & 2.57E-04    &
 1.30   \\
  7  &    3.44   & 2.34E-04 &   2.15  & 1.79E-02 & 1.00E-03  & 2.72E-02  &
 1.32E-05  & 4.97E-02 &   5.71E-04    &  9.89E-05  & 1.17E-03 & 2.99E-04    &
 1.30   \\
\hline
\multicolumn{14}{c}{}\\
\multicolumn{14}{c}{{\bf $Z$=0.05}}\\
\hline
 $M$ & \Hydrogen & \Hetre   & \Helium & \Carbon  & \Ctredici & \Nitrogen &
\Nquindici & \Oxygen  & \Odiciassette & \Odiciotto &  \Neon   & \Neventidue &
$M_{r}$ \\
\hline
  6  &    2.49   & 3.37E-04 &   1.97  & 4.03E-02 & 2.08E-03  & 4.39E-02  &
 3.02E-05  & 0.111    &   1.66E-03    &  2.21E-04  & 2.41E-03 & 6.42E-04    &
 1.30   \\
  7  &    2.88   & 3.59E-04 &   2.53  & 4.61E-02 & 2.51E-03  & 6.13E-02  &
 3.38E-05  & 0.129    &   1.99E-03    &  2.53E-04  & 2.92E-03 & 7.49E-04    &
 1.30   \\
\hline
\end{tabular}
\end{center}
\normalsize
\end{minipage}
\end{table*}


\subsection{The adopted $M_{He}(M)$ and $M_{CO}(M)$ relations}

\noindent
In Fig.~\ref{MHefig} we plot the relation between the initial mass $M$ and the
corresponding He--core mass \MHe\ at the end of the evolutionary track. For the
sake of comparison, asterisks indicate \MHe\ values given by Woosley \& Weaver
1986 for pre--SN models calculated at constant mass. Up to $\sim$ 30~\Msol\
the He--cores are more massive in our models due to overshooting, while for
larger masses the effects of mass loss dominate. Indeed, in the range 10 $\div$
30~\Msol\ stellar evolution is more sensitive to the treatment of
convection, while for larger masses the assumptions about mass loss have an
overwhelming influence (CM86). 

In Fig.~\ref{MCOfig} we plot \MCO\ at the end of the track
versus $M$; in our models, we define \MCO\ as the mass interior to the outer
edge of the He--burning region. Asterisks indicate \MCO\ values deduced from
the recent SN models by WW95 (see App.~A), calculated from pre--SN
structures with no mass loss and with no overshooting.

In Tab.~\ref{MHetab} and Tab.~\ref{MCOtab} we list the values of \MHe\ and of
\MCO\ respectively, for our quasi-massive and massive stars 
($M>M_{up}=$5~\Msol). In models with
overshooting all these stars eventually give SN\ae,  and
their core masses are prerequisite to characterize the outcoming supernova and
its ejecta. 


\subsection{Electron capture supernov\ae}

\noindent
Quasi-massive stars (8~$\div$~10~\Msol\ in standard models and 
6~$\div$~8~\Msol\ in models with overshoot) generally
develop a degenerate O-Ne-Mg core after C--burning and eventually
explode as ``electron capture'' supernov\ae\ (EC SN\ae), 
leaving a neutron star of
$\sim$1.3~\Msol\ as a remnant and expelling very few heavy elements (Nomoto
1984, 1987; Mayle \& Wilson 1988). Recent calculations seem to confirm that a
degenerate O-Ne-Mg core reaching 1.375~\Msol, the limit mass for electron
captures on \Magnesium\ and \Neon, develops density conditions extreme enough
that it collapses to a neutron star, rather than undergo a thermonuclear
explosion due to disruptive O-Ne burning in degenerate material (Gutierrez \etal
1996). 

But the final stages of stars in this mass range are still debated. These
stars develop a double--shell structure similar to that of lower mass stars,
and experience a thermally pulsing, mass losing phase as well. Nomoto's models
didn't include thermal pulses nor mass loss, but the detailed evolution of a
model 10~\Msol\ star has been recently followed by Garcia--Berro \& Iben (1994),
Ritossa \etal (1996). Their models suggest that such a star might be
unable to reach the limit core mass for electron capture and explode in an EC
SN: it might rather get rid of the overlying layers through a superwind phase
typical
of TP-AGB stars and result in a massive (1.26~\Msol) O-Ne-Mg white dwarf.

By comparing the size of our stellar cores with those of Nomoto (1984, 1987)
and Ritossa \etal (1996), in our models with convective
overshooting such a behaviour is expected between 6 and 8~\Msol\ (\MCO\ $\sim$
1.0 $\div$ 1.3~\Msol\ at the beginning of core C--burning). For our models in
this mass range, we assume that the inner 1.3~\Msol\ remain locked in the
remnant, either a neutron star or a white dwarf, while the overlying layers are
expelled, either by a final explosion or by a superwind during the TP-AGB
phase. The resulting ejecta  for this mass range are presented in
Tab.~\ref{ejquasimassivetab}. 

We thus neglect the products of explosive nucleosynthesis in
case of SN explosion, which is reasonable since EC SN\ae\ are believed to
produce negligible amounts of heavy elements (not more than 0.002~\Msol, Mayle
\& Wilson 1988). More important, we are neglecting the effects of thermal
pulses and of the III dredge-up in the nucleosynthetic yields of SAGB stars
(SAGB = Super-Asymptotic Giant Branch, see Garcia-Berro \& Iben 1994, Ritossa
\etal 1996). No full calculations of stellar yields for this kind of stars
exist yet in literature, neither for standard models nor for models with
overshooting. Further investigation of nucleosynthesis and mass loss in this
mass range should be done. 

\begin{table*}[t]
\begin{minipage}{18truecm}
\caption{Remnant mass $M_{r}$ and ejecta $E_{i}^{CO}$ (in \Msol) of CO--cores
for different elements, as deduced from WW95}
\label{WMCOej}
\scriptsize
\begin{center}
\begin{tabular}{|r|c|c|c|c|c|c|c|c|c|c|}
\multicolumn{11}{c}{$Z$=\Zsol (case S in WW95)} \\
\hline
\MCO & $M_{r}$ & \Carbon & \Nquindici & \Oxygen & \Neon & \Magnesium &
\Silicon & \Sulfur & \Calcium & \Fe \\
\hline
1.49 & 1.26 & 3.13E-02 & 1.29E-05 & 6.47E-02 & 1.58E-02 & 4.34E-03 &
1.55E-02 & 5.94E-03 & 7.50E-04 & 6.93E-02\\
1.72 & 1.30 & 5.59E-02 & 3.77E-05 & 0.130 & 6.55E-03 & 2.92E-03 & 8.42E-02
& 7.16E-02 & 1.38E-02 & 4.34E-02\\
1.95 & 1.40 & 8.95E-02 & 4.01E-06 & 0.192 & 2.67E-02 & 1.07E-02 & 5.13E-02
& 2.16E-02 & 2.95E-03 & 0.133\\
2.53 & 1.40 & 0.134 & 9.20E-05 & 0.591 & 9.08E-02 & 2.03E-02 & 0.102 & 
5.85E-02 & 1.03E-02 & 0.115\\
3.54 & 1.70 & 0.215 & 9.34E-05 & 1.02 & 0.254 & 4.78E-02 & 0.128 & 4.89E-02
& 5.95E-03 & 6.59E-02\\
4.16 & 1.90 & 0.250 & 6.83E-05 & 1.32 & 8.20E-02 & 1.75E-02 & 0.267 & 0.123
& 1.26E-02 & 0.100\\
4.73 & 2.00 & 0.178 & 8.48E-05 & 1.83 & 8.03E-02 & 2.34E-02 & 0.278 & 0.146
& 1.34E-02 & 8.84E-02\\
5.37 & 1.90 & 0.203 & 1.10E-04 & 2.26 & 4.33E-02 & 3.30E-02 & 0.345 & 0.165
& 1.63E-02 & 0.205\\
6.59 & 1.90 & 0.281 & 1.64E-04 & 3.12 & 3.64 & 9.65E-02 & 0.303 & 0.134 &
1.56E-02 & 0.129\\
8.67 & 4.10 & 0.241 & 9.62E-05 & 3.51 & 0.346 & 0.201 & 8.38E-02 & 5.35E-03
& 3.25E-05 & 4.80E-05\\
11.12 & 7.20 & 0.253 & 5.13E-05 & 2.92 & 0.476 & 0.111 & 2.60E-02 & 1.34E-03
& 2.97E-05 & 0\\
15.41 & 12.30 & 0.267  & 2.07E-05 & 2.21 & 0.408 & 5.53E-02 & 8.64E-03 &
4.62E-04 & 6.95E-06 &0\\
\hline
\multicolumn{11}{c}{} \\
\multicolumn{11}{c}{$Z$=0.1 \Zsol (case P in WW95)} \\
\hline
\MCO & $M_{r}$ & \Carbon & \Nquindici & \Oxygen & \Neon & \Magnesium &
\Silicon & \Sulfur & \Calcium & \Fe \\
\hline
1.81 & 1.30 & 8.66E-02 & 9.74E-06 & 0.137 & 1.04E-02 & 4.95E-03 & 2.95E-02
& 1.32E-02 & 2.78E-03 & 0.177\\     
2.01 & 1.20 & 0.106 & 1.65E-05 & 0.282 & 4.04E-02 & 1.63E-02 & 6.43E-02 &
3.11E-02 & 5.16E-03 & 0.180\\          
2.52 & 1.40 & 0.139 & 4.79E-05 & 0.546 & 7.26E-02 & 2.11E-02 & 7.03E-02 &
3.35E-02 & 5.83E-03 & 0.194\\          
3.43 & 1.60 & 0.222 & 5.82E-05 & 0.983 & 0.204 & 7.71E-02 & 0.126 & 4.74E-02
& 7.75E-03 & 0.136\\          
4.32 & 1.90 & 0.242 & 7.17E-05 & 1.51 & 9.36E-02 & 5.07E-02 & 0.231 & 0.128
& 1.86E-02 & 0.121\\          
5.11 & 2.00 & 0.228 & 1.01E-04 & 2.11 & 3.32E-02 & 1.64E-02 & 0.325 & 0.183
& 2.27E-02 & 0.123\\          
6.06 & 1.90 & 0.263 & 1.25E-04 & 2.89 & 4.13E-02 & 2.95E-02 & 0.394 & 0.216
& 2.16E-02 & 0.202\\          
8.06 & 2.60 & 0.299 & 1.28E-04 & 4.10 & 0.634 & 0.319 & 7.06E-02 & 1.95E-03
& 5.06E-05 & 0\\     
10.55 & 6.50 & 0.314 & 5.89E-05 &  3.08 & 0.488 & 0.144 & 2.28E-02 & 8.32E-04
& 5.06E-05 & 0\\     
12.65 & 8.90 & 0.332 & 3.55E-05 & 2.70 & 5.99 & 7.61E-02 & 6.02E-03 &
5.27E-04 & 5.51E-05 & 0\\     
\hline
\multicolumn{11}{c}{} \\
\multicolumn{11}{c}{$Z$=0.01 \Zsol (case T in WW95)} \\
\hline
\MCO & $M_{r}$ & \Carbon & \Nquindici & \Oxygen & \Neon & \Magnesium &
\Silicon & \Sulfur & \Calcium & \Fe \\
\hline
1.80 & 1.30 & 8.98E-02 & 6.65E-06 & 0.141 & 9.46E-03 & 5.75E-03 & 3.12E-02
& 1.40E-02 & 2.61E-03 & 0.152\\
2.01 & 1.40 & 0.109 & 1.01E-05 & 0.209 & 2.48E-02 & 1.05E-02 & 4.14E-02 &
2.18E-02 & 4.89E-03 & 0.191\\
2.42 & 1.50 & 0.149 & 2.14E-05 & 0.422 & 4.00E-02 & 2.12E-02 & 7.32E-02 &
3.46E-02 & 6.46E-03 & 0.171\\
3.18 & 1.50 & 0.194 & 4.76E-05 & 0.951 & 0.185 & 3.58E-02 & 0.101 & 5.60E-02
& 9.94E-03 & 0.135\\
4.31 & 1.90 & 0.208 & 8.71E-05 & 1.62 & 5.46E-02 & 1.66E-02 & 0.244 & 0.148
& 1.84E-02 & 7.30E-02\\
4.83 & 1.90 & 0.248 & 9.57E-05 & 1.94 & 4.11E-02 & 5.73E-02 & 0.311 & 0.148
& 1.93E-02 & 9.88E-02\\
5.84 & 1.70 & 0.279 & 1.27E-04 & 2.83 & 4.37E-02 & 3.95E-02 & 0.385 & 0.205
& 1.94E-02 & 0.187\\
8.10 & 3.10 & 0.315 & 1.35E-04 & 3.98 & 0.357 & 0.255 & 0.116 & 3.24E-03 & 0
& 0\\
10.20 & 5.20 & 0.343 & 8.29E-05 & 3.67 & 0.711 & 0.180 & 1.78E-02 & 1.39E-04
& 0 & 0\\
12.39 & 8.90 & 0.340 & 3.82E-05 & 2.61 & 0.532 & 7.41E-02 & 3.60E-03 & 0 & 0
&  0\\
\hline
\end{tabular}
\end{center}
\normalsize
\end{minipage}
\end{table*}


\subsection{Iron--core collapse supernov\ae}
\label{ironSNae}

\noindent
In the range $M>$8~\Msol, our models go through the whole sequence of nuclear
burning stages and eventually explode after iron--core collapse. Very few
models explode as ``pair creation'' SN\ae\ (PC SN\ae) before building an iron
core (see \S~\ref{PCSN}). To get the final outcome of iron core
collapse SN\ae, we link our models to the recent SN models by WW95, which cover
a wide range of masses and metallicities (from 11 to 40~\Msol\ and from $Z$=0
to $Z$=\Zsol=0.02). In WW95, pre-SN models are calculated at constant mass and
convective regions are treated adopting the Schwarzschild criterion and
semiconvection; the models are exploded by letting a piston move outward,
tuning it so
that the final ejecta gain a typical kinetic energy at infinity (1.2 $\times$
10$^{51}$ erg, although for M $>$ 30~\Msol\ higher energies are also considered
in cases B and C). 

\begin{figure*}
\psfig{file=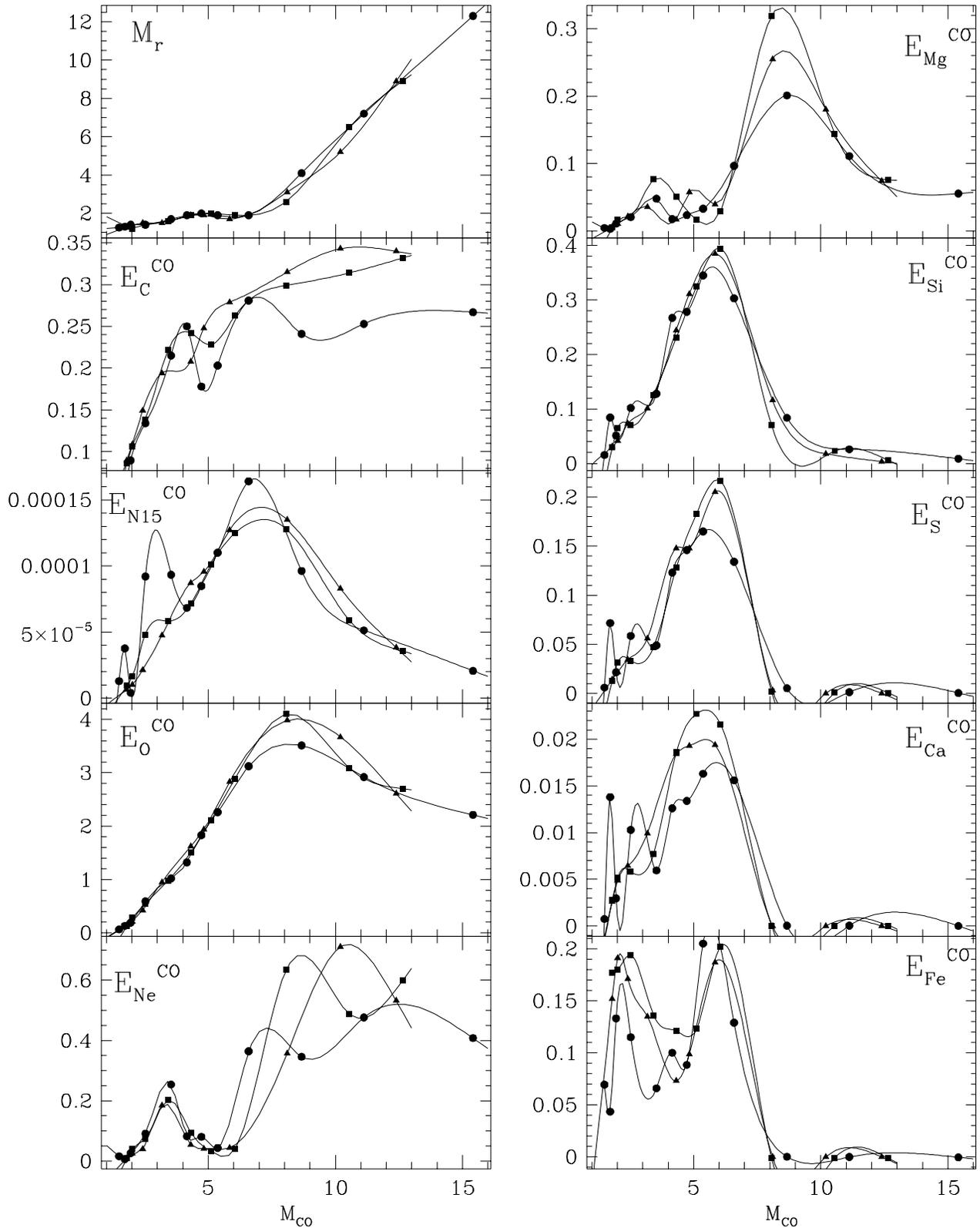,width=18truecm}
\begin{minipage}{18truecm}
\caption{Remnant mass and ejecta of the CO--core as functions of \MCO\ for three
different metallicities, as deduced from WW95 (see text). Circles: case
$Z$=\Zsol; squares: case $Z$=0.1 \Zsol; triangles: case $Z$=0.01 \Zsol.}
\label{WMCOfig}
\end{minipage}
\end{figure*}

WW95 give the total ejecta, i.e. the ejecta relevant to the whole star. In
order to link the outcome of these SN\ae\ to our models, we need to
discriminate the contribution of the CO core alone in the models of WW95.
We consider the following elemental species: H, $^{4}$He, $^{12}$C, $^{13}$C,
$^{14}$N, $^{15}$N, $^{16}$O, $^{17}$O, $^{18}$O, $^{20}$Ne, $^{22}$Ne,
$^{24}$Mg, $^{28}$Si, $^{32}$S, $^{40}$Ca, $^{56}$Fe. By means of the method
described in App.~A, for  various metallicities referred to in WW95 
we derive suitable relations between the mass of the CO--core
and the amount of different elements expelled by the core in the explosion.
These $E_{i}^{CO}$---\MCO\ relations
are shown in Tab.~\ref{WMCOej} and in Fig.~\ref{WMCOfig}. We notice quite a
regular behaviour of the remnant mass and of the ejecta of most elements as a
function of \MCO, even with varying initial metallicity. This confirms that
\MCO\ is a good parameter for our link (but see also App.~B). 

By means of a linear interpolation with respect to \MCO\ and to metallicity in
such relations, we get the ejecta $E_{i}^{CO}$ from the CO--cores of {\it our}
models. Then we add (1) the composition of the overlying layers still left in
our models at the time of explosion, and (2) the ejecta $E_{i}^{wind}$ from the
stellar wind (Tab.~4), to get our total ejecta as the sum:

\begin{equation}
\label{totalejecta}
E_{i}=E_{i}^{CO}+E_{i}^{over}+E_{i}^{wind}
\end{equation}

\subsection{Pair creation supernov\ae}
\label{PCSN}

Massive stars with larger cores (\MHe $\gsim$ 35~\Msol) are known to undergo
pair creation events during O--burning. This ``pair creation'' instability
(Fowler \& Hoyle 1964) may lead either to violent pulsational instability
with ejection of some external layers and later iron core collapse (35~\Msol
$\lsim$ \MHe $\lsim$ 60~\Msol), or to complete thermonuclear explosion
(60~\Msol $\lsim$ \MHe $\lsim$ 110~\Msol), or to collapse to black hole after
He--exhaustion (\MHe $\gsim$ 110~\Msol); for details refer to Woosley (1986).
In terms of CO--core masses, up to \MCO $\sim$ 57~\Msol\ we expect pulsational
instability, in the range \MCO=57 $\div$ 106~\Msol\ a complete disruption is
supposed to occur, while larger cores collapse to black holes.

Because of mass loss, in our stellar models we seldom face such large cores and
``pair creation'' supernov\ae\ (PC SN\ae) are likely to occur only in the case
of the higher mass, low metallicity stars ($M$=100--120~\Msol, 
$Z$=0.0004--0.004), as mass loss in not very efficient. These stars have
He--cores in the range 35~$\div$~60~\Msol\ (Tab.~\ref{MHetab}); following
Woosley (1986) we assume that they expel some of the external layers when pair
instability develops, while whatever remains bound  finally collapses to a
black hole. In order to establish what is to be expelled in our CO--cores and
what is to be locked in the remnant, we interpolate between the cases
\MHe=45~\Msol\ and 55~\Msol\ of Woosley (1986), corresponding to 
\MCO~$\sim$~41~\Msol\ and \MCO~$\sim$~52~\Msol, respectively. 
For \MCO=41~\Msol\ only
overlying layers are expelled, while for the \MCO=52~\Msol\ the inner 27~\Msol\
remain bound. Detailed ejecta of this case, as deduced from Tab.~18 of Woosley
(1986), are shown in Tab.~\ref{PCSNtab}. 
The total ejecta of our stellar models experiencing a PC SN event are
calculated according to Eq.~(\ref{totalejecta}).

\begin{table}[hb]
\caption{1$^{st}$ column: \MHe\ values of PC SN\ae\ models of Woosley (1986).
2$^{nd}$ column: corresponding CO-core masses. 3$^{rd}$ to 6$^{th}$ column:
ejecta from the CO--core for significant elemental species. 7$^{th}$ column:
remnant mass. All quantities are expressed in \Msol}
\label{PCSNtab}
\scriptsize
\begin{center}
\begin{tabular}{|c|c|c|c|c|c|c|}
\hline
 \MHe  &  \MCO & \Carbon & \Oxygen & \Neon & \Magnesium  & $M_{r}$ \\
\hline
  45 &    41  &   0 	&   0	  &   0	  &    0	&  41  \\
  55 &    52  &   1.1	&   21    &   1.5 &    1.1	&  27  \\
\hline
\end{tabular}
\end{center}
\normalsize
\end{table}


\section{The total ejecta of massive stars}

\begin{figure}[t]
\psfig{file=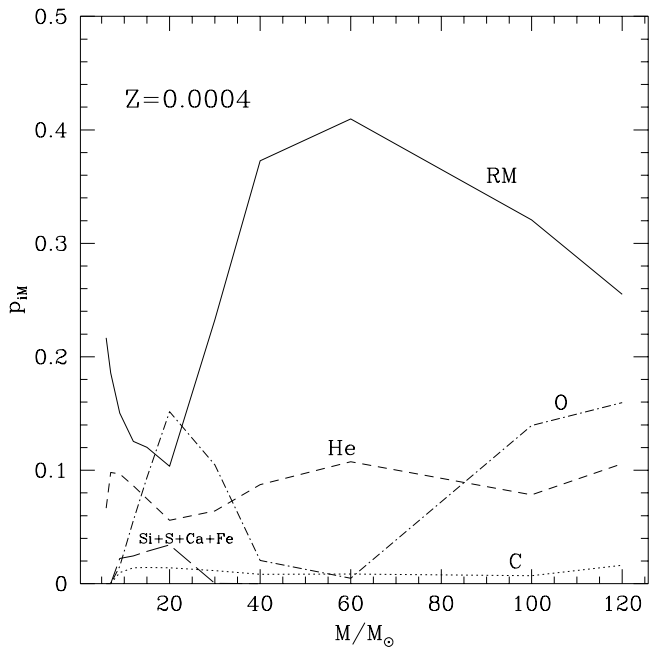,width=9truecm}
\caption{Fractional remnant mass RM=$M_{r}/M$ and total yields $p_{iM}$ for
some elements vs.\ the initial stellar mass $M$, for the $Z$=0.0004 set.
Solid line: RM; dashed line: helium yields; dotted line: carbon yields;
dash-dotted line: oxygen yields;long-dashed line: heavy elements yields.}
\label{yplotZ0004}
\end{figure}

\begin{figure}[ht]
\psfig{file=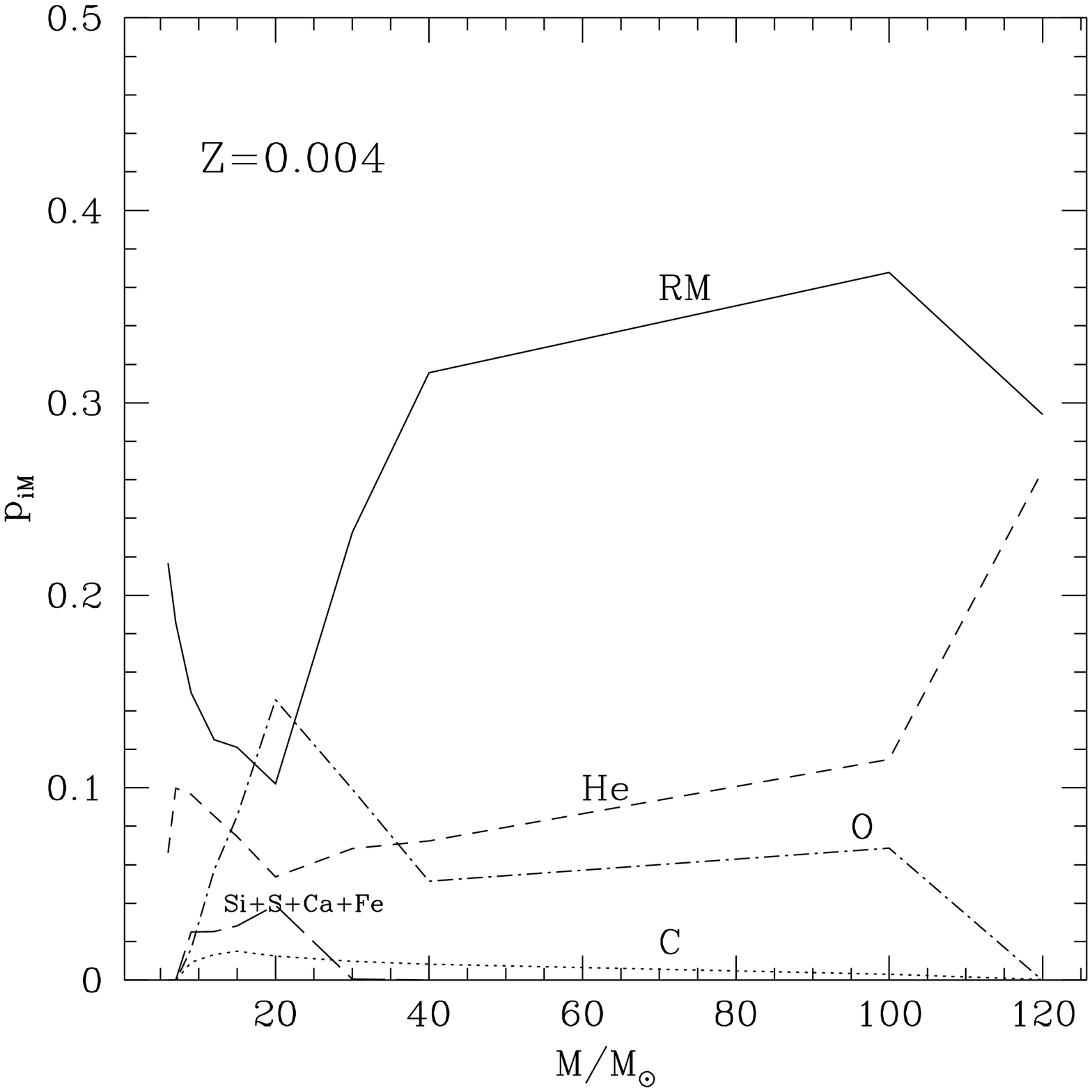,width=9truecm}
\caption{Same as previous figure, for the $Z$=0.004 set}
\label{yplotZ004}
\end{figure}

\begin{figure}[ht]
\psfig{file=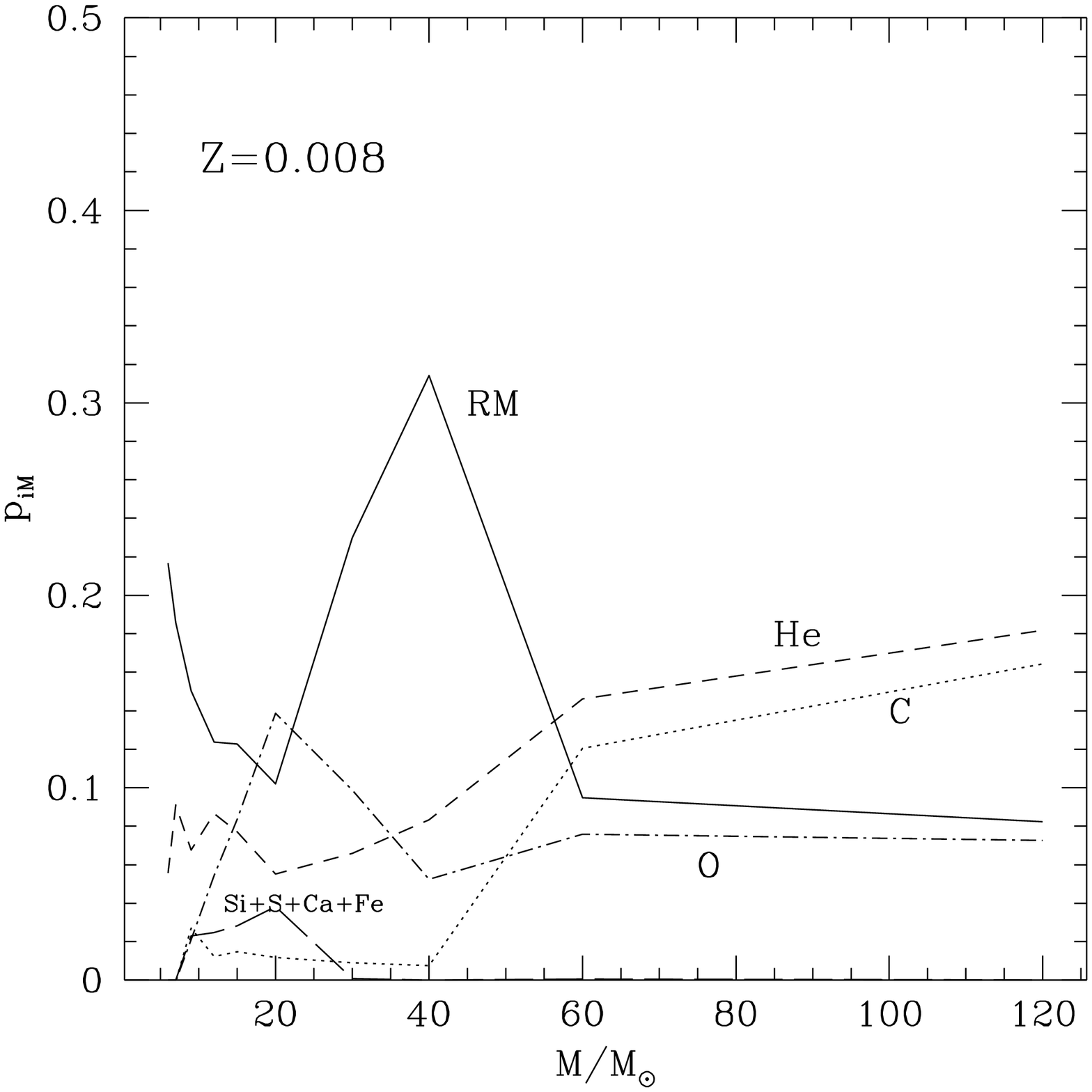,width=9truecm}
\caption{Same as previous figures, for the $Z$=0.008 set}
\label{yplotZ008}
\end{figure}

\begin{figure}[ht]
\psfig{file=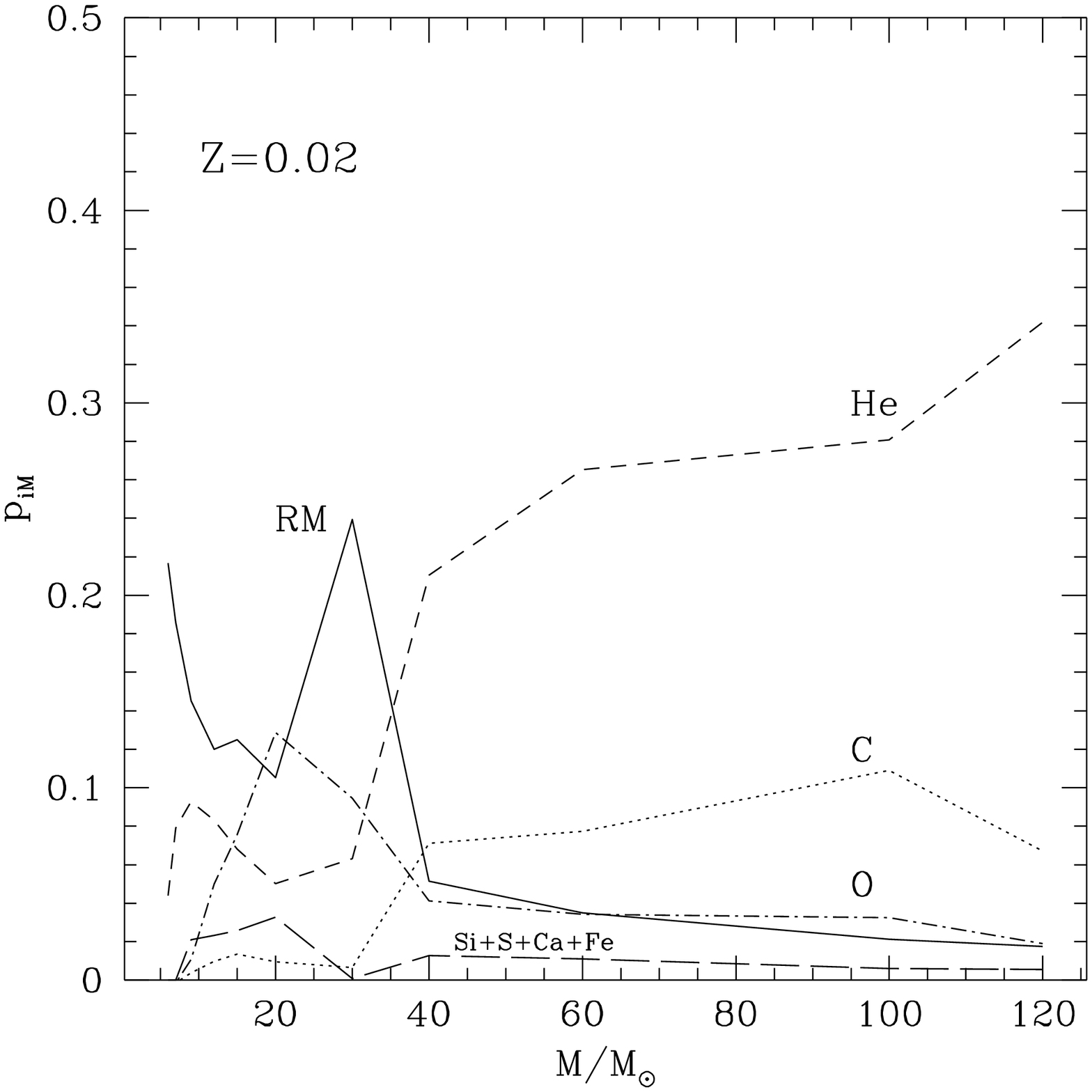,width=9truecm}
\caption{Same as previous figure, for the $Z$=0.02 set}
\label{yplotZ02}
\end{figure}

\begin{figure}[ht]
\psfig{file=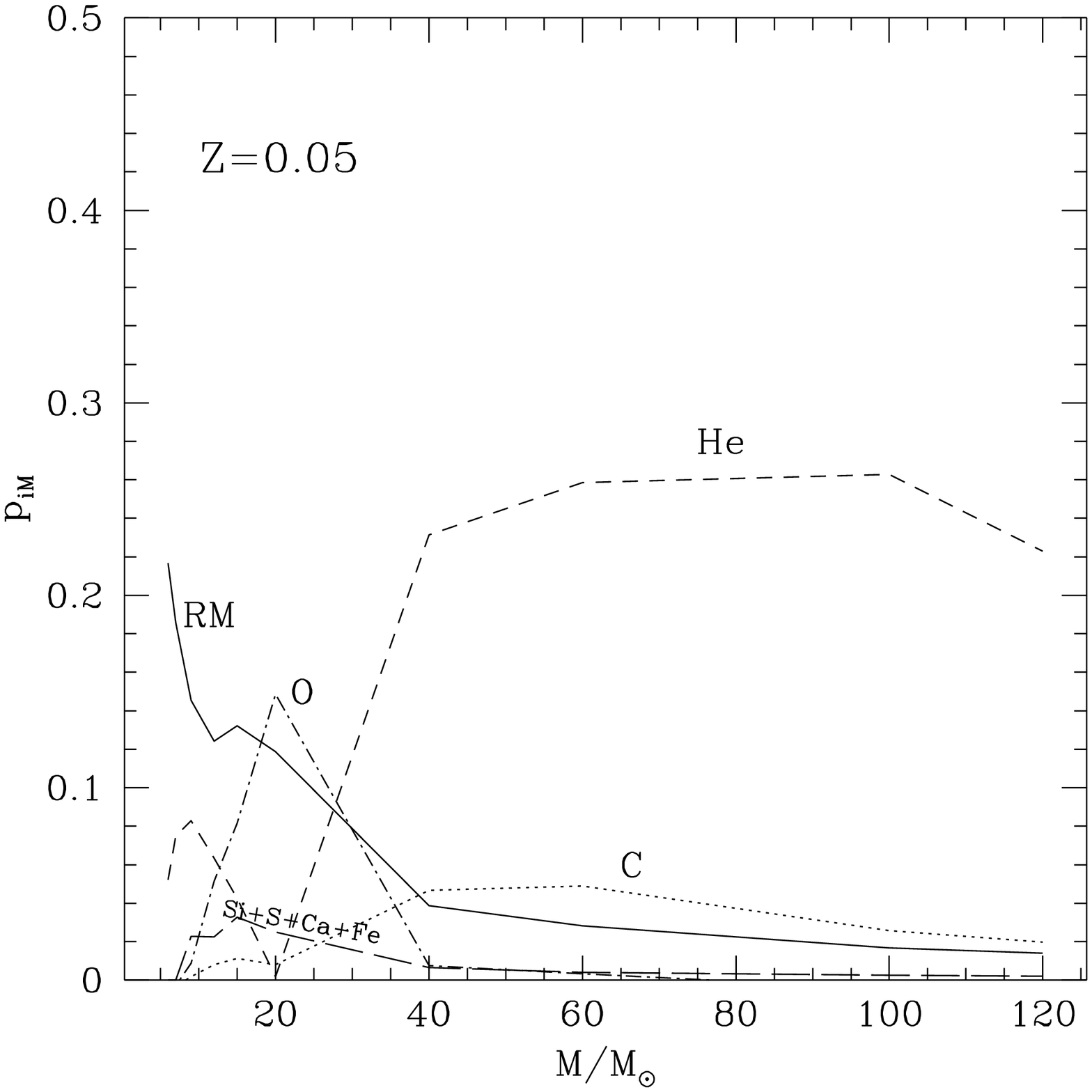,width=9truecm}
\caption{Same as previous figures, for the $Z$=0.05 set}
\label{yplotZ05}
\end{figure}

\begin{table*}[ht]
\begin{minipage}{18truecm}
\caption{Total ejecta (in \Msol) for model stars between 9 and 120~\Msol\ for
different metallicities}
\label{ejmassivetab}
\tiny
\begin{center}
\begin{tabular}{|r |c |c |c |c |c |c |c |c |c |c |c|}
\multicolumn{12}{c}{{\bf $Z$=0.0004}}\\
\hline
  M & \Hydrogen & \Hetre & \Helium & \Carbon & \Ctredici & \Nitrogen &
\Nquindici & \Oxygen & \Odiciassette & \Odiciotto &\\
\hline
&&&    &\Neon & \Neventidue & \Magnesium & \Silicon & \Sulfur & \Calcium & \Fe & $M_{r}$ \\
\hline
   9   & 4.57  & 1.22E-04 & 2.63  & 8.97E-02 & 2.27E-05 &
9.31E-04& 7.16E-06 & 0.140 & 4.78E-06 & 2.43E-05&\\
&&&    & 9.43E-03 & 2.96E-05 & 5.61E-03 & 3.08E-02 & 1.38E-02 & 2.60E-03 & 0.155     & 1.35  \\
  12   & 5.77  & 9.50E-05 & 3.45  & 0.170     & 2.91E-05 & 1.28E-03
& 3.58E-05 & 0.671 & 5.09E-06 & 2.97E-05&\\
&&&    & 0.108 & 3.82E-05 & 2.91E-02 & 8.63E-02 & 4.39E-02 & 7.91E-03 & 0.157     & 1.50  \\
  15   & 6.86  & 8.02E-05 & 4.16  & 0.214     & 3.58E-05 & 1.50E-03 & 7.45E-05 & 1.41  & 5.80E-06 & 7.57E-05&\\
&&&    & 9.92E-02 & 6.44E-05 & 2.68E-02 & 0.200     & 0.118     & 1.57E-02 & 9.59E-02 & 1.80  \\
  20   & 8.51  & 5.83E-05 & 5.24  & 0.284     & 4.21E-05 & 2.18E-03 & 1.29E-04 & 3.03  & 5.66E-06 & 3.01E-06&\\
&&&    & 9.99E-02 & 5.64E-05 & 7.59E-02 & 0.340     & 0.172     & 1.63E-02 & 0.157     & 2.07  \\
  30   & 11.54 & 2.59E-05 & 7.22  & 0.350     & 4.49E-05 & 3.33E-03 & 6.14E-05 & 3.15  & 1.03E-05 & 1.34E-04&\\
&&&    & 0.616     & 2.25E-04 & 0.130     & 1.21E-02 & 2.90E-04 & 2.42E-05 & 4.39E-04 & 6.98  \\
  40   & 14.33 & 1.73E-05 & 9.26  & 0.339     & 5.00E-05 & 4.86E-03 & 6.15E-07 & 0.824 & 1.71E-05 & 3.60E-06&\\
&&&    & 0.271     & 9.69E-05 & 5.43E-02 & 3.08E-04 & 1.87E-04 & 2.83E-05 & 5.52E-04 & 14.91 \\
  60   & 20.01 & 1.09E-05 & 14.60 & 0.510     & 8.04E-05 & 7.57E-03 & 8.86E-07 & 0.288 & 3.08E-05 & 3.31E-06&\\
&&&    & 1.14E-03 & 4.64E-04 & 3.34E-04 & 4.63E-04 & 2.81E-04 & 4.24E-05 & 8.29E-04 & 24.58 \\
 100   & 28.07 & 5.20E-05 & 23.46 & 0.719     & 2.20E-04 & 1.08E-02 & 1.21E-06 & 13.95 & 1.52E-05 & 7.60E-06&\\
&&&    & 1.00     & 2.12E-04 & 0.718     & 6.73E-0.3 & 4.08E-0.3 & 6.18E-0.4 & 1.21E-03 & 32.0.6 \\
 120   & 32.80 & 5.05E-05 & 33.25 & 1.97      & 1.77E-04 & 1.58E-02 & 1.86E-06 & 19.16 & 5.69E-05 & 8.80E-06&\\
&&&    & 1.33     & 3.03E-04 & 0.867     & 9.21E-04 & 5.58E-04 & 8.44E-05 & 1.65E-03 & 30.60 \\
\hline
\multicolumn{12}{c}{}\\
\multicolumn{12}{c}{{\bf $Z$=0.004}}\\
\hline
  M & \Hydrogen & \Hetre & \Helium & \Carbon & \Ctredici & \Nitrogen & \Nquindici & \Oxygen & \Odiciassette & \Odiciotto &\\
\hline
&&& &\Neon & \Neventidue & \Magnesium & \Silicon & \Sulfur & \Calcium & \Fe & $M_{r}$ \\
\hline
   9   & 4.44  & 1.40E-04 & 2.71  & 9.03E-02 & 2.49E-04 & 8.60E-03 & 1.45E-05 & 0.162 & 4.00E-05 & 2.19E-05 &\\
&&&    & 1.54E-02 & 2.94E-04 & 6.65E-03 & 3.74E-02 & 1.90E-02 & 3.71E-03 & 0.169     & 1.35  \\
  12   & 5.59  & 1.15E-04 & 3.54  & 0.171     & 3.23E-04 & 1.16E-02 & 5.92E-05 & 0.707 & 6.26E-05 & 4.81E-04 &\\
&&&    & 0.121     & 3.86E-04 & 3.92E-02 & 9.22E-02 & 4.08E-02 & 6.86E-03 & 0.169     & 1.50  \\
  15   & 6.71  & 1.00E-04 & 4.28  & 0.240     & 3.73E-04 & 1.52E-02 & 7.12E-05 & 1.31  & 8.36E-05 & 2.95E-05 &\\
&&&    & 0.141     & 4.53E-04 & 5.86E-02 & 0.194     & 9.69E-02 & 1.40E-02 & 0.125     & 1.82 \\
  20   & 8.40  & 1.07E-04 & 5.38  & 0.271     & 4.89E-04 & 1.87E-02 & 1.33E-04 & 2.95  & 7.66E-05 & 3.98E-05 &\\
&&&    & 9.31E-02 & 5.68E-04 & 4.77E-02 & 0.374     & 0.200     & 2.02E-02 & 0.192     & 2.04 \\
  30   & 11.39 & 8.51E-05 & 7.58  & 0.319     & 4.35E-04 & 3.13E-02 & 5.96E-05 & 3.03  & 3.20E-04 & 4.12E-05 &\\
&&&    & 0.510     & 7.82E-04 & 0.131     & 2.32E-02 & 2.38E-03 & 2.77E-04 & 4.46E-03 & 6.98 \\
  40   & 14.41 & 7.77E-05 & 9.46  & 0.361     & 5.72E-04 & 4.06E-02 & 8.17E-06 & 2.12  & 3.60E-04 & 3.31E-05 &\\
&&&    & 0.743     & 9.84E-04 & 0.223     & 3.13E-03 & 1.90E-03 & 2.87E-04 & 5.61E-03 & 12.6 \\
 100   & 28.17 & 1.26E-04 & 26.66 & 0.375     & 9.20E-04 & 0.124     & 1.52E-05 & 6.98  & 4.04E-04 & 6.04E-05 &\\
&&&    & 0.515     & 2.26E-03 & 0.379     & 7.19E-03 & 4.36E-03 & 6.60E-04 & 1.29E-02 & 36.7 \\
 120   & 32.05 & 1.20E-04 & 52.10 & 0.127     & 1.63E-03 & 0.224     & 2.04E-05 & 0.147 & 3.15E-04 & 6.12E-05 &\\
&&&    & 2.59E-02 & 3.79E-03 & 8.00E-03 & 1.11E-02 & 6.71E-03 & 1.01E-03 & 1.98E-02 & 35.2 \\
\hline
\multicolumn{12}{c}{}\\
\multicolumn{12}{c}{{\bf $Z$=0.008}}\\
\hline
  M & \Hydrogen & \Hetre & \Helium & \Carbon & \Ctredici & \Nitrogen & \Nquindici & \Oxygen & \Odiciassette & \Odiciotto &\\
\hline
&&&    &\Neon & \Neventidue & \Magnesium & \Silicon & \Sulfur & \Calcium & \Fe & $M_{r}$ \\
\hline
   9   & 4.40  & 1.60E-04 & 2.52  & 0.259     & 5.06E-04 & 1.47E-02 & 2.21E-05 & 0.220 & 1.69E-04 & 5.25E-04 &\\
&&&    & 8.34E-03 & 3.25E-03 & 3.68E-03 & 4.12E-02 & 2.81E-02 & 5.68E-03 & 0.139     & 1.35 \\
  12   & 5.50  & 1.80E-04 & 3.67  & 0.168     & 6.69E-04 & 2.32E-02 & 7.17E-05 & 0.696 & 2.28E-05 & 1.91E-04 &\\
&&&    & 0.119     & 7.62E-04 & 3.42E-02 & 9.48E-02 & 4.50E-02 & 7.55E-03 & 0.158     & 1.48 \\
  15   & 6.49  & 1.78E-04 & 4.45  & 0.248     & 7.84E-04 & 2.92E-02 & 7.79E-05 & 1.31  & 2.79E-05 & 6.39E-05 &\\
&&&    & 0.143     & 9.13E-04 & 5.18E-02 & 0.204     & 9.92E-02 & 1.33E-02 & 0.120     & 1.84 \\
  20   & 8.27  & 1.55E-04 & 5.59  & 0.271     & 1.03E-03 & 3.67E-02 & 1.39E-04 & 2.85  & 2.42E-04 & 7.94E-05 &\\
&&&    & 0.107     & 1.15E-03 & 4.42E-02 & 0.372     & 0.195     & 2.01E-02 & 0.194     & 2.04 \\
  30   & 11.23 & 1.14E-04 & 7.75  & 0.313     & 8.34E-04 & 5.79E-02 & 6.87E-05 & 3.06  & 4.84E-04 & 5.71E-05 &\\
&&&    & 0.502     & 1.57E-03 & 0.133     & 2.93E-02 & 4.14E-03 & 5.03E-04 & 8.96E-03 & 6.90 \\
  40   & 13.90 & 2.19E-04 & 10.19 & 0.352     & 1.02E-03 & 7.69E-02 & 3.36E-05 & 2.21  & 4.89E-04 & 1.19E-04 &\\
&&&    & 0.656     & 1.99E-03 & 6.74E-03 & 6.35E-03 & 4.04E-03 & 6.25E-04 & 1.14E-02 & 12.5 \\
  60   & 18.91 & 2.67E-04 & 22.35 & 7.33      & 2.02E-03 & 0.133     & 1.04E-04 & 4.78  & 7.08E-04 & 2.33E-03 &\\
&&&    & 0.517     & 9.59E-03 & 0.184     & 5.35E-02 & 9.73E-03 & 1.24E-03 & 2.34E-02 & 5.69 \\
 120   & 30.08 & 3.51E-04 & 49.35 & 19.95     & 4.85E-03 & 0.349     & 8.52E-05 & 9.17  & 8.51E-04 & 1.93E-04 &\\
&&&    & 0.638     & 0.300     & 8.17E-02 & 3.34E-02 & 1.75E-02 & 2.60E-03 & 4.99E-02 & 9.89 \\
\hline
\multicolumn{12}{c}{}\\
\multicolumn{12}{c}{{\bf $Z$=\Zsol=0.02}}\\
\hline
  M & \Hydrogen & \Hetre & \Helium & \Carbon & \Ctredici & \Nitrogen & \Nquindici & \Oxygen & \Odiciassette & \Odiciotto &\\
\hline
&&&    &\Neon & \Neventidue & \Magnesium & \Silicon & \Sulfur & \Calcium & \Fe & $M_{r}$ \\
\hline
   9   & 4.18  & 2.28E-04 & 2.99  & 7.27E-02 & 1.29E-03 & 4.02E-02 & 4.85E-05 & 0.178 & 0.748E-04 & 1.18E-04 &\\
&&&    & 1.95E-02 & 1.50E-03 & 6.64E-03 & 7.62E-02 & 6.22E-02 & 1.18E-02 & 5.68E-02 & 1.31 \\
  12   & 5.20  & 2.45E-04 & 3.95  & 0.170     & 1.65E-03 & 5.46E-02 & 1.12E-04 & 0.712 & 0.827E-04 & 1.49E-04 &\\
&&&    & 0.123     & 1.92E-03 & 2.78E-02 & 0.111     & 6.11E-02 & 1.04E-02 & 0.121     & 1.44 \\
  15   & 6.18  & 2.65E-04 & 4.70  & 0.266     & 1.95E-03 & 6.68E-02 & 1.02E-04 & 1.273 & 0.882E-04 & 1.66E-04 &\\
&&&    & 0.175     & 2.28E-03 & 3.62E-02 & 0.213     & 9.46E-02 & 1.03E-02 & 9.79E-02 & 1.87 \\
  20   & 7.75  & 3.23E-04 & 6.01  & 0.279     & 2.06E-03 & 8.84E-02 & 1.65E-04 & 2.76  & 1.04E-03 & 1.99E-04 &\\
&&&    & 0.214     & 2.89E-03 & 6.91E-02 & 0.335     & 0.156     & 1.68E-02 & 0.186     & 2.11 \\
  30   & 10.30 & 3.96E-04 & 8.28  & 0.307     & 2.01E-03 & 0.128     & 9.46E-05 & 3.08  & 1.45E-03 & 2.32E-04 &\\
&&&    & 0.500     & 3.89E-03 & 0.124     & 4.07E-02 & 9.02E-03 & 1.17E-03 & 2.22E-02 & 7.18 \\
  40   & 12.52 & 4.72E-04 & 19.05 & 3.03      & 3.61E-03 & 0.211     & 1.35E-04 & 2.05  & 3.37E-03 & 7.00E-03 &\\
&&&    & 0.135     & 0.282     & 3.86E-02 & 0.298     & 0.154     & 1.53E-02 & 0.133     & 2.06 \\
  60   & 16.49 & 5.27E-04 & 32.14 & 4.93      & 5.41E-03 & 0.343     & 1.81E-04 & 2.66  & 3.84E-03 & 2.89E-04 &\\
&&&    & 0.138     & 0.336     & 5.60E-02 & 0.360     & 0.181     & 1.87E-02 & 0.234     & 2.09 \\
 100   & 23.97 & 5.51E-04 & 55.48 & 11.40     & 9.79E-03 & 0.682     & 2.55E-04 & 4.27  & 5.58E-03 & 3.04E-04 &\\
&&&    & 0.455     & 0.613     & 0.130     & 0.371     & 0.176     & 2.13E-02 & 0.251     & 2.12 \\
 120   & 28.72 & 1.37E-03 & 74.03 & 8.63      & 1.16E-02 & 1.01      & 3.25E-04 & 3.52  & 2.17E-03 & 7.53E-04 &\\
&&&    & 0.349     & 0.542     & 0.113     & 0.402     & 0.197     & 2.28E-02 & 0.307     & 2.11 \\
\hline
\multicolumn{12}{c}{}\\
\multicolumn{12}{c}{{\bf $Z$=0.05}}\\
\hline
  M & \Hydrogen & \Hetre & \Helium & \Carbon & \Ctredici & \Nitrogen & \Nquindici & \Oxygen & \Odiciassette & \Odiciotto &\\
\hline
&&&    & \Neon & \Neventidue & \Magnesium & \Silicon & \Sulfur & \Calcium & \Fe & $M_{r}$ \\
\hline
   9   & 3.45  & 4.06E-04 & 3.45  & 0.111     & 3.19E-03 & 8.79E-02 & 7.73E-05 & 0.283 & 2.52E-03 & 2.14E-03 &\\
&&&    & 3.49E-02 & 3.74E-03 & 1.16E-02 & 9.26E-02 & 7.55E-02 & 1.42E-02 & 6.61E-02 & 1.31 \\
  12   & 4.25  & 4.92E-04 & 4.46  & 0.225     & 3.81E-03 & 0.116     & 1.44E-04 & 0.895 & 2.77E-03 & 5.47E-03 &\\
&&&    & 0.168     & 4.73E-03 & 3.83E-02 & 0.124     & 6.51E-02 & 1.05E-02 & 0.129     & 1.49 \\
  15   & 5.02  & 5.74E-04 & 5.22  & 0.329     & 4.79E-03 & 0.136     & 1.30E-04 & 1.57  & 3.37E-03 & 8.71E-03 &\\
&&&    & 0.123     & 5.63E-03 & 3.05E-02 & 0.285     & 0.135     & 1.43E-02 & 0.131     & 1.98 \\
  20   & 6.22  & 7.36E-04 & 6.24  & 0.376     & 4.79E-03 & 0.174     & 2.34E-04 & 3.44  & 4.24E-03 & 1.79E-03 &\\
&&&    & 0.412     & 6.75E-03 & 0.124     & 0.299     & 0.132     & 1.57E-02 & 0.152     & 2.38 \\
  40   & 9.98  & 1.17E-03 & 22.79 & 2.34      & 7.69E-03 & 0.484     & 2.43E-04 & 1.32  & 1.40E-02 & 7.61E-03 &\\
&&&    & 0.301     & 0.646     & 7.56E-02 & 0.173     & 9.11E-02 & 1.40E-02 & 0.203     & 1.55 \\
  60   & 14.06 & 6.47E-04 & 36.03 & 3.66      & 1.14E-02 & 1.06      & 2.19E-04 & 1.73  & 3.39E-02 & 3.67E-04 &\\
&&&    & 0.448     & 0.562     & 0.111     & 0.217     & 0.106     & 1.50E-02 & 0.238     & 1.70 \\
 100   & 26.74 & 8.60E-04 & 60.90 & 3.79      & 1.94E-02 & 1.79      & 2.78E-04 & 2.14  & 5.52E-02 & 3.23E-02 &\\
&&&    & 0.594     & 1.27      & 0.157     & 0.281     & 0.146     & 2.12E-02 & 0.357     & 1.68 \\
 120   & 38.07 & 1.15E-03 & 68.39 & 3.82      & 2.61E-02 & 2.11      & 3.20E-04 & 2.60  & 6.31E-02 & 0.177     &\\
&&&    & 0.669     & 1.27      & 0.180     & 0.314     & 0.166     & 2.42E-02 & 0.416     & 1.68 \\
\hline
\end{tabular}
\end{center}
\normalsize
\end{minipage}
\end{table*}

\noindent
The total ejecta for our stellar models, obtained as outlined in the previous
sections, are listed in Tab.~\ref{ejmassivetab}. 

In Figures~\ref{yplotZ0004} to~\ref{yplotZ05} we plot the fractional remnant
mass $M_{r}/M$ and the total yields $p_{iM}$ for some representative elements
vs.\ the stellar mass $M$, for our five sets of metallicity. A trend with
metallicity is evident. 

In the lower metallicity sets ($Z$=0.0004 and $Z$=0.004) remnant masses are
very large due to lower mass-loss efficiency: massive CO--cores are built in
the pre-SN phase, resulting in high remnant masses (i.e.\ black holes, cfr.\
Tab.~\ref{WMCOej}). We notice an anti--correlation between the remnant mass and
the oxygen yield: for instance, in the case $Z$=0.0004, $M$=40 and
60~\Msol\ (\MCO $\sim$ 18--24~\Msol) almost the whole CO--core collapses to a
black hole and a very small amount of new oxygen is released. For
larger masses PC SN\ae\ are found and oxygen yields increase again, while
remnant masses decrease. This is in line with what Langer \& Woosley (1996)
suggest: at low metallicity (i.e.\ in the early phases of galactic evolution)
stars in the range 30 $\div$ 100~\Msol\ may form black holes and have a
secondary role in chemical enrichment, while higher masses may give an
important contribution through PC SN\ae, also depending on the
IMF in the early epochs. For solar metallicity, on the contrary,
mass loss inhibits black hole formation and PC SN\ae\ events. 

In the case $Z$=0.008, the effects of mass loss are getting evident: for large
masses ($M \gsim$ 40~\Msol) remnant masses are sensibly smaller than in the
previous sets, while helium and carbon yields are larger because of the
contribution from the wind.

These effects are exalted in the higher metallicity sets ($Z$=\Zsol=0.02 and
$Z$=0.05), where large initial masses end as WR stars with small masses and
small cores (cfr.\ Tab.~\ref{Mfintab} and Tab.~\ref{MCOtab}), resulting in
relatively low oxygen yields but very large helium yields. Carbon yields
decrease again moving from the $Z$=0.008 to the $Z$=0.05 case, both because a
higher mass loss rate is able to take more helium away
before it is turned to carbon, and because with an increasing metallicity an
increasing fraction of the original carbon is turned to \Nitrogen\ in the CNO
cycle.

\newpage
\clearpage 

\noindent
As far as heavy elements (\Silicon\ + \Sulfur\ + \Calcium\ + \Fe) are concerned,
the bulk of contribution seems to come from stars between 10 and 20~\Msol. For
larger masses large remnants are found, so that most heavy elements produced
remain locked in the remnant, at least for $Z \leq$ 0.008. In the case
of solar or super-solar metallicity, also stars with $M >$ 20~\Msol\ result in
low CO--cores and low-mass remnants and can release some heavy elements.


\section{Low and intermediate mass stars}

For low and intermediate mass stars ($M \leq$~5~\Msol) we adopt the yields by
Marigo \etal (1996, 1997), which are calculated basing on the same series of
stellar tracks we adopt for massive stars. Specifically, by means of a
semi-analytical model Marigo \etal (1996) extend the tracks of the Padua
library, which stop at the end of the E-AGB phase, to the TP-AGB and III
dredge-up phase. Further improvements of the semi-analytical formulation also
allows to properly take into account the effects of envelope burning on the
yields of stars between 4 and 5~\Msol (Marigo \etal 1997). The related
production of primary \Nitrogen\ has remarkable effects upon the predictions of
chemical models for the Solar Neighbourhood (\S~\ref{abundanceratios}). 

Thus we extend our grid of stellar ejecta to the range of low and
intermediate mass stars and provide a complete data-base for the
chemical models of galaxies. This data-base is referred to a unique grid of
stellar models for the whole range of stellar masses, which gives it
coherence and homogeneity of basic physical prescriptions. 


\section{Very Massive Objects}

The structure and evolution of very massive massive objects (VMOs), i.e.\
stars in the mass range 120$< M <$1000~\Msol,  was explored in past years,
mainly in the framework of a primeval Population III of very massive stars
formed by metal--free gas (Bond \etal 1984; Carr \etal 1984; El~Eid \etal 1983;
Ober \etal 1983; Woosley \& Weaver 1982). Population III was invoked to
solve the G-dwarf problem and to explain the non-zero metallicity of Population
II stars, to provide a substantial amount of dark matter by means of black hole
formation, to account for black holes in AGNs, to explain the reionization of
the Universe, to produce primordial helium and to account for some difficulties
in Big Bang nucleosynthesis; for a review see Bond (1984). 

Some recent interest toward stars more massive than the canonical limit
of say 120~\Msol\  come from studies of chemical evolution and
population synthesis  in galaxies under unconventional IMFs (Chiosi
\etal 1997)   such as that proposed by  Padoan \etal (1997) 
which formally allows for  a non--zero, though small, probability
that objects more massive than 120~\Msol\ can form.

We are not going to discuss here the possible existence of VMOs in present-day
or primeval galaxies, nor the cosmological
consequences of Population III; rather, we want to make reasonable
assumptions about nucleosynthesis and mass loss in VMOs and
to extend our five grids of ejecta up to 1000~\Msol, for the purpose of
future use if required by the particular problem under examination. Notice
that we do not include chemical enrichment from VMOs in our chemical model
of the Solar Neighbourhood (see Sect.~9).

\begin{table}[t]
\caption{Lifetimes, H--burning core masses and CO--core masses for Very Massive 
Objects}
\label{VMOcoretab}
\begin{small}
\begin{center}
\begin{tabular}{|l|r|r|r|c|}
\hline
~		 &  $M$ & $\tau_{H}$ & $M_{Hb}$ & \MCO \\
\hline
{\bf $Z$=0.0004} &  150 & 3.0 10$^{6}$ &  75 & PC SN \\
~		 &  200 & 2.7 10$^{6}$ & 100 & PC SN \\
~		 &  300 & 2.5 10$^{6}$ & 150 & 18    \\
~		 &  500 & 2.0 10$^{6}$ & 250 & 21    \\
~		 & 1000 & 2.0 10$^{6}$ & 500 & 21    \\
\hline
{\bf $Z$=0.004}  &  150 & 3.0 10$^{6}$ &  75 & 35 \\
~		 &  200 & 2.7 10$^{6}$ & 100 & 35 \\
~		 &  300 & 2.5 10$^{6}$ & 150 &  8 \\
~		 &  500 & 2.0 10$^{6}$ & 250 & 10 \\
~		 & 1000 & 2.0 10$^{6}$ & 500 & 10 \\
\hline
{\bf $Z$=0.008}  &  150 & 3.0 10$^{6}$ &  75 & 13  \\
~		 &  200 & 2.7 10$^{6}$ & 100 & 13  \\
~		 &  300 & 2.5 10$^{6}$ & 150 &  7  \\
~		 &  500 & 2.0 10$^{6}$ & 250 &  8  \\
~		 & 1000 & 2.0 10$^{6}$ & 500 &  8  \\
\hline
{\bf $Z$=\Zsol=0.02} &  150 & 3.0 10$^{6}$ &  75 & 6 \\
~		      &  200 & 2.7 10$^{6}$ & 100 & 5 \\
~		      &  300 & 2.5 10$^{6}$ & 150 & 5 \\
~		      &  500 & 2.0 10$^{6}$ & 250 & 5 \\
~		      & 1000 & 2.0 10$^{6}$ & 500 & 5 \\
\hline
{\bf $Z$=0.05}&  150 & 3.0 10$^{6}$ &  75 & 3.5 \\
~	      &  200 & 2.7 10$^{6}$ & 100 & 3.5 \\
~	      &  300 & 2.5 10$^{6}$ & 150 & 3.5 \\
~	      &  500 & 2.0 10$^{6}$ & 250 & 3.5 \\
~	      & 1000 & 2.0 10$^{6}$ & 500 & 3.5 \\
\hline
\end{tabular}
\end{center}
\end{small}
\end{table}

\begin{table*}[t]
\begin{minipage}{18truecm}
\caption{Total ejecta in \Msol\ for VMOs (150 $\div$ 1000~\Msol) of different
metallicities}
\label{ejVMOtab}
\scriptsize
\begin{center}
\begin{tabular}{|r |c |c |c |c |c |c |c |c |c |c |c |c|}
\multicolumn{12}{c}{{\bf $Z$=0.0004}}\\
\hline
  M & \Hydrogen & \Hetre & \Helium & \Carbon & \Ctredici & \Nitrogen & \Nquindici & \Oxygen & \Odiciassette & \Odiciotto &\\
\hline
&&& &\Neon & \Neventidue & \Magnesium & \Silicon & \Sulfur & \Calcium & \Fe & $M_{r}$ \\
\hline
 150 & 59.51 & 6.60E-05 & 20.95 & 1.31      & 8.88E-05 & 2.67E-03 & 7.32E-06 & 38.02     & 6.30E-06 & 3.63E-05 &\\ 
&&&  & 1.80      & 3.08E-04 & 1.60      & 16.00     &  8.90     &  1.30     & 0.612     &   0   \\
 200 & 79.82 & 8.80E-05 & 28.90 & 0.840     & 1.18E-04 & 1.46E-02 & 9.76E-06 & 34.02     & 8.40E-06 & 4.84E-05 &\\ 
&&&  & 1.50      & 4.10E-04 & 1.50      & 23.00     &  15.00    &  2.40     & 13.00     &   0   \\
 300 & 152.7 & 1.32E-04 & 129.1 & 1.48E-02 & 1.78E-04 & 4.48E-02 & 1.46E-05 & 3.17E-02 & 1.26E-05 & 7.27E-05 &\\ 
&&&  & 8.61E-03 & 1.16E-03 & 2.66E-03 & 3.68E-03 & 2.23E-03 & 3.38E-04 & 6.60E-03 & 18.00 \\
 500 & 292.2 & 2.20E-04 & 186.6 & 2.47E-02 & 2.96E-04 & 7.75E-02 & 2.44E-05 & 5.28E-02 & 2.10E-05 & 1.21E-04 &\\ 
&&&  & 1.46E-02 & 1.97E-03 & 4.52E-03 & 6.26E-03 & 3.79E-03 & 5.74E-04 & 1.12E-02 & 21.00 \\
1000 & 596.2 & 4.40E-04 & 382.4 & 4.94E-02 & 5.92E-04 & 0.162     & 4.88E-05 & 0.106     & 4.20E-05 & 2.42E-04 &\\ 
&&&  & 2.99E-02 & 4.02E-03 & 9.24E-03 & 1.28E-02 & 7.75E-03 & 1.17E-03 & 2.29E-02 & 21.00 \\
\hline
\multicolumn{12}{c}{}\\
\multicolumn{12}{c}{{\bf $Z$=0.004}}\\
\hline
  M & \Hydrogen & \Hetre & \Helium & \Carbon & \Ctredici & \Nitrogen & \Nquindici & \Oxygen & \Odiciassette & \Odiciotto &\\
\hline
&&& &\Neon & \Neventidue & \Magnesium & \Silicon & \Sulfur & \Calcium & \Fe & $M_{r}$ \\
\hline
 150 & 56.67 & 6.60E-04 & 57.84 & 7.41E-02 & 8.88E-04 & 0.153     & 7.32E-05 & 0.159     & 6.30E-05 & 3.63E-04 &\\ 
&&&  & 3.51E-02 & 4.72E-03 & 1.09E-02 & 1.50E-02 & 9.11E-03 & 1.38E-03 & 2.69E-02 & 35.00 \\
 200 & 75.57 & 8.80E-04 & 88.74 & 9.89E-02 & 1.18E-03 & 0.227     & 9.76E-05 & 0.211     & 8.40E-05 & 4.84E-04 &\\ 
&&&  & 5.04E-02 & 6.77E-03 & 1.56E-02 & 2.15E-02 & 1.31E-02 & 1.98E-03 & 3.86E-02 & 35.00 \\
 300 & 159.2 & 1.32E-03 & 131.6 & 0.442     & 1.78E-03 & 0.480     & 2.73E-04 & 4.34      & 1.26E-04 & 7.27E-04 &\\ 
&&&  & 0.692     & 1.20E-02 & 0.330     & 0.126     & 3.01E-02 & 4.10E-03 & 7.29E-02 & 2.68  \\
 500 & 295.9 & 2.20E-03 & 192.0 & 0.551     & 2.96E-03 & 0.810     & 3.18E-04 & 3.82      & 2.10E-04 & 1.21E-03 &\\ 
&&&  & 0.658     & 2.01E-02 & 0.226     & 9.95E-02 & 4.01E-02 & 5.92E-03 & 0.115     & 5.67  \\
1000 & 591.9 & 4.40E-03 & 394.0 & 0.798     & 5.92E-03 & 1.65      & 5.62E-04 & 4.35      & 4.20E-04 & 2.42E-03 &\\ 
&&&  & 0.811     & 4.02E-02 & 0.273     & 0.165     & 7.97E-02 & 1.19E-02 & 0.232     & 5.67  \\
\hline
\multicolumn{12}{c}{}\\
\multicolumn{12}{c}{{\bf $Z$=0.008}}\\
\hline
  M & \Hydrogen & \Hetre & \Helium & \Carbon & \Ctredici & \Nitrogen & \Nquindici & \Oxygen & \Odiciassette & \Odiciotto &\\
\hline
&&&  &\Neon & \Neventidue & \Magnesium & \Silicon & \Sulfur & \Calcium & \Fe & $M_{r}$ \\
\hline
 150 & 55.54 & 1.32E-03 & 80.25 & 0.460     & 1.78E-03 & 0.423     & 1.75E-04 & 2.86      & 1.26E-04 & 7.27E-04 &\\ 
&&&  & 0.636     & 1.12E-02 & 9.16E-02 & 4.42E-02 & 2.22E-02 & 3.32E-03 & 6.41E-02 & 9.57  \\
 200 & 80.98 & 1.76E-03 & 104.4 & 0.509     & 2.37E-03 & 0.592     & 2.24E-04 & 2.97      & 1.68E-04 & 9.69E-04 &\\ 
&&&  & 0.667     & 1.53E-02 & 0.101     & 5.73E-02 & 3.02E-02 & 4.52E-03 & 8.75E-02 & 9.57  \\
 300 & 159.1 & 2.64E-03 & 131.9 & 0.572     & 3.35E-03 & 0.968     & 4.31E-04 & 3.81      & 2.52E-04 & 1.45E-03 &\\ 
&&&  & 0.433     & 2.41E-02 & 0.158     & 0.383     & 0.197     & 2.28E-02 & 0.279     & 2.15  \\
 500 & 292.9 & 4.40E-03 & 195.0 & 0.778     & 5.92E-03 & 1.63      & 6.13E-04 & 4.92      & 4.20E-04 & 2.42E-03 &\\ 
&&&  & 0.841     & 4.04E-02 & 0.361     & 0.227     & 9.48E-02 & 1.35E-02 & 0.244     & 2.96  \\
1000 & 582.7 & 8.80E-03 & 401.1 & 1.27      & 1.18E-02 & 3.31      & 1.10E-03 & 5.97      & 8.40E-04 & 4.84E-03 &\\ 
&&&  &  1.14     & 8.14E-02 & 0.456     & 0.358     & 0.174     & 2.55E-02 & 0.478     & 2.96  \\
\hline
\multicolumn{12}{c}{}\\
\multicolumn{12}{c}{{\bf $Z$=\Zsol=0.02}}\\
\hline
  M & \Hydrogen & \Hetre & \Helium & \Carbon & \Ctredici & \Nitrogen & \Nquindici & \Oxygen & \Odiciassette & \Odiciotto &\\
\hline
&&&  &\Neon & \Neventidue & \Magnesium & \Silicon & \Sulfur & \Calcium & \Fe & $M_{r}$ \\
\hline
 150 & 52.39 & 3.30E-03 & 88.62 & 0.614     & 4.44E-03 & 1.17      & 5.04E-04 & 3.50      & 3.15E-04 & 1.82E-03 &\\ 
&&&  & 0.429     & 2.95E-02 & 0.134     & 0.417     & 0.206     & 2.46E-02 & 0.334     & 2.11  \\
 200 & 96.71 & 4.40E-03 & 94.51 & 0.672     & 5.92E-03 & 1.61      & 5.73E-04 & 2.89      & 4.20E-04 & 2.42E-03 &\\ 
&&&  & 0.379     & 4.01E-02 & 0.116     & 0.406     & 0.223     & 2.51E-02 & 0.317     & 2.09  \\
 300 & 155.3 & 6.60E-03 & 133.4 & 0.919     & 8.88E-03 & 2.45      & 8.17E-04 & 3.42      & 6.30E-04 & 3.63E-03 &\\ 
&&&  & 0.531     & 6.06E-02 & 0.613     & 0.471     & 0.263     & 3.11E-02 & 0.434     & 2.09  \\
 500 & 281.4 & 1.10E-02 & 202.9 & 1.44      & 1.48E-02 & 4.12      & 1.33E-03 & 4.90      & 1.05E-03 & 6.05E-03 &\\ 
&&&  & 0.799     & 0.101     & 0.266     & 0.668     & 0.361     & 4.59E-02 & 0.784     & 2.09  \\
1000 & 553.8 & 2.20E-02 & 420.2 & 2.67      & 2.96E-02 & 8.32      & 2.55E-03 & 7.54      & 2.10E-03 & 1.21E-02 &\\ 
&&&  &  1.56     & 0.204     & 0.502     & 0.994     & 0.559     & 7.59E-02 &  1.37     & 2.09  \\
\hline
\multicolumn{12}{c}{}\\
\multicolumn{12}{c}{{\bf $Z$=0.05}}\\
\hline
  M & \Hydrogen & \Hetre & \Helium & \Carbon & \Ctredici & \Nitrogen & \Nquindici & \Oxygen & \Odiciassette & \Odiciotto &\\
\hline
&&& &\Neon & \Neventidue & \Magnesium & \Silicon & \Sulfur & \Calcium & \Fe & $M_{r}$ \\
\hline
 150 & 44.58 & 8.25E-03 & 94.29 & 1.14      & 1.11E-02 & 3.02      & 1.01E-03 & 3.00      & 7.87E-04 & 4.54E-03 &\\ 
&&&  & 0.813     & 7.51E-02 & 0.221     & 0.367     & 0.194     & 2.79E-02 & 0.494     & 1.74  \\
 200 & 87.44 & 1.10E-02 & 98.84 & 1.45      & 1.48E-02 & 4.07      & 1.31E-03 & 3.66      & 1.05E-03 & 6.06E-03 &\\ 
&&&  &  1.00     & 0.101     & 0.280     & 0.449     & 0.243     & 3.54E-02 & 0.641     & 1.74  \\
 300 & 140.9 & 1.65E-02 & 140.2 & 2.07      & 2.22E-02 & 6.17      & 1.92E-03 & 4.98      & 1.58E-03 & 9.08E-03 &\\ 
&&&  &  1.39     & 0.152     & 0.398     & 0.612     & 0.342     & 5.03E-02 & 0.933     & 1.74  \\
 500 & 246.6 & 2.75E-02 & 224.1 & 3.30      & 3.70E-02 & 10.37     & 3.14E-03 & 7.62      & 2.62E-03 & 1.51E-02 &\\ 
&&&  &  2.15     & 0.255     & 0.634     & 0.938     & 0.540     & 8.03E-02 &  1.52     & 1.74  \\
1000 & 478.2 & 5.50E-02 & 466.6 & 6.39      & 7.40E-02 & 20.88     & 6.19E-03 & 14.23     & 5.25E-03 & 3.03E-02 &\\ 
&&&  &  4.06     & 0.511     &  1.22     &  1.75     &  1.04     & 0.155     &  2.98     & 1.74  \\
\hline
\end{tabular}
\end{center}
\normalsize
\end{minipage}
\end{table*}

\noindent
VMOs are pulsationally unstable during H-burning, so once they form
they are likely to undergo violent mass loss until they fall below
the critical mass for pulsational stability; this limit is generally around 
100~\Msol, though depending on metallicity. We assume that VMOs lose mass at a
rate of 10$^{-3}$~\Msol/yr (comparable to the extreme mass loss rate adopted in
our stellar tracks beyond the de Jager limit), independently of metallicity
since mass loss here is driven by instability rather than by radiation
pressure. We stop this phase of paroxysmal mass loss when the mass has
decreased down to 150~\Msol. 

Meanwhile, we assume that the star has been burning hydrogen in the inner 50\%
of its mass (He--cores of VMOs involve $\sim$0.56 of the initial mass, see the
references quoted above) and that the hydrogen content $X$ in the core has
been linearly decreasing with time, consistently with the behaviour of our
high-mass stellar tracks. The hydrogen content in the core at any time $t$ is
given by: 

\begin{equation}
\label{Xcore}
X(t) = X^{0} \left( 1 - \frac{t}{\tau_{H}} \right)
\end{equation}

\noindent
where $X^{0}$ is the initial hydrogen abundance in the star and $\tau_{H}$ is
the H--burning timescale, assumed following the references above and listed in
the second column of Tab.~\ref{VMOcoretab}. 

At the end of the violent mass loss phase (i.e. when the current mass
falls below 120~\Msol) we adopt a rate for
radiation pressure driven mass loss derived from Lamers \& Cassinelli (1996): 

\begin{equation}
\dot{M} = 2.11 \times  10^{-14} \times M^{-4.421}
\nonumber
\end{equation}

\noindent
scaled with metallicity in conformity with our
stellar tracks for massive stars ($\dot{M} \propto Z^{0.5}$).
This mass loss rate holds until the star
becomes a WR star; in case it enters this stage, for the rest of its lifetime 
we adopt the mass loss rate by Langer (1989), scaled with metallicity.
For the sake of consistency with our
stellar tracks, we assume that the WR stage begins if the surface hydrogen
content falls below $X$=0.3, i.e.\ when two conditions are fulfilled: (1) the
H--burning or He--core is revealed on the surface (the stellar mass must have
fallen below 1/2 of its initial value) and (2) the hydrogen content in the
core, given by Eq.~(\ref{Xcore}), is lower than $X$=0.3. 

In the papers referenced above about VMOs, later burning processes build a
CO--core which almost fills the whole He--core (\MCO$\sim$\MHe). This 
result cannot hold here, since we include strong mass-loss and a possible 
WR stage which sensitively depletes the mass. The CO--core is assumed to fill
the total remaining mass at the end of the lifetime. The assumed 
lifetimes and H--burning core masses, and the obtained CO--core masses, are 
listed in Tab.~\ref{VMOcoretab}. We notice that, for the higher metallicity
cases, the final masses of the CO--cores tend to converge to the same small
value for all the VMOs, due to a prolonged WR stage with very efficient mass
loss.

To derive the composition of the ejected material, we assume that the
layers over the H--burning core have not altered their
initial composition, i.e.\ we neglect any convective dredge-up of
processed material and a possible recession of the convective core from initial
masses larger than 1/2 the total mass. When the core is revealed on the
surface, we assume that all initial CNO isotopes have been converted to
\Nitrogen, that the hydrogen content is given by Eq.~(\ref{Xcore}) at any time 
$t$ and that the remaining fraction is \Helium.

The size of the CO--core eventually drives the SN explosion, as we
have assumed to hold for lower masses (\S~\ref{SNae}). In most cases the final
outcome is an iron--core collapse SN, whose ejecta are calculated as in
\S~\ref{ironSNae}. The only exceptions are the cases $M$=150 and 200~\Msol,
$Z$=0.0004, whose CO--cores are marked ``PC SN'' in Tab.~\ref{VMOcoretab}.
These objects never lose enough mass to let the core uncovered; they build a
75~\Msol\ and a 100~\Msol\ He--core, respectively, which eventually explode as
PC SN\ae. The ejecta of such cores are taken directly from Tab.~18 of Woosley
(1986), but for those of \Nitrogen\ which are to be scaled with metallicity.
In fact, Woosley's PC SN\ae\ are
calculated for solar metallicity and their initial abundances of CNO isotopes,
which are turned to \Nitrogen\ during H--burning, are solar; for our 
$Z$=0.0004 case, the initial CNO content is much lower. 

The resulting ejecta for the whole set of VMOs and for the five metallicity
cases included in our grid are listed in Tab.~\ref{ejVMOtab}.


\section{The chemical evolution model}

Our results about stellar nucleosynthesis and ejecta described in previous
sections were inserted in a model of chemical evolution, which was applied to
the analysis of the Solar Neighbourhood. In this section we outline
the main features and equations of the model.

We adopt an open model with continuous {\it infall} of primordial gas that
builds the disk gradually, as suggested by dynamical studies (Larson
1976; Burkert \etal 1992; Carraro \etal 1997). Infall also provides for the
better solution to the ``G--dwarf problem'' (Lynden-Bell 1975;  Tinsley
1980; Chiosi 1978, 1980; Pagel
1989; Matteucci 1991). Our formulation follows the one framed by
Talbot \& Arnett (1971, 1973, 1975) and adapted to open models by
Chiosi (1980). This formulation, particularly suitable for galactic
discs, is also currently used by other authors (e.g.\ Matteucci \& Fran\c{c}ois
1989; Timmes \etal 1995). 

The Galactic disc is divided into concentric cylindrical shells, 2 kpc wide
each, which evolve independently, neglecting any possible radial
flows of gas or stars. Each ring consists of a homogeneous
mixture of gas and stars, so that for each ring the only independent variable
is time $t$ ({\it one--zone} formulation, Talbot \& Arnett 1971). Galactic
discs are comfortably described in terms of surface mass density
$\sigma(r,t)$, which depends both on the galactocentric radius $r$ of the ring 
and
also on time $t$, since in each ring the surface density is growing in time due
to gradual infall of gas. Since in this paper the model is applied only to the
Solar Neighbourhood ($r=r_{\odot}$), for the
sake of simplicity we drop the dependence on $r$ in the following equations.
If we indicate with $\sigma_{g}(t)$ the surface gas density, the gas fraction
in the ring at any time $t$ is the ratio: 

\[ \frac{\sigma_{g}(t)}{\sigma(t)} \]

\noindent
while the surface star density is:

\[ \sigma_{s}(t) = \sigma(t) - \sigma_{g}(t) \]

\noindent
In closed models the total surface density $\sigma$ is constant and the other
quantities can be normalized with respect to it; in open models it is
suitable to normalize with respect to the total surface density at the present
age of the Galaxy 

\newpage
\clearpage

\noindent
$t_{G}$, i.e.\ at the final age of the model. So we
introduce the normalized surface gas density as:

\[ G(t) = \frac{\sigma_{g}(t)}{\sigma(t_{G})} \]

\noindent
In each ring the gas is assumed to be chemically homogeneous, and the
normalized gas density for each chemical species $i$ is: 

\[ G_{i}(t) = X_{i}(t)\,G(t) \]

\noindent
where $X_{i}$ is the fractionary mass abundance of species $i$; $\sum_{i}
X_{i}$=1 by definition. The chemical evolution of the ISM is
the evolution of the set of the $G_{i}$'s, described by: 

\begin{eqnarray}
\label{dGi/dt}
\frac{d}{dt} \, G_{i}(t) & = & - \, X_{i}(t)\Psi(t)\, +\, \nonumber \\ 
 & & + \int_{M_{l}}^{M_{u}} \Psi(t-\tau_{M})\,R_{Mi}(t-\tau_{M}) 
\Phi(M)\,dM\:+ \nonumber \\
 & & + \left[\frac{d}{dt} G_{i}(t) \right]_{inf}
\end{eqnarray}

\noindent
where $\Psi(t)$ is the SFR, $\Phi(M)$ is the 
IMF, $R_{Mi}(t)$ is the mass fraction of a star of mass $M$ ejected
into the ISM in the form of element $i$, $M_{l}$ and
$M_{u}$ are the lower and upper limit for stellar mass respectively, and
$\tau_{M}$ is the lifetime of a star of mass $M$. The first term on the right
side represents the depletion of species $i$ from the ISM due
to star formation; the second term represents the amount of species $i$ put in
the ISM by stellar ejecta; the third term is the contribution of the infalling
gas. We describe the various ingredients in the following subsections. 


\subsection{The infall term}

In open models the surface mass density $\sigma(r,t)$ increases by slowly 
accreting gas at a rate
$\dot{\sigma}_{inf}(r,t)$, until it reaches the observed present values. An
infall rate exponentially decreasing in time with a timescale $\tau$: 

\begin{equation}
\label{eqinfall}
\dot{\sigma}_{inf}(r,t) = A(r) \, e^{-\frac{t}{\tau}}
\end{equation}

\noindent
well reproduces the results of dynamical models (Larson 1976; Burkert
\etal 1992; Carraro \etal 1997), with the exception of radial flows. $A(r)$ is
obtained by integrating upon time and by imposing that at the
age $t_{G}$ the observed present surface mass density
$\sigma(r,t_{G})$ is matched: 

\begin{equation}
 A(r) \left( 1-e^{-\frac{t_{G}}{\tau}} \right)\,\tau =\sigma(r,t_{G}) 
\nonumber
\end{equation}

\noindent
In the case of the solar ring:

\begin{equation}
\label{eqinfallnorm}
A_{\odot} \left( 1-e^{-\frac{t_{G}}{\tau}} \right)\,\tau =
\sigma(r_{\odot},t_{G})
\end{equation}

\noindent
The contribution of the infalling gas to the evolution of the (normalized) gas 
fraction $G(t)$ is:

\begin{equation}
 \left[ \frac{d}{dt}G(t) \right]_{inf} = \frac{\dot{\sigma}_{inf}(t)}
                                         {\sigma(t_{G})} 
\nonumber
\end{equation}

\noindent
and, for each single chemical species $i$:

\begin{equation}
 \left[ \frac{d}{dt} G_{i}(t) \right]_{inf}\,=\,
\frac{\dot{\sigma}_{inf}(t)\,X_{i,inf}}{\sigma(t_{G})} 
\nonumber
\end{equation}

\noindent
which gives the third term on the right side of Eq.~(\ref{dGi/dt}).


\subsection{The Star Formation Rate}
\label{SFlaw}

We adopt the formulation of the SFR for the Galactic disk given by Talbot \& 
Arnett (1975):

\begin{equation}
 \frac{d}{dt}\,\sigma_{g}(r,t) = \nu\:\left[ \frac{\sigma(r,t) \,
\sigma_{g}(r,t)} {\tilde{\sigma}(\tilde{r},t)} \right] ^{\kappa-1}\: 
\sigma_{g}(r,t) 
\nonumber
\end{equation}

\noindent
Here, $\kappa$ is the exponent of the Schmidt (1959) law for star formation,
$\Psi \propto \rho^{\kappa}$; plausible values  for $\kappa$ range from 1 to 2
(proportional to gas density or proportional to cloud--cloud collision events
respectively, Larson 1991); $\tilde{\sigma}(\tilde{r},t)$ is the surface mass
density at a given galactocentric distance $\tilde{r}$, adopted as a
normalization factor; $\nu$ is a parameter for the star formation efficiency,
related to the choice of $\tilde{r}$. 

Talbot \& Arnett's SFR is based on Schmidt's law, but it also takes into
account that the cooling due to gas accretion onto the equatorial plane is
balanced by the heating due to the feed--back of massive stars.
The SFR turns out to be related to the dynamical timescale, shorter where 
the mass density is larger,
while the timescale of gas accretion onto the equatorial plane is longer than
the purely dynamical one due to the feed--back from star formation. As a caveat,
we remind that the formula
above holds in the plane--parallel approximation, which may not be a good one
in the case of open models. 

In terms of the formalism introduced in previous sections, this adopted SFR 

\begin{equation}
 \Psi(r,t)=\nu\, \left[\frac{\sigma(r,t)}{\tilde{\sigma}(\tilde{r},t)}
     \right]^{2(\kappa-1)}\, \left[\frac{\sigma(r,t_{G})}{\sigma(r,t)}\right]
     ^{\kappa-1}\, G^{\kappa}(r,t) 
\nonumber
\end{equation}

\noindent
Limiting to the solar ring and taking $\tilde{r}=r_{\odot}$ we get (dropping 
the dependence on $r$):

\begin{equation}
 \Psi(t)=\nu\, \left[\frac{\sigma(t_{G})}{\sigma(t)}\right]
                 ^{\kappa-1}\, G^{\kappa}(t) 
\nonumber
\end{equation}

\noindent
where $\nu$ is to be fixed so as to reproduce the features of the Solar 
Neighbourhood.

\begin{table*}[ht]
\begin{minipage}{18truecm}
\caption{The $Q_{ij}$ matrix}
\label{Qijtab}
\tiny
\begin{center}
\begin{tabular}{|l|c c c c c c c c c c c c c|}
\hline
 & $H$ & $^{4}He$ & $^{12}C$ & $^{13}C$ & $^{14}N$ & $^{16}O$ & nr &
$^{20}Ne$ & $^{24}Mg$ & $^{28}Si$ & $^{32}S$ & $^{40}Ca$ & $^{56}Fe$\\
 & (1) & (2) & (3) & (4) & (5) & (6) & (7) & (8) & (9) & (10) & (11) & (12) &
(13)\\
\hline
 & & & & & & & & & & & & & \\
~(1) $H$ & $1-q_{4}$ & & & & & & & & & & & & \\
 & & & & & & & & & & & & & \\
~(2) $^{4}He$ & $q_{4}-q_{C}$ & $1-q_{C}$ & & & & & & & & & & & \\
 & & & & & & & & & & & & & \\
~(3) $^{12}C$ & $\chi_{C} w_{C}$ & $\chi_{C} w_{C}$ & $1-q_{C13s}$ & & & & &
 & & & & & \\
 & & & & & & & & & & & & & \\
~(4) $^{13}C$ & $\chi_{C13} w_{C}$ & $\chi_{C13} w_{C}$ & $q_{C13s}-q_{Ns}$ &
$1-q_{Ns}$ & & & & & & & & & \\
 & & & & & & & & & & & & & \\
~(5) $^{14}N$ & $\chi_{N} w_{C}$ & $\chi_{N} w_{C}$ & $q_{Ns}-q_{C}$ &
$q_{Ns}-q_{C}$ & $1-q_{C}$ & $q_{Ns}-q_{C}$ & & & & & & & \\
 & & & & & & & & & & & & & \\
~(6) $^{16}O$ & $\chi_{O} w_{C}$ & $\chi_{O} w_{C}$ & & & & $1-q_{Ns}$ & & &
 & & & & \\
 & & & & & & & & & & & & & \\
~(7) nr & & & $w_{C}$ & $w_{C}$ & $w_{C}$ & $w_{C}$ & $1-d$ & & & & & & \\
 & & & & & & & & & & & & & \\
~(8) $^{20}Ne$ & $\chi_{Ne} w_{C}$ & $\chi_{Ne} w_{C}$ & & & & & & $1-d$ & &
 & & & \\
 & & & & & & & & & & & & & \\
~(9) $^{24}Mg$ & $\chi_{Mg} w_{C}$ & $\chi_{Mg} w_{C}$ & & & & & & & $1-d$ &
 & & & \\
 & & & & & & & & & & & & & \\
(10) $^{28}Si$ & $\chi_{Si} w_{C}$ & $\chi_{Si} w_{C}$ & & & & & & & & $1-d$ &
 & & \\
 & & & & & & & & & & & & & \\
(11) $^{32}S$ & $\chi_{S} w_{C}$ & $\chi_{S} w_{C}$ & & & & & & & & & $1-d$ &
 & \\
 & & & & & & & & & & & & & \\
(12) $^{40}Ca$ & $\chi_{Ca} w_{C}$ & $\chi_{Ca} w_{C}$ & & & & & & & & & &
$1-d$ & \\
 & & & & & & & & & & & & & \\
(13) $^{56}Fe$ & $\chi_{Fe} w_{C}$ & $\chi_{Fe} w_{C}$ & & & & & & & & & & &
$1-d$\\
\hline
\end{tabular}
\end{center}
\normalsize
\end{minipage}
\end{table*}


\subsection{The Initial Mass Function}

We adopt a Salpeter--like IMF:

\begin{equation}
 \Phi(M)\,dM=C\,M^{-\mu}\,dM 
\nonumber
\end{equation}

\noindent
where $\mu$=1.35 in Salpeter's (1955) law.

Once the slope $\mu$ and the limiting masses $M_{l}$ and $M_{u}$ are chosen,
the coefficient $C$ is fixed by normalizing the IMF over the
whole mass interval:

\begin{equation}
 \int_{M_{l}}^{M_{u}} \Phi(M)\,dM\,=1 
\nonumber
\end{equation}

\noindent
Since the bulk of chemical enrichment is due to stars with $M \ge$1~\Msol, it
is meaningful to fix the fraction $\zeta$ of the total stellar mass 
distributed in stars above 1~\Msol, which is equivalent to fixing $M_{l}$
(see below). Then we get the normalization condition: 

\begin{equation}
 \zeta\,=\,\int_{M_{1}}^{M_{u}} \Phi(M)\,dM \,=\,
C \int_{M_{1}}^{M_{u}} M^{-\mu}\,dM 
\nonumber
\end{equation}

\noindent
where $M_{1}$=1~\Msol\ and $M_{u}$ is chosen to be 100~\Msol\ (no 
VMOs are supposed to be present).
The normalization condition fixes $M_l$ once $\zeta$ is given:

\begin{equation}
 \int_{M_{l}}^{M_{1}} \Phi(M)\,dM + \zeta \,=1 
\nonumber
\end{equation}

\noindent
The slope $\mu$ of the IMF may not be constant over the whole range of stellar
masses (Miller \& Scalo 1980; Scalo 1986). The model includes also the 
possibility
of a variable $\mu$ over different mass ranges; in this case, beside the 
normalization condition, we impose that the IMF is continuous 
where the different mass ranges connect and thus determine
the normalization coefficients $C_{1}$,......$C_{n}$ for the different ranges.
See, for instance, Chiosi \& Matteucci (1982).


\subsection{The contribution of stellar ejecta}
\label{Qij}

\noindent
We calculate the fraction
of a star of initial mass $M$ that is ejected back
in form of chemical species $i$ 

\begin{equation}
 R_{Mi} = \frac{E_{iM}}{M} 
\nonumber
\end{equation}

\noindent
according to our results on stellar
ejecta (discussed in previous sections) and by means of the ``$Q_{ij}$ matrix''
formalism, first introduced by Talbot \& Arnett (1973) and later adopted by many
authors (Chiosi \& Matteucci 1982; Matteucci \&
Fran\c{c}ois 1989; Ferrini \etal 1992). Each matrix element $Q_{ij}$ is defined
as the mass fraction of a star originally in form of species $j$ which has been
processed and ejected as species $i$. $R_{Mi}$ is the sum
over all ``fuels'' $j$: 

\begin{equation}
\label{RMi}
R_{Mi} = \sum_{j=1}^{N}Q_{ij}(M)\,X_{j}
\end{equation}

\noindent
Each matrix element is defined as:

\begin{equation}
 Q_{ij} = \frac{M_{ij,exp}}{X_{j}M}
\nonumber
\end{equation}

\noindent
where $X_{j}$ is the initial mass abundance of species $j$ and
$M_{ij,exp}$ is the amount of species $i$ synthesized starting from $j$ and
eventually expelled. The diagonal elements $Q_{ii}$ represent the unprocessed
fraction of species $i$ that is eventually re-ejected. By summing upon all the
sources of species $i$, one must get the total ejecta of $i$: 

\begin{equation}
\label{Qijnorm}
\sum Q_{ij} X_{j} M= E_{iM}
\end{equation}

\noindent
Each stellar mass $M$ corresponds to a different $Q_{ij}$ matrix, which 
generally depends also on the metallicity of the star.

The $Q_{ij}$ matrix of Talbot \& Arnett originally treated \Hydrogen, $^{2}$H,
\Hetre, \Helium, \Carbon--\Oxygen, \Nitrogen, neutron rich isotopes ({\it nr})
and the bulk of heavy elements ({\it h}). Later, the dimensions of the matrix
were extended to include a higher number of chemical species (Ferrini \etal
1992). The formalism we use here is analogous to that in Ferrini
\etal (1992), although the definitions of some matrix elements have been
revised 
In our model we follow the evolution of 13 elements: \Hydrogen,
\Helium, \Carbon, \Ctredici, \Nitrogen, \Oxygen, neutron rich isotopes ($nr$),
\Neon, \Magnesium, \Silicon, \Sulfur, \Calcium, \Fe. 

The set of non--zero $Q_{ij}$ elements is displayed in Tab.~\ref{Qijtab}. Here
we list the various quantities entering the matrix elements: 
\medskip

\begin{tabular}{c p{5.5truecm}}
$d$ & mass fraction eventually locked in the remnant \\
$q_{4}$ & mass fraction involved in H--burning \\
$q_{C}$ & mass fraction involved in He--burning \\
$w_{C}=q_{C}-d$ & mass fraction that has been processed by He--burning
and has been ejected \\
$q_{Ns}$ & mass fraction where \Carbon, \Ctredici\ and \Oxygen\ have been
turned to secondary \Nitrogen\ by the CNO cycle \\ 
$q_{C13s}$ & mass fraction where \Carbon\ is turned to secondary \Ctredici\ by 
the CNO cycle \\
$\chi_{i}$ & fractionary mass abundance within $w_{C}$ of {\it newly
synthesized} species $i$ \\ 
\end {tabular}

\medskip \noindent
The meaning and the calculation of the non--zero elements of the $Q_{ij}$ 
matrix are discussed in App.~C.

\begin{table*}[ht]
\begin{minipage}{18truecm}
\caption{Stellar lifetimes in years for stars from 0.6 to 120~\Msol, 
for different metallicities}
\label{lifetimetab}
\begin{scriptsize}
\begin{center}
\begin{tabular}{|c|c|c|c|c|c|}
\hline
 M  &  Z=0.0004  &  Z=0.004   &  Z=0.008   &  Z=0.02    &   Z=0.05\\
\hline
 0.6 & 4.28E+10 & 5.35E+10 & 6.47E+10 & 7.92E+10 & 7.18E+10\\
 0.7 & 2.37E+10 & 2.95E+10 & 3.54E+10 & 4.45E+10 & 4.00E+10\\
 0.8 & 1.41E+10 & 1.73E+10 & 2.09E+10 & 2.61E+10 & 2.33E+10\\
 0.9 & 8.97E+09 & 1.09E+10 & 1.30E+10 & 1.59E+10 & 1.42E+10\\
 1.0 & 6.03E+09 & 7.13E+09 & 8.46E+09 & 1.03E+10 & 8.88E+09\\
 1.1 & 4.23E+09 & 4.93E+09 & 5.72E+09 & 6.89E+09 & 5.95E+09\\
 1.2 & 3.08E+09 & 3.52E+09 & 4.12E+09 & 4.73E+09 & 4.39E+09\\
 1.3 & 2.34E+09 & 2.64E+09 & 2.92E+09 & 3.59E+09 & 3.37E+09\\
 1.4 & 1.92E+09 & 2.39E+09 & 2.36E+09 & 2.87E+09 & 3.10E+09\\
 1.5 & 1.66E+09 & 1.95E+09 & 2.18E+09 & 2.64E+09 & 2.51E+09\\
 1.6 & 1.39E+09 & 1.63E+09 & 1.82E+09 & 2.18E+09 & 2.06E+09\\
 1.7 & 1.18E+09 & 1.28E+09 & 1.58E+09 & 1.84E+09 & 1.76E+09\\
 1.8 & 1.11E+09 & 1.25E+09 & 1.41E+09 & 1.59E+09 & 1.51E+09\\
 1.9 & 9.66E+08 & 1.23E+09 & 1.25E+09 & 1.38E+09 & 1.34E+09\\
 2.0 & 8.33E+08 & 1.08E+09 & 1.23E+09 & 1.21E+09 & 1.24E+09\\
 2.5 & 4.64E+08 & 5.98E+08 & 6.86E+08 & 7.64E+08 & 6.58E+08\\
  3  & 3.03E+08 & 3.67E+08 & 4.12E+08 & 4.56E+08 & 3.81E+08\\
  4  & 1.61E+08 & 1.82E+08 & 1.93E+08 & 2.03E+08 & 1.64E+08\\
  5  & 1.01E+08 & 1.11E+08 & 1.15E+08 & 1.15E+08 & 8.91E+07\\
  6  & 7.15E+07 & 7.62E+07 & 7.71E+07 & 7.45E+07 & 5.67E+07\\
  7  & 5.33E+07 & 5.61E+07 & 5.59E+07 & 5.31E+07 & 3.97E+07\\
  9  & 3.42E+07 & 3.51E+07 & 3.44E+07 & 3.17E+07 & 2.33E+07\\
 12  & 2.13E+07 & 2.14E+07 & 2.10E+07 & 1.89E+07 & 1.39E+07\\
 15  & 1.54E+07 & 1.52E+07 & 1.49E+07 & 1.33E+07 & 9.95E+06\\   
 20  & 1.06E+07 & 1.05E+07 & 1.01E+07 & 9.15E+06 & 6.99E+06\\      
 30  & 6.90E+06 & 6.85E+06 & 6.65E+06 & 6.13E+06 & 5.15E+06\\      
 40  & 5.45E+06 & 5.44E+06 & 5.30E+06 & 5.12E+06 & 4.34E+06\\      
 60  & 4.20E+06 & 4.19E+06 & 4.15E+06 & 4.12E+06 & 3.62E+06\\      
 100 & 3.32E+06 & 3.38E+06 & 3.44E+06 & 3.39E+06 & 3.11E+06\\   
 120 & 3.11E+06 & 3.23E+06 & 3.32E+06 & 3.23E+06 & 3.11E+06\\      
\hline
\end{tabular}
\end{center}
\end{scriptsize}
\end{minipage}
\end{table*}


\subsection{Stellar lifetimes}

To follow the temporal behaviour of different chemical species and
isotopes, we drop the ``instantaneous recycling approximation'' by
taking into account the role of finite lifetimes for stars of different masses.
Our stellar models of different metallicities allow us to 
consider the effects of a different metal content not only on the ejecta
but also on the lifetimes $\tau_{M}=\tau_{M}(Z)$. The
dependence of lifetimes on metallicity can be 
up to a factor of 2 for stars around 1~\Msol. 

The lifetimes we adopt are calculated as the sum $\tau=t_{H}+t_{He}$ of the
H--burning and He--burning timescales of the stellar tracks of the Padua
library, and are listed in Tab.~\ref{lifetimetab}. The trend of
$\tau_{M}$ with metallicity is not univocal: up to solar metallicity,
$\tau_{M}$ increases with $Z$ due to the effect of increasing opacity, while
for super--solar metallicity ($Z=$0.05) $\tau_{M}$ decreases because the helium
content $Y$ increases, due to the assumed ratio $\Delta Y/\Delta Z$=2.5. When
$Y$ increases sensitively, (1) the hydrogen content $X=1-Y-Z$, i.e.\ the
available ``fuel'', decreases and (2) the average molecular weight
$\mu$ increases, which leads to a higher luminosity ($L \sim \mu^{7.4}$). For
super--solar metallicities the effects of $Y$ overcome those of opacity,
giving shorter lifetimes with increasing $Z$. 

In the chemical model, for each mass $M$ and metallicity $Z$ we calculate
the corresponding lifetime $\tau_{M}(Z)$ by interpolating within the 
logarithmic 
relation log($M$)---log($t$) for the tabulated metallicities, and then by 
interpolating with respect to $Z$.

Around $\sim$1~\Msol\ the influence of metallicity reduces lifetimes of a factor
of 2 for low metallicities with respect to the solar case: low and intermediate
mass stars born in early galactic stages eject their nucleosynthetic products
in shorter times.

We assume that each star expels its ejecta all at once at the end
of its lifetime, and that the ejected material is immediately mixed in the
ISM, which remains always homogeneous. This ``instantaneous mixing
approximation'' is suitable to reproduce the average trends of the
age--metallicity relation, of abundance ratios and so on, while it can't model
the observed scatter of the data around the average trend. Only few models
relaxing the instantaneous mixing approximation can be found in literature
(Malinie \etal 1991, 1993; Pilyugin \& Edmunds 1996;
van den Hoek \& de Jong 1997). 


\subsection{The numerical solution}
\label{risolvente}

For the set of equations~\ref{dGi/dt} we adopt the numerical solution 
suggested by Talbot \& Arnett (1971), with the addition of the infall
term (Chiosi 1980).  Each equation is treated as a linear
differential equation, whose analytical solution can be written as:

\begin{equation}
 G_{i}(t_{2}) = G_{i}(t_{1})\,e^{-\chi(t_{1},t_{2})}\,+\,
\int_{t_{1}}^{t_{2}}W_{i}(t)e^{-\chi(t,t_{2})}\,dt 
\nonumber
\end{equation}

\noindent
where:

\[ \chi(t_{1},t_{2}) \equiv \int_{t_{1}}^{t_{2}} \eta(t)\,dt \:, \]

\[\eta(t) \equiv \frac{B(t)}{G(t)} = \frac{\Psi(t)}{G(t)} \]

\noindent
and 
\begin{eqnarray}
\label{Wimass}
W_{i}(t) & \equiv & \int_{M_{l}}^{M_{u}} \Psi(t-\tau_{M}) \Phi(M)
		    R_{Mi}(t-\tau_{M}) \,dM \,+ \nonumber \\ 
	 &	  & + \left[ \frac{d}{dt}G_{i}(t) \right]_{inf}
\end{eqnarray}

\noindent
We perform the integration over a timestep $\Delta t=t^{n+1}-t^{n}$ by
approximating $\eta$ and $W_{i}$ over the whole timestep with a constant value
(the one estimated at time $t^{n+\frac{1}{2}}=t^{n}+\frac{1}{2}\Delta t$):

\begin{equation}
 G_{i}(t^{n+1}) = G_{i}(t^{n}) e^{-\eta\Delta t}\,+\,\frac{W_{i}}{\eta}
   \left[ 1-e^{-\eta\Delta t} \right] 
\nonumber
\end{equation}

\noindent
Since $\eta$ and $W_{i}$ are not
strictly constant over $\Delta t$, an iteration is required to provide with
convergency. As suggested by Talbot \& Arnett (1971), we choose to iterate
only with respect to $\eta$, because $W_{i}$ is an integrated quantity 
involving all past value of
$G(t)$ and is not so sensitive to the exact value of $G(t^{n+1}$), and also
because the resolving expression depends exponentially on $\eta$, while
only linearly on $W_{i}$.
If $G_{i}^{(k)}$ is the estimate of $G_{i}(t^{n+1})$ 
at the $k$th iteration, we proceed through successive corrections:

\begin{equation}
 G_{i}^{(k+1)} = G_{i}^{(k)} \left[ 1+\delta_{i}^{(k+1)} \right] 
\nonumber
\end{equation}

\noindent
where

\begin{equation}
 \delta_{i}^{(k+1)}=\frac{G_{i}^{(k)}-A_{i}}{\beta_{i}-G_{i}^{(k)}} 
\nonumber
\end{equation}

\begin{equation}
 A_{i}=e^{-\eta \Delta t} \left[ G_{i}(t^{n})-\frac{W_{i}}{\eta} \right]\,+
\,\frac{W_{i}}{\eta} 
\nonumber
\end{equation}

\begin{displaymath}
 \beta_{i}=\frac{1}{2} \frac{\partial \ln \eta}{\partial \ln G}
                         \times 
\end{displaymath}
\begin{equation} 
~~~~~\times \left[\left( e^{-\eta \Delta t}-1 \right) \frac{W_{i}}{\eta}-\eta \Delta t
e^{-\eta \Delta t} \left( G_{i}(t^{n})-\frac{W_{i}}{\eta} \right) \right] 
\nonumber
\end{equation}

\noindent
and $\eta$, $W_{i}$ are estimated at $t^{n+\frac{1}{2}}$. $G_{i}^{(k)}$ is
updated until the correction $\delta_{i}^{(k+1)}$ is smaller than a chosen
limit $\delta_{max}$. 

The timestep $\Delta t$ of the model is chosen to be the minimum value between:
(1) a timestep $\Delta t_{1}$ which guarantees that the relative variation 
of the
$G_{i}(t)$'s is lower than a fixed $\epsilon$; (2) a timestep $\Delta t_{2}$
which guarantees that the surface mass density $\sigma(t)$ increases no more than
5\%; (3) a timestep $\Delta t_{3}$ which is twice the previous timestep of the
model, in order to speed up the calculation when possible. In this way, the
infall term in the $W_{i}$'s can be assumed to be constant within $\Delta t$,
and the formulation by Talbot \& Arnett (1971), suited to close models, can be
applied also to open models. 

As for the initial conditions, we assume $\sigma(0)$ to be very small, although 
non--zero to avoid mathematical infinities for $t$=0. At 
the beginning the disk is formed by gas only, while star formation is not 
active yet: $\sigma_{g}(0)=\sigma(0)$. With the adopted normalization, this 
translates in $G(0)=G_{i}(0)=G_{s}(0)\simeq 0$.

The $W_{i}$'s are calculated by integrating with respect to time, rather than 
with respect to mass; Eq.~(\ref{Wimass}) is re-written as:

\begin{displaymath}
W_i(t) =  
\end{displaymath}
\begin{displaymath}
~~~~~  \int_{0}^{t-\tau_{M_u}} \Psi(t') \left[ \Phi(M) R_{Mi}(t')
	     \left(-\frac{dM}{d\tau_M} \right) \right]_{M(t-t')}dt' +
\end{displaymath}
\begin{equation}	     
~~~~~           + \left[ \frac{d}{dt}G_i(t) \right]_{inf}
\label{Witime}
\end{equation}

\noindent
where $M(\tau)$ is the mass of a star of lifetime $\tau$. We need to integrate
with respect to time when introducing the dependence of metallicity, since all
quantities then depend on $Z(t)$ as well as on $M$: the integral on the right 
hand is to be
calculated on the path of the $(M,t)$ plane fixed by the relation between mass
and lifetime, or better by the $M(\tau,Z(t-\tau))$ vs.\ $\tau$ relation,
where $Z(t)$ is built by the on--going model itself.  Also
the ``restitution fractions'' $R_{Mi}$ depend on $Z(t)$ (\S~\ref{Qij});
therefore, all the stored values of $G_{i}(t^{n})$, $X_{i}(t^{n})$ and
$Z(t^{n})$ of all the timesteps $t^{n}$ of the model enter the evaluation of
$\Psi(t')$, $R_{Mi}(t')$ and $dM/d\tau_{M}$ for each $t'$ in the integral. 

Operatively, the integration is turned to a summation over a series of time
intervals $[t^{(k-1)},t^{(k)}]$, 
with $t^{(0)} = 0$ e $t^{(k_{max})} = t$ (the
current age of the model). Within each time interval, the integrand is assumed
to be constant and is estimated at the middle point
$(t^{(k-1)}+t^{(k)})/2$. The time intervals $[t^{(k-1)},t^{(k)}]$ are fixed as
follows. The whole mass range $[M_{l},M_{u}]$ is divided in 200 intervals
$[M^{(p)},M^{(p+1)}]$, where the $M^{(p)}$ are equidistant in logarithmic
scale. At any age $t$ of the model, we calculate the corresponding birthtimes
$t^{(p)}=t-\tau_{M^{(p)}}$. The $t^{(k)}$'s defining the grid for the 
integration are
given by the $t^{(p)}$, implemented with the previous timesteps $t^{n}$ of the
model. 


\subsection{Inserting Type Ia supernov\ae}
\label{SNaeIa}

Up to now we have discussed the ejecta of single stars of different mass
ranges, but in a chemical model we also need to include Type Ia
supernov\ae\ (SN\ae\ Ia), that originate in close binary systems and contribute
an important fraction of heavy elements, especially iron. We need to introduce
SN\ae\ Ia especially to explain the observed evolution of abundance
ratios of $\alpha$--elements with respect to iron (Sneden \etal 1979; Matteucci
\& Greggio 1986). 

Prescriptions for the rate of SN\ae\ Ia and for the composition of their
ejecta are needed. We adopt the rate suggested by
Greggio \& Renzini (1983), which assumes the scenario of Whelan \& Iben 
(1973): SN\ae\ Ia, are due to the explosion of a CO white dwarf that
reaches the Chandrasekhar limit by accreting material from a giant companion 
filling its Roche lobe. An upper limit to the mass of the primary is fixed by 
the requirement that it builds a degenerate CO--core before filling its Roche
lobe, so that it doesn't explode as a SN II. For models with convective 
overshooting, this translates in 

\[ M_1 \leq 6 M_{\odot} \]

\noindent
where $M_1$ is the mass of the primary ($M_1 > M_2$); therefore, the
upper limit for the total mass $M_B=M_1+M_2$ of a binary system able to
produce a SN Ia is: 

\[ M_{B,u} = 12 M_{\odot} \]

Since the primary needs to accrete enough mass to reach the
Chandrasekhar mass, a minimum total mass for the system is also introduced,
generally $M_{B,l} \sim$3~\Msol. The distribution function of the fractionary
mass of the secondary $\mu=M_2/M_B$, $\mu \leq 1/2$ is (Greggio \& Renzini
1983): 

\[ F(\mu)=24 \mu^{2} \]

\noindent
which is normalized between 0 and 0.5 and favours systems whose components have 
similar masses ($M_1 \sim M_2$, Tutukov \& Yungelson 1980).

As the star formation rate for single stars of mass $M$ is $\Psi(t) \Phi(M)$,
we assume that the SFR for binary system precursors of SN\ae\ Ia is
$A \, \Psi(t) \Phi(M)$. This means that a fraction $A$ of stars between
$M_{B,l}$ and
$M_{B,u}$ is assumed to form binaries (with such characteristics that a
SN Ia is eventually produced), rather than single stars. $A$ is a parameter to
be fixed so to match the observed rate of SN\ae\ Ia.

In this scenario, the typical timescale for the explosion of a SN Ia is fixed
by the lifetime $\tau_{M_2}$ of the secondary. The explosion rate (by number)
of SN\ae\ Ia is expressed as (Greggio \& Renzini 1983): 

\begin{displaymath}
 R_{SNI}(t)= 
\end{displaymath}
\begin{equation}
~~~ A \int_{M_{B,l}}^{M_{B,u}} \frac{\Phi(M_B)}{M_B}
		\left[ \int_{\mu_m}^{0.5} F(\mu) \Psi(t-\tau_{M_2}) d\mu
		\right] dM_B 
\nonumber
\end{equation}

\noindent
where $\mu_m$ is the minimum mass fraction contributing to the rate of SN\ae\
Ia at time $t$: 

\[
\mu_m = \max \left\{\frac{M_2(t)}{M_B},\,\frac{M_B-0.5 M_{B,u}}{M_B} \right\}
\]

\noindent
For Type Ia SN\ae\ we adopt the
ejecta $E_{SNI}$ of the W7 model of Nomoto \etal (1984), in the updated version
by Thielemann \etal (1993). Due to the homogeneity of SN\ae\ Ia, we can safely
assume that the same set of ejecta holds for all of these objects. 

To include the contribution of SN\ae\ Ia, we need to change the
formulation of the $W_{i}$'s. Following Matteucci \& Greggio (1986) with some
slight changes, in the range $M_{B,l}$---$M_{B,u}$ we distinguish the
contribution of single stars --- a fraction $(1-A)$ of the whole ---
from that of
binaries originating SN\ae\ Ia --- a fraction $A$ of the whole. In the case of
the binaries, we assume that their ejecta are released in two steps:
after a time $\tau_{M_1}$ the primary expels its products behaving just like
a single star (i.e.\ according to the $Q_{ij}$ matrix suited to its mass and
metallicity), while after a time $\tau_{M_2}$ the secondary pours mass on the
companion originating the SN explosion. Therefore, introducing the notation:

\[ \Psi \Phi R_i (M,t) = \Psi(t-\tau_M) \Phi(M) R_{Mi}(t-\tau_M) \]

\noindent
we can write: 

\begin{displaymath}
W_i(t) =   \int_{M_l}^{M_{B,l}} \Psi \Phi R_i (M,t) \, dM \,+\, 
\end{displaymath}
\begin{displaymath}
 ~~~~ + \,(1-A) \int_{M_{B,l}}^{M_{B,u}} \Psi \Phi R_i (M,t) \,dM \,+
\end{displaymath}	      
\begin{displaymath}
 ~~~~ + \int_{M_{B,u}}^{M_u} \Psi \Phi R_i (M,t) \, dM \,+ \nonumber \\
\end{displaymath}
\begin{displaymath}
 ~~~~ + \, A \int_{M_{B,l}}^{M_{B,u}} \frac{\Phi(M_B)}{M_B}
		\left[ I_1(t,M_B) + I_2(t,M_B) \right] dM_B + \nonumber \\
\end{displaymath}
\begin{equation}
 ~~~~ + \left[ \frac{d}{dt}G_{i}(t) \right]_{inf} 
\nonumber
\label{WiSNImass}
\end{equation}

\noindent
indicating with:

\[ I_1(t,M_B)  =  \int_{0}^{0.5} F(\mu) \Psi(t-\tau_{M_1})
		  R_{M_1i}(t-\tau_{M_1}) M_1 d\mu \, \] 

\[ M_1=(1-\mu)M_B \]

\[ I_2 (t,M_B) = \int_{\mu_m}^{0.5} F(\mu) \,\Psi(t-\tau_{M_2}) \,E_{SNIi} \,d\mu, \]

\[ M_2 = \mu M_B ~. \]

\noindent
In Eq.~(\ref{WiSNImass}), the first three terms on the right hand side 
represent the contribution of
single stars, the fourth term represents that of binaries becoming
SN\ae~Ia and the fifth term is the contribution of infall. Since the ejecta
$E_{SNIi}$ of SN\ae\ Ia are assumed to be independent of $M_{B}$ or
$\mu$, the term describing binaries can also be written as: 

\[
A \int_{M_{B,l}}^{M_{B,u}} \frac{\Phi(M_B)}{M_B} \, I_1(t,M_B) \, dM_B \,+\,
R_{SNI}(t) E_{SNIi}
\]

\noindent
This formulation better shows how the primary star is assumed to evolve like
a single star, unaffected by its companion as far as nucleosynthesis is
concerned and releasing all its ejecta away from the system, while the
secondary pours all its ejecta to produce the SN Ia, without any direct
contribution to the enrichment of the ISM. 

The term expressing the ejecta of the primary stars can also be integrated with 
respect to the mass of the primary $M_1$, if we write it as: 

\[ A \int_{M_{1,min}}^{M_{1,max}} \Psi(t-\tau_{M_1}) \, R_{M_1i}(t-\tau(M_1) \,
{\cal F}(M_1) \, dM_{1} \]

\noindent
where:

\[ M_{1,min}=\frac{M_{B,l}}{2}~~~~~~~~~~~~~M_{1,max}=M_{B,u} \]

\[ {\cal F}(M_1) = \int_{\nu_{min}}^{\nu_{max}} f(\nu) \,\,
\Phi\left(\frac{M_{1}}{\nu}\right) \, d\nu \]

\[ \nu=\frac{M_{1}}{M_{B}}=1-\mu~~~~~~~~~~~~f(\nu)=24 (1-\nu)^{2},~~~
   \nu \in \left[ 0.5,1 \right] \]

\[ \nu_{min} = \max \left\{ 0.5,\frac{M_{1}}{M_{B,u}} \right\}~~~~~~~~~~
   \nu_{max} = \min \left\{ 1,\frac{M_{1}}{M_{B,l}} \right\} \]

\noindent
Integrating the $W_{i}$'s with respect to time and introducing the notation:

\[ \Phi R_i dM (t,t') = \left[ \Phi(M) R_{Mi}(t') \left(-\frac{dM}{d\tau_M} 
		        \right) \right]_{M(t-t')} \]

\[ {\cal F} R_i dM_1 (t,t') = \left[ {\cal F}(M_1) R_{M_1i}(t') 
			      \left(-\frac{dM_1}{d\tau_{M_1}} \right)
			      \right]_{M_1(t-t')} \]

\noindent
when we insert SN\ae\ Ia Eq.~(\ref{Witime}) becomes:

\begin{eqnarray}
W_{i}(t) & = & \int_{0}^{t-\tau_{M_{B,l}}} \Psi(t') \, \Phi R_i dM (t,t') 
	       \,dt' \,+ \nonumber \\
         &   & + \, (1-A) \int_{t-\tau_{M_{B,l}}}^{t-\tau_{M_{B,u}}} \Psi(t')
	       \, \Phi R_i dM (t,t') \,dt' \,+ \nonumber \\
	 &   & + \int_{t-\tau_{M_{B,u}}}^{t-\tau_{M_u}} \Psi(t')
	       \, \Phi R_i dM (t,t') \,dt' \,+ \nonumber \\
	 &   & + \, A \int_{t-\tau_{M_{1,min}}}^{t-\tau_{M_{1,max}}} \Psi(t') 
	       \, {\cal F} R_i(t') dM_1 (t,t') \, dt' \,+ \nonumber \\
	 &   & + R_{SNI} E_{SNIi} \,+ \nonumber \\
	 &   & + \left[ \frac{d}{dt}G_i(t) \right]_{inf}
\end{eqnarray}

\noindent
The integration is performed as described in \S~\ref{risolvente}. Again,
here we need to integrate with respect to time rather than with respect to
mass if we want to include the implicit dependence on the metallicity $Z(t)$. 


\section{Chemical evolution in the Solar Neighbourhood}

As a first application of our model, we analyze the chemical
evolution in the Solar Neighbourhood. Standard observational
counterparts for chemical models, such as 
(1) the current gas fraction, 
(2) the rate of Type~I and Type~II SN\ae\ 
(3) the age--metallicity relation, 
(4) the past and current estimated SFR, 
(5) the distribution of long--lived stars in metallicity (G--dwarf problem), 
are used to calibrate the free parameters of the model. The abundance
ratios of different elements observed in the atmospheres of nearby stars 
are used as a test
for our nucleosynthesis prescriptions, since model predictions for most
abundance ratios depend mainly on the adopted yields and IMF, and
only slightly on other model parameters.

\noindent
Here we list the parameters of the model:

\medskip
\begin{tabular}{|c|p{6cm}|}
\hline
$\nu$ & star formation efficiency \\
$\kappa$ & exponent of the star formation law \\
$\mu$ & exponent of the IMF \\
$M_{u}$ & higher limit for the IMF \\
$\zeta$ & mass fraction of the IMF in stars with $M \geq$1~\Msol \\
$\tau$ & infall timescale \\
$t_G$ & age of the system \\
$A$ & ``amplitude'' factor for SN\ae\ Ia \\
$r_{\odot}$ & galactocentric radius of the Sun \\
$\sigma(r_{\odot})$ & surface mass density in the Solar Neighbourhood \\
\hline
\end{tabular}

\medskip
\noindent
Some of these parameters are directly fixed by observational determinations.
We adopt $r_{\odot} \sim$~8~kpc, as indicated by recent estimates of the 
galactocentric distance of the Sun (Reid 1993; Paczynski \& Stanek 1997).
From a collection of independent measurements of the local surface mass
density of the disc 
Sackett (1997) derives a fiducial estimate of 53$\pm$13~\Msol/pc$^2$;
in this paper we therefore assume $\sigma(r_{\odot})$=50~\Msol/pc$^2$.

For the local IMF, we assume a Salpeter (1955) slope $\mu$=1.35 for
$M <$2~\Msol\ and, following Scalo (1986), $\mu$=1.7 for $M >$2~\Msol.
Actually, the assumed slope below 1~\Msol\ 
is relatively unimportant for our models, since these stars give little
contribution to the chemical enrichment and only act as a sink of matter
from the gaseous phase. What actually matters in the model is the fraction
of mass that each stellar generation stores in such low-mass, long lived stars,
which is formulated in term of a free parameter as (1-$\zeta$).

We further assume the infalling material to be protogalactic gas with
primordial chemical composition: 76\% \Hydrogen\ and 24\% \Helium\
(Walker \etal 1991); the abundance of all heavier elements is assumed
to be zero in the infalling gas (in practice, we assume a negligible
abundance of 10$^{-11}$ to avoid mathematical infinities at the
beginning of the model).

On the other hand, $\nu$, $\kappa$, $\tau$, $\zeta$ and $A$ are treated as free
parameters to be calibrated in order to match the observational
constraints. In the following we discuss the space of model parameter with
respect to the available empirical constraints.


\subsection{The current infall rate}

In principle, an observational estimate of the current infall rate on the
galactic disc could fix the infall timescale $\tau$, since
combining Eqs.~(\ref{eqinfall}) and Eq.~(\ref{eqinfallnorm}) we get

\begin{equation}
\frac{\dot{\sigma_{inf}}}{\sigma}(t_G) = 
\left[ \tau \left( e^{\frac{t_G}{\tau}} -1 \right) \right]^{-1} 
\nonumber
\end{equation}

\noindent
independently of other model parameters.

\begin{figure}[t]
\psfig{file=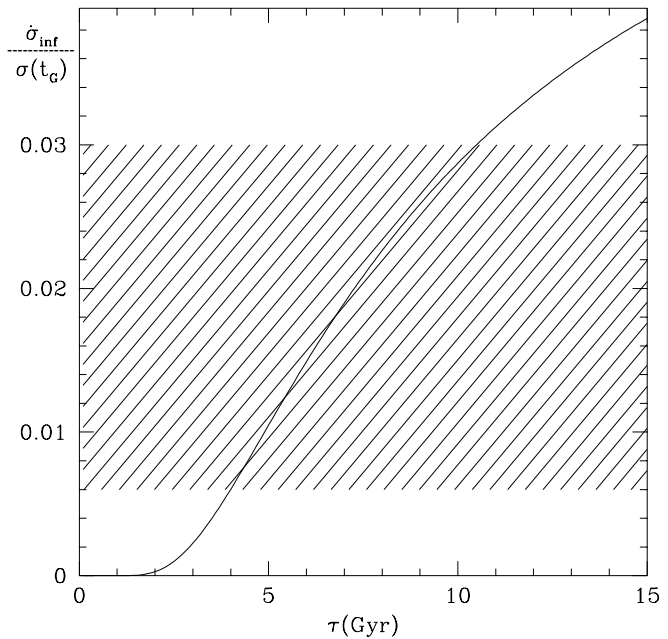,width=9truecm}
\caption{Current normalized infall rate vs.\ infall timescale, as predicted
by model equations (see text). The shaded area traces the observational
estimates.}
\label{infallfig}
\end{figure}

Unluckily, current data cannot provide tight constraints in practice.
Observational evidence of infall comes from High Velocity Clouds (Oort 1970)
and Very High Velocity Clouds, infalling on the disk at a rate of 
$\sim$0.2 and 1.0~\Msol/yr, 
respectively (Tosi 1988 and references therein). While there is general 
agreement on the primordial nature of VHVCs, the metal content displayed
by some HVCs 
suggests a Galactic origin for at least a part of them (Schwarz \etal 1995).
So, we can take the numbers above as the uncertainty range for the current 
global infall rate on the galactic disc. Assuming that such infall is
uniform over the disc and that the radius of the disc is $\sim$15~kpc,
we get a local infall rate of 0.3--1.5~\Msol/pc$^2$/Gyr, or
$\dot{\sigma}_{inf}/\sigma (r_{\odot},t_G) \sim 6 \, 10^{-3} - 3 \, 10^{-2}$
in the Solar Neighbourhood, taking $\sigma(r_{\odot})$=50~\Msol/pc$^2$.
As displayed in Fig.~\ref{infallfig}, any
$\tau \geq 4$ (or even less since our estimate is rather crude)
is basically allowed within the uncertainties, and $\tau$ is not properly
constrained.

Better hints about the infall timescale come from the metallicity distribution
of local long-lived stars: recent data indicate infall
timescales longer that $\sim$5~Gyrs (see \S~\ref{Gdwarfs}). 
Therefore, we start our investigation of the
parameter space by selecting the range $\tau=5 \div 9$~Gyr.


\subsection{The current gas fraction}

\begin{figure}[t]
\psfig{file=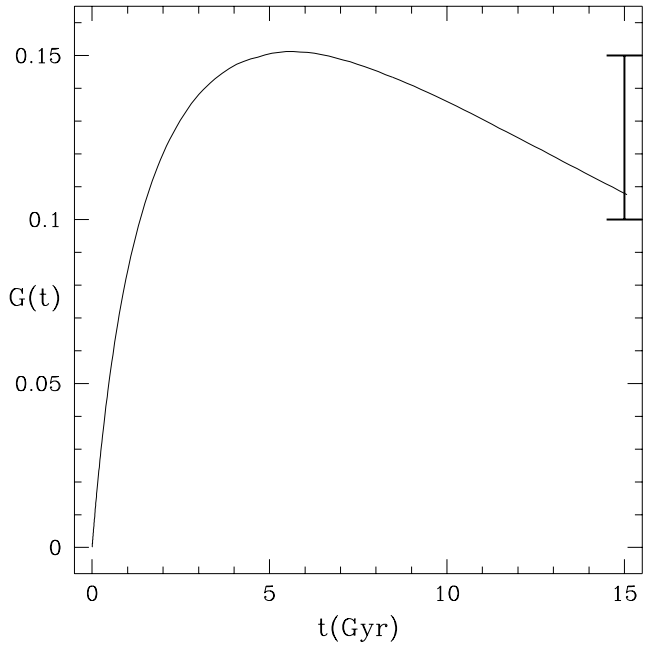,width=9truecm}
\caption{Evolution of the normalized gas fraction G(t) for our model B.
The vertical bar at $t =$15~Gyrs shows the observational constraint
on the current gas fraction}
\label{tGOfig}
\end{figure}

The current gas fraction predicted by our models
is to be compared to observational estimates.
From a compilation of data on the HI and H$_2$ distribution, Rana
(1991) quotes as the current surface gas density in the Solar Neighbourhood 
$\sigma_g(r_{\odot}) \sim$5.7--7~\Msol/pc$^2$, which
implies a gas fraction $\sim$0.1--0.15 of the total surface mass density
($\sim$50~\Msol/pc$^2$).

The gas fraction $G(t_G)$ predicted by our models at the present age
$t_G$=15~Gyr turns out to depend mainly on the parameters $\kappa$ and $\nu$
of the star formation law, while being only slightly dependent on other model
parameters. Since $\kappa$ is limited to the range 1$\div$2 (see 
\S~\ref{SFlaw}), we considered the extreme cases $\kappa=$1, $\kappa=$2 and
the intermediate case $\kappa=$1.5. To fulfill the constraint set by the
current gas fraction, $\nu$ needs to fall within a suitable
range of values, which increase with $\kappa$. For a given $\kappa$, 
the ``good'' values
of $\nu$ slightly increase with the infall timescale $\tau$ and with the
mass fraction $\zeta$ of short-lived stars. Here below we list our
detailed results on the suitable ranges for $\nu$ as a function of
$\kappa$, $\tau$ and $\zeta$.
As an example, Fig.~\ref{tGOfig} shows the evolution of the gas fraction 
in time for our Model B ($\kappa =$1.5, $\zeta =$0.3, $\tau =$9,
$\nu =$1.2, see \S~\ref{Gdwarfs}).

\begin{small}
\begin{tabular}{|l l|c|c|c|}
\hline
 & & $\tau=$5 & $\tau=$7 & $\tau=$9 \\
\hline
$\kappa$=2.0 & $\zeta$=0.2 & {\small 1.2--2.5} & {\small 1.5--3} & 
{\small 2--4} \\
 & $\zeta$=0.3 & {\small 1.2--2.5} & {\small 1.5--3.5} & {\small 2--4} \\
 & $\zeta$=0.4 & {\small 1.5--3} & {\small 2--4} & {\small 2--4.5} \\
 & $\zeta$=0.5 & {\small 1.5--3} & {\small 2--4} & {\small 2--5} \\
\hline
$\kappa$=1.5 & $\zeta$=0.2 & {\small 0.5--0.8} & {\small 0.6--1} & 
{\small 0.7--1.2} \\
 & $\zeta$=0.3 & {\small 0.6--1} & {\small 0.7--1.2} & {\small 0.7--1.3} \\
 & $\zeta$=0.4 & {\small 0.6--1} & {\small 0.7--1.2} & {\small 0.8--1.5} \\
 & $\zeta$=0.5 & {\small 0.7--1.2} & {\small 1--1.5} & {\small 1--1.7} \\
\hline
$\kappa$=1.0 & $\zeta$=0.2 & {\small 0.25--0.3} & {\small 0.35--0.4} & 
{\small 0.3--0.4} \\
 & $\zeta$=0.3 & {\small 0.28--0.35} & {\small 0.35--0.4} & 
{\small 0.35--0.5} \\
 & $\zeta$=0.4 & {\small 0.3--0.4} & {\small 0.4--0.5} & {\small 0.4--0.5} \\
 & $\zeta$=0.5 & {\small 0.35--0.45} & {\small 0.4--0.5} & {\small 0.45--0.6}\\
\hline
\end{tabular}
\end{small}


\subsection{The supernova rates}

\begin{figure}[t]
\psfig{file=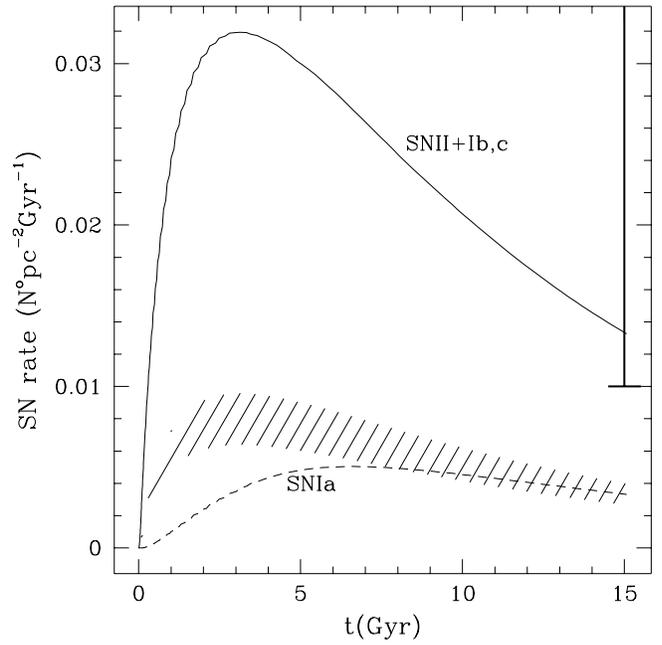,width=9truecm}
\caption{Evolution of the SN~II+Ib,c (solid line) and SN~Ia rate (dashed line)
for our model B, in SN\ae/pc$^2$/Gyr.
The vertical bar at $t=$15~Gyr shows the observational constraints on the
local rate of SN\ae~II+Ib,c. The shaded area traces a factor 0.2$\div$0.3
of the rate of SN\ae~II+Ib,c in time. The observational
constraints on the relative rate of SN\ae~Ia require that at the present age
$t_G$=15~Gyrs the dashed line falls within the shaded area.}
\label{rateSNfig}
\end{figure}

\noindent
The most recent empirical estimates of the Galactic SN rate give
14$\pm$6 (Cappellaro \etal 1997) and 21$\pm$5 (Tamman \etal 1994)
SN events originating by
massive stars --- namely, Type II + Type Ib,c SN\ae\ --- per 1000 yrs.
Taking a disc radius of $\sim$15~kpc, this roughly translates into 
(2$\pm$1)10$^{-2}$ and (3$\pm$1)10$^{-2}$ SN\ae/pc$^2$/Gyr.
We therefore adopt (1$\div$4)10$^{-2}$ SN\ae/pc$^2$/Gyr as the observational
constraint to compare with the rate of SN\ae~II predicted by our models.

As far as SN\ae~Ia are concerned, Cappellaro \etal (1997) give a Galactic
rate of 2$\pm$1 SN Ia per millennium, which leads to a ratio of 0.3$\pm$0.2
of Type Ia versus Type II+Ib,c SN\ae. Tamman \etal (1994) estimate that 
$\sim$85\% of the SN\ae\ exploding in a galaxy like our own come from massive 
stars, which gives a value of $\sim$0.2 for the above mentioned ratio.
We therefore require the number of SN\ae~Ia to be a factor 0.2$\div$0.3 of
the number of SN\ae\ originated by massive stars.
In our models, once $M_{B,l}$ is fixed (see \S~\ref{SNaeIa})
the efficiency with which binaries produce to SN\ae~Ia is
driven by the parameter $A$. For $M_{B,l}$=3~\Msol, $A$ is
constrained to be in the range

\[ A = 0.05 \div 0.08 \]

\noindent
independently of other model parameters. 
Choosing $M_{B,l}$=2~\Msol, $A$ would be constrained in the range
0.03$\div$0.05, but this alternative choice of the ($M_{B,l}$,$A$) has little
effect on any practical outcome of other model predictions. Therefore, in the
following we will limit our discussion to models with $M_{B,l}$=3~\Msol,
$A$=0.05$\div$0.08.

As an example, Fig.~\ref{rateSNfig} shows the evolution
of the local rate of SN\ae\ II+Ib,c and SN\ae\ Ia (solid and dashed line,
respectively) as predicted by our Model B
($A=$0.07), together with the relevant observational constraints. The observed
present relative rate of SN\ae\ Ia is reproduced if, at the present age
$t_G$=15~Gyrs, the dashed line falls within the shaded area, which represents
a factor 0.2$\div$0.3 of the SN~II rate. We notice that this constraint does
not sensitively depend on the assumed age of the disc, since the requirement
remains actually fulfilled over a long period (from 8 to 15~Gyrs).


\subsection{The age--metallicity relation}
\label{AMR}

The age--metallicity relation (AMR) for stars in the Solar Neighbourhood was
first plotted by Twarog (1980). The data showed a large dispersion, therefore
age--bins and the average metallicity per bin were used to trace the local AMR,
which turned out to increase rapidly from 13 to 5 Gyrs ago, and more slowly
afterwards. Part of Twarog's data were re-examined by Carlberg \etal (1985),
who did not include in their sample about fifty of the lowest metallicity stars
and therefore obtained a shallower slope for the AMR in the early phases,
suggesting that the disc evolution started with a high initial metallicity. The
bulk of Twarog's data was later analysed again by Meusinger \etal (1991), and
their AMR confirmed the original result of a steep slope at old ages. 

\begin{figure}[t]
\psfig{file=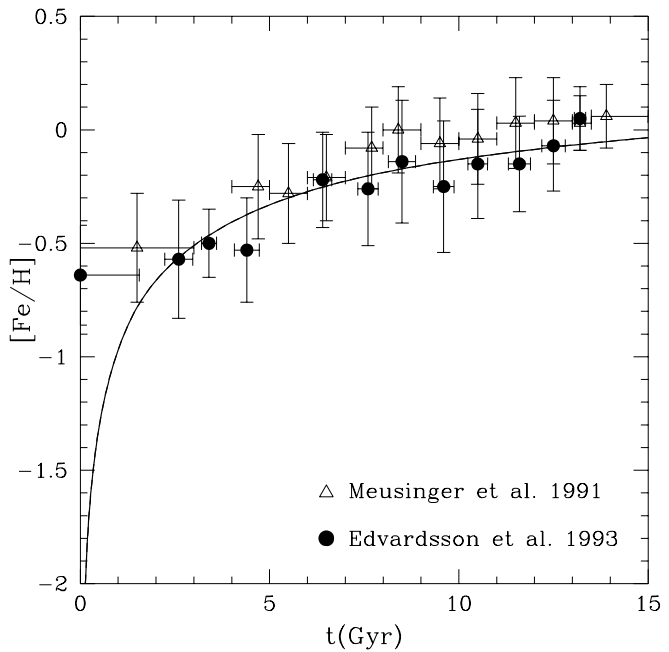,width=9truecm}
\caption{Observational data on the local age-metallicity relation shown
together with the predictions of our Model~B.}
\label{AMRfig}
\end{figure}

A more recent study about the local AMR, based on a new sample of 189 nearby F
and G dwarfs, was performed by Edvardsson \etal (1993) using high resolution
spectra with theoretical LTE atmospheres to derive the chemical compositions,
and photometric fits with Vandenberg (1985) isochrones to derive ages. The
resulting binned, average AMR is in good agreement with that by Meusinger \etal
(Fig.~\ref{AMRfig}), but the dispersion of the data about the average is so
large that the AMR is not a tight constraint for chemical models. Ng \&
Bertelli (1997) have lately re-examined the Edvardsson \etal dataset, by means
of modern isochrones including new opacity tables, of new distances from the
Hipparcos dataset and giving a higher weight to stars with most reliable age
determination. The resulting AMR shows a slope of $\sim$0.07 dex/Gyr, in
agreement with Edvardsson et~al., but still with a large scatter. 

Additional information on the AMR comes from open clusters, whose age estimates
are much more reliable than those for isolated field stars, but with the
counterargument of the uncertainty on the correction for the radial 
metallicity gradient
and the different galactocentric distance of clusters. See Carraro \etal (1997)
for a more comprehensive discussion of the disc AMR.

To reproduce the large scatter in the local AMR, more complex models would be
required, including various possible mechanisms responsible for the scatter
(orbital diffusion, non-instantaneous mixing of enriched material, 
self-propagation of star formation, local infall
episodes and so forth). Our model applies the standard ``one-zone'' scheme,
therefore it is aimed at reproducing average features and can only be compared
to the average trend of the AMR, regardless of the scatter.

The comparison with the observed binned AMR allows us to single out
a limited range of values for the parameter $\zeta$.
Even including the highest possible contribution on
iron enrichment from SN\ae~Ia ($A$=0.08), models with $\zeta<$0.3 still
predict too low metallicities for all the suitable combinations of
$\kappa$, $\nu$ and $\tau$, and are therefore ruled out. At the other end,
models with $\zeta=$0.5 are compatible with the observed AMR only if combined
with the lowest allowed relative rate of SN\ae~Ia ($A$=0.05);
larger values for $\zeta$ are hence ruled out, since such models
would predict too high metallicities. Therefore, from now on we will only
consider models with 

\[ \zeta=0.3 \div 0.5 \]


\begin{figure}[t]
\psfig{file=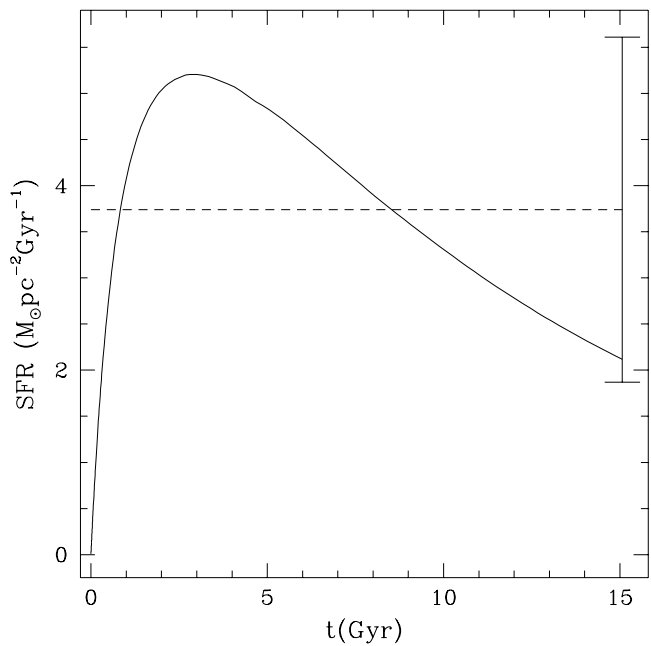,width=9truecm}
\caption{Evolution of the SFR for our model B.
The dashed horizontal line shows the average SFR over the lifetime of the disc,
and the vertical bar at $t =$15~Gyrs shows the range 0.5$\div$1.5 times the
average SFR. Observational constraint (see taxt) require that the current SFR
at $t=t_G$ falls within the range indicated by the bar, as well as within
the observational estimates of 2--10~\Msol~pc$^{-2}$/Gyr$^{-1}$}
\label{SFRfig}
\end{figure}


\subsection{The Star Formation Rate}

The present SFR in the Solar Neighbourhood is estimated to be 
$\sim$2--10~\Msol/pc$^2$/Gyr (G\"usten \& Mezger 1982).
Besides, Scalo (1986) estimates that the present SFR is within
a factor 0.5$\div$1.5 of the average SFR in the past over the whole
lifetime of the disk:

\[ \frac{\Psi(t_G)}{\frac{1}{t_G} \int_0^{t_G} \Psi(t) dt} = 0.5 \div 1.5 \]

\noindent
By imposing these two observational constraints on the predicted
SF history, we can rule out some models and further
restrict the range of ``good'' combinations of the parameters $\kappa$, $\nu$,
$\tau$ and $\zeta$. Notably, if we assume $\tau$=5~Gyr, models
with $\kappa$=2 or 1.5 are unable to fulfill the above requirements
on the SFR.

\noindent
Thus we further filtered the ranges of values for $\nu$ listed here below:

\medskip \noindent
\begin{small}
\begin{center}
\begin{tabular}{|l l|c|c|c|}
\hline
 & & $\tau=$5 & $\tau=$7 & $\tau=$9 \\
\hline
$\kappa$=2.0 & $\zeta$=0.3 & --- & --- & {\small 2--4} \\
 & $\zeta$=0.4 & --- & {\small $\sim$2} & {\small 2--4.5} \\
 & $\zeta$=0.5 & --- & {\small 2--3} & {\small 2--5} \\
\hline
$\kappa$=1.5 & $\zeta$=0.3 & --- & {\small $\sim$0.7} & 
{\small 0.8--1.3} \\
 & $\zeta$=0.4 & --- & {\small 0.7--1.2} & {\small 0.8--1.5}\\
 & $\zeta$=0.5 & {\small $\sim$0.7} & {\small 1--1.5} & 
{\small 1--1.7} \\
\hline
$\kappa$=1.0 & $\zeta$=0.3 & {\small $\sim$0.28} & {\small 0.35--0.4} 
& {\small 0.35--0.5} \\
 & $\zeta$=0.4 & {\small $\sim$0.3} & {\small 0.4--0.5} &
{\small 0.4--0.5} \\
 & $\zeta$=0.5 & {\small 0.35--0.45} & {\small 0.4--0.5} &
{\small 0.45--0.6} \\
\hline
\end{tabular}
\end{center}
\end{small}

\medskip \noindent
As an example, Fig.~\ref{SFRfig} shows the evolution of the SFR
in time for our Model B, with the related observational constraints.


\subsection{The G-dwarf distribution}
\label{Gdwarfs}

\begin{figure}[t]
\psfig{file=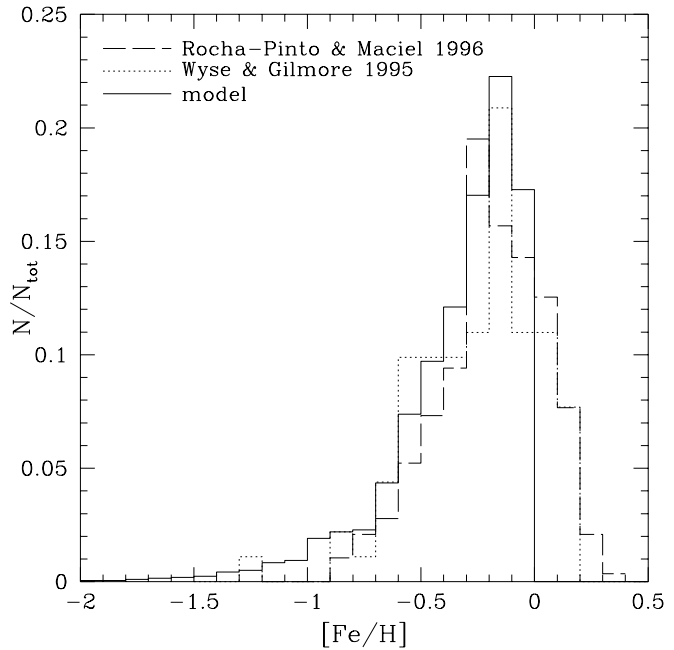,width=9truecm}
\caption{Observational data on the local metallicity distribution,
in relative number, of long-lived
stars (dashed and dotted line) shown together with the predictions of our 
Model~B (solid line).}
\label{Gdwarfsfig}
\end{figure}

\noindent
The G-dwarf problem, i.e.\ the paucity of metal--poor stars in the Solar
Vicinity with respect to the predictions of the ``simple closed--box model'',
is usually interpreted as a consequence of the progressive building-up
of the Galaxy through
accretion of primordial gas, hence the failure of closed models in this respect
(Lynden-Bell 1975; Tinsley 1980).
Other solutions have actually been suggested:
prompt initial enrichment from the halo or bulge providing a finite initial
metallicity in the ISM, higher yields due to an IMF skewed toward massive stars 
in the early galactic phases, more effective star
formation in high-metallicity inhomogeneities of the ISM (see Pagel 1997
for a review); but open models with progressive inflow of primordial material
remain the best scheme to explain the G-dwarf problem, as well as being
consistent with dynamical simulations of the formation of 
galactic discs and with the observation of infall of High and Very High 
Velocity Clouds.

Basing on a complete volume--limited sample of nearby G-dwarfs,
Pagel \& Patchett (1975) provided a reference dataset (later
revised by Pagel 1989 and by Sommer--Larsen 1991), which was
was nicely reproduced by
chemical models with infall timescales of 3--4 Gyrs (e.g.\
Matteucci \& Fran\c{c}ois, 1989). The latest compilations of G--dwarf
metallicities (Wyse \& Gilmore 1995; Rocha--Pinto \& Maciel 1996) show instead
a narrower distribution with a prominent peak around [Fe/H]$\sim$-0.2, and
seem therefore to favour longer timescales (e.g.\ Chiappini
et~al.\ 1997 and Fig.~\ref{Gdwarfsfig}). This is also consistent with dynamical
studies, which indicate
that a rather long time is needed to build up a galactic disc ($\sim$6~Gyrs,
Burkert \etal 1992), especially when one takes into account that primordial or
low-metallicity gas in the early galactic phases has a very low cooling
efficiency (Carraro \etal 1997). 

In our models, the peak metallicity of the predicted distribution increases
with increasing $\zeta$ and $A$. On the other hand, the height of the peak
mainly depends on other parameters: the distribution gets narrower with
increasing $\tau$, $\kappa$, and $\nu$.

All calculated models with $\tau$=5~Gyrs and $\tau$=6~Gyr (and with the
suitable combinations of $\kappa$ and $\nu$ as discussed above) predict
metallicity distributions of stars which are too broad with respect to
the observed one. Therefore, we can rule out all these models and set
the limit

\[ \tau \geq 7 \]

\noindent
On the other hand, models with $\tau \geq$10 tend to predict too narrow
distributions, therefore form now on we will consider only models with
$\tau =$7--9~Gyrs.

If $\zeta$=0.4--0.5, whatever the iron contribution of SN\ae~Ia
(with $A$=0.05 to 0.08) the predicted G-dwarf distribution remaines
peaked around [Fe/H]$\geq$0, not consistent with observations.
We rule these models out and set the limit

\[ \zeta < 0.4 \]

\noindent
In the following, we discuss models with $\zeta$=0.3 as a representative case.
Here below we list, for $\zeta$=0.3, the sets of model parameters which are
able to predict a metallicity distribution in agreement
with the observed one, as well as being in agreement with all the other
constraints.

\medskip \noindent
\begin{small}
\begin{tabular}{|l|l l l l|}
\hline
             & $\tau =$7 & $\kappa =$1.5 & $\nu \sim$0.7 & $A =$0.06--0.07 \\
 & & & & \\
             & $\tau =$8 & $\kappa =$2   & $\nu =$2--3     & $A =$0.07 \\
             &           &               & $\nu \sim$3     & $A =$0.06 \\
             &           & $\kappa =$1.5 & $\nu \sim$1.2   & $A =$0.06 \\
$\zeta =$0.3 &           &               & $\nu \sim$0.7   & $A =$0.07 \\
             &           & $\kappa =$1.0 & $\nu \sim$0.45  & $A =$0.06 \\
 & & & & \\
             & $\tau =$9 & $\kappa =$2   & $\nu \sim$2     & $A =$0.06--0.08 \\
             &           & $\kappa =$1.5 & $\nu =$0.8--1.3 & $A =$0.07 \\
             &           &               & $\nu \sim$0.8   & $A =$0.06-0.08 \\
             &           & $\kappa =$1.0 & $\nu \sim$0.35--0.5 & $A =$0.06 \\
\hline
\end{tabular}
\end{small}

\medskip \noindent
We take as representative of the three cases $\kappa =$1, 1.5 and 2 the
following models:

\medskip 
\noindent
\begin{center}
\begin{small}
\begin{tabular}{|c|c c c c c|}
\hline
 model & $\kappa$ & $\zeta$ & $\tau$ & $\nu$ & $A$ \\
\hline
   A   &    2.0   &   0.3   &    9   &  2.0  & 0.08 \\
   B   &    1.5   &   0.3   &    9   &  1.2  & 0.07 \\
   C   &    1.0   &   0.3   &    9   &  0.5  & 0.06 \\
\hline
\end{tabular}
\end{small}
\end{center}

\medskip 
\noindent
We chose to use model B in Figures~\ref{tGOfig} to~\ref{Gdwarfsfig}
as the representative model to plot versus the observational constraints.
In Fig.~\ref{Gdwarfsfig}, the model well reproduces the location and height
of the peak, while it cannot reproduce the higher-metallicity tail of
the observed distribution. In fact, since our one-zone model cannot
reproduce scatter around the average AMR, no stars are predicted to form
with a metallicity higher than the average one reached at $t_G =$15~Gyrs
(see also Fig~\ref{AMRfig}).


\subsection{Abundance ratios}
\label{abundanceratios}

\begin{figure}
\psfig{file=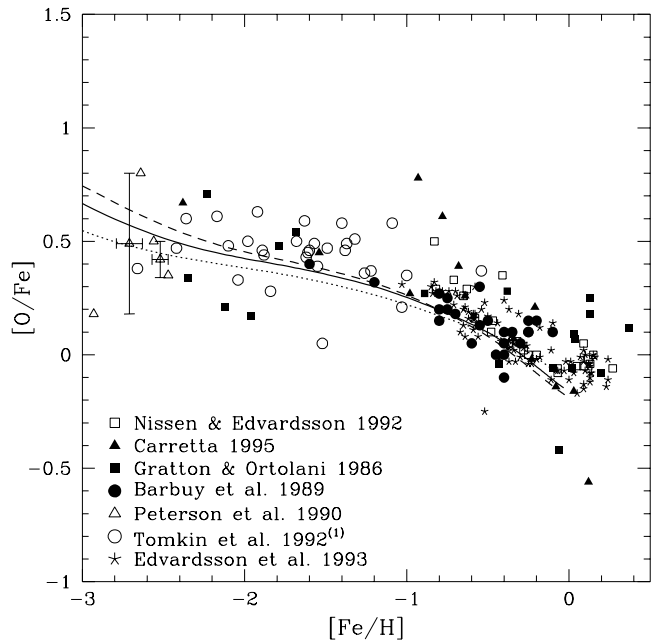,width=9truecm}
\caption{Observational data on the [O/Fe] ratio vs.\ [Fe/H] as observed
in the atmospheres of nearby stars. Predictions of model A (dotted line),
B (solid line) and C (short-dashed line) are shown for comparison. $^{(1)}$The
data from Tomkin et al.\ (1992) are taken as revised by Carretta (1995)}
\label{OsuFefig}
\end{figure}

\begin{figure}[t]
\psfig{file=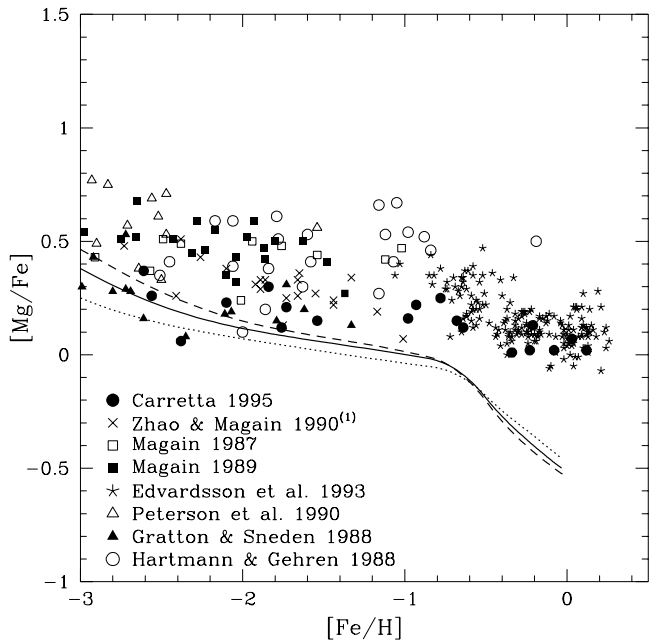,width=9truecm}
\caption{Observational data on the [Mg/Fe] ratio vs.\ [Fe/H]. Model predictions
are plotted as in Fig.~16. $^{(1)}$The data from Zhao \& Magain (1990) are
taken as revised by Carretta (1995)}
\label{MgsuFefig}
\end{figure}

\begin{figure}
\psfig{file=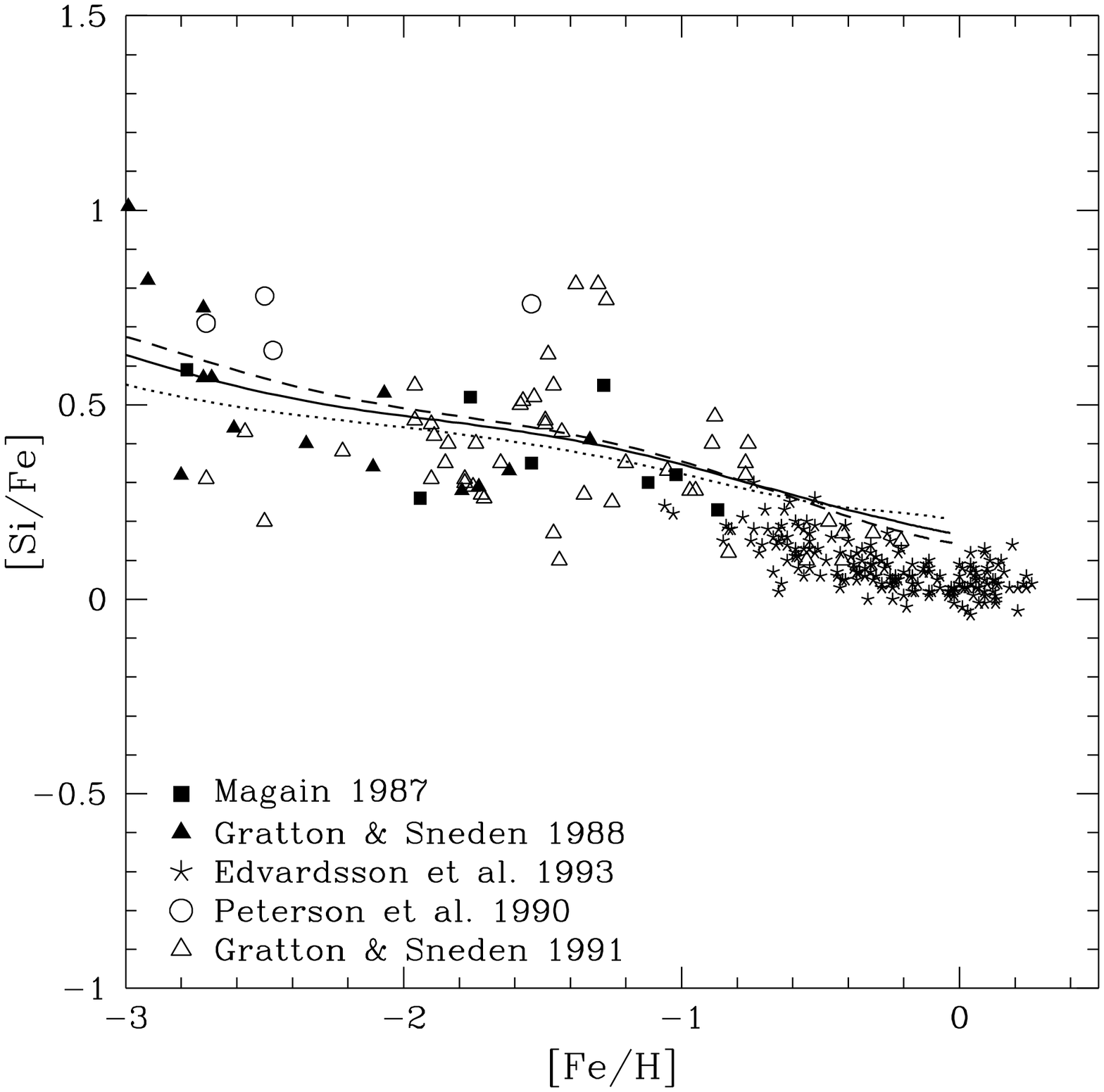,width=9truecm}
\caption{Observational data on the [Si/Fe] ratio vs.\ [Fe/H]. Model predictions
are plotted as in Fig.~16}
\label{SisuFefig}
\end{figure}

\begin{figure}
\psfig{file=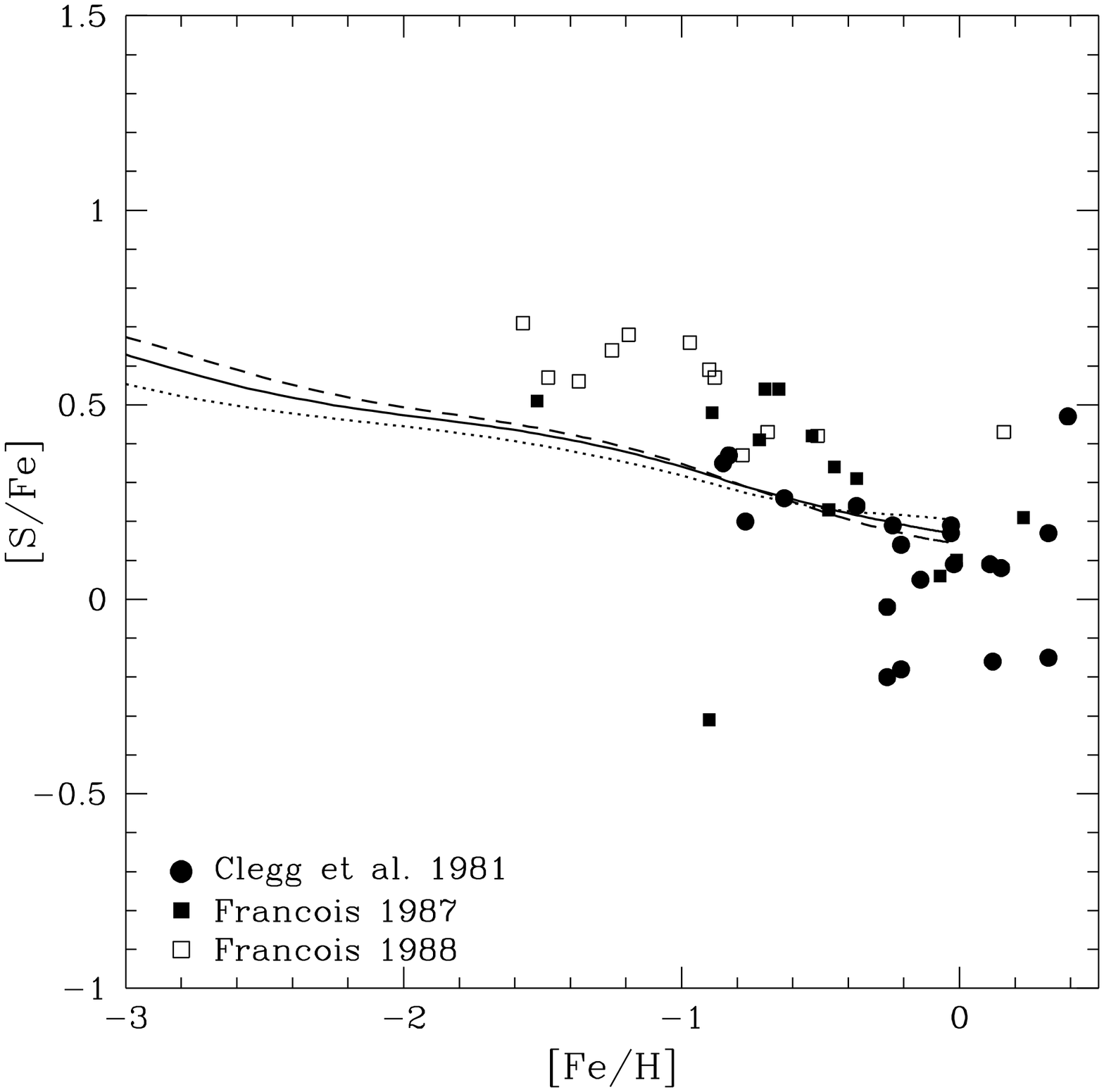,width=9truecm}
\caption{Observational data on the [S/Fe] ratio vs.\ [Fe/H]. Model predictions
are plotted as in Fig.~16}
\label{SsuFefig}
\end{figure}

\begin{figure}
\psfig{file=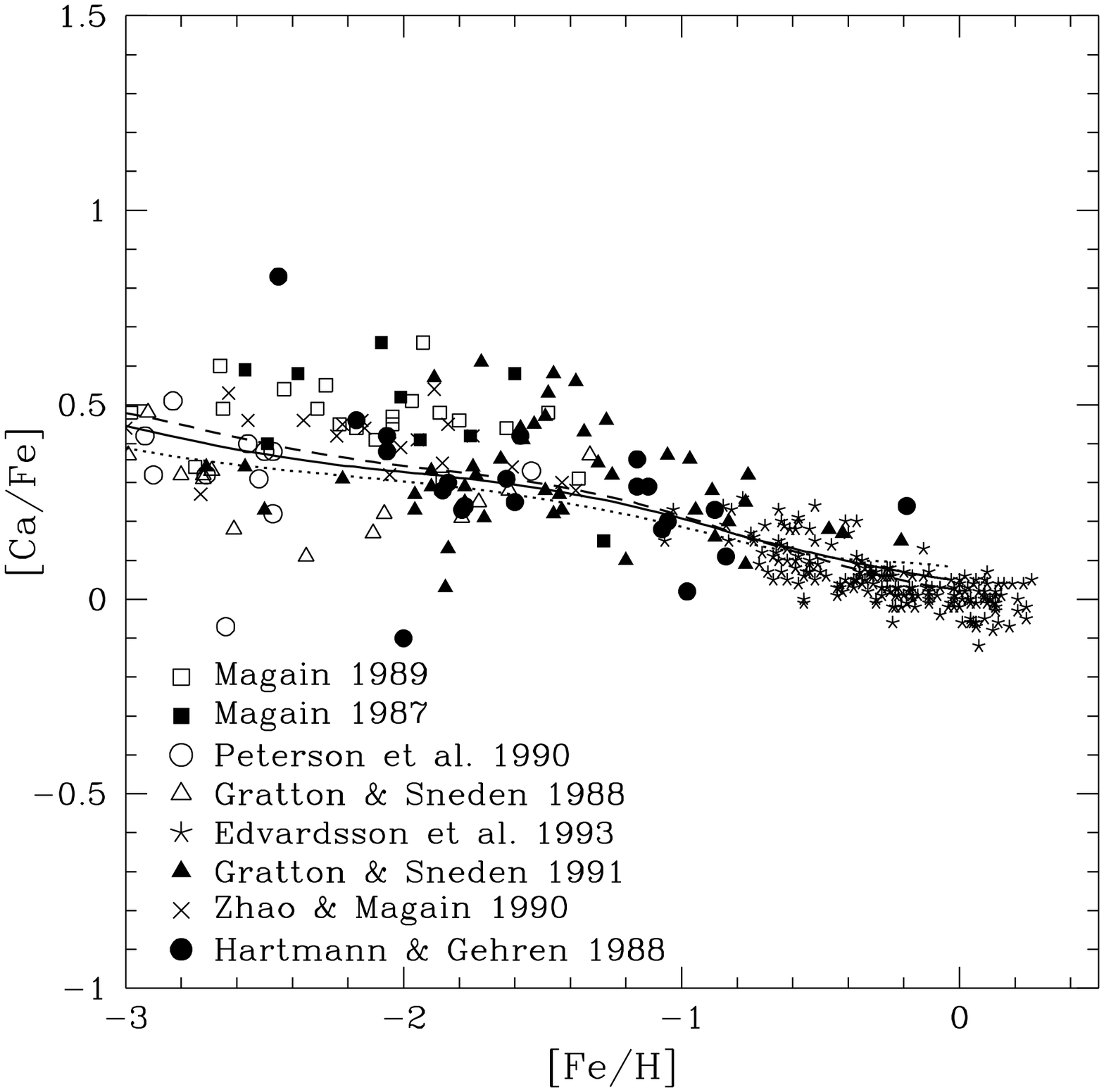,width=9truecm}
\caption{Observational data on the [Ca/Fe] ratio vs.\ [Fe/H]. Model predictions
are plotted as in Fig.~16}
\label{CasuFefig}
\end{figure}

\begin{figure}
\psfig{file=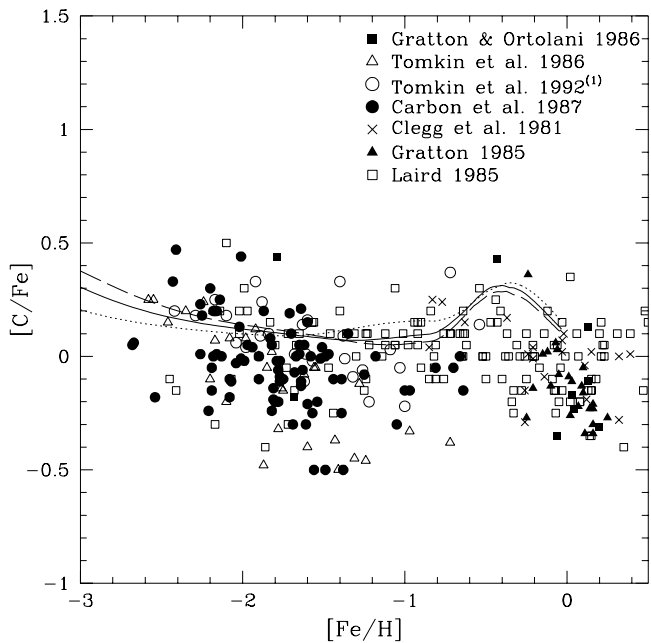,width=9truecm}
\caption{Observational data on the [C/Fe] ratio vs.\ [Fe/H]. Model predictions
are plotted as in Fig.~16. $^{(1)}$The data from Tomkin et al.\ (1992) are
taken as revised by Carretta (1995)}
\label{CsuFefig}
\end{figure}

\begin{figure}
\psfig{file=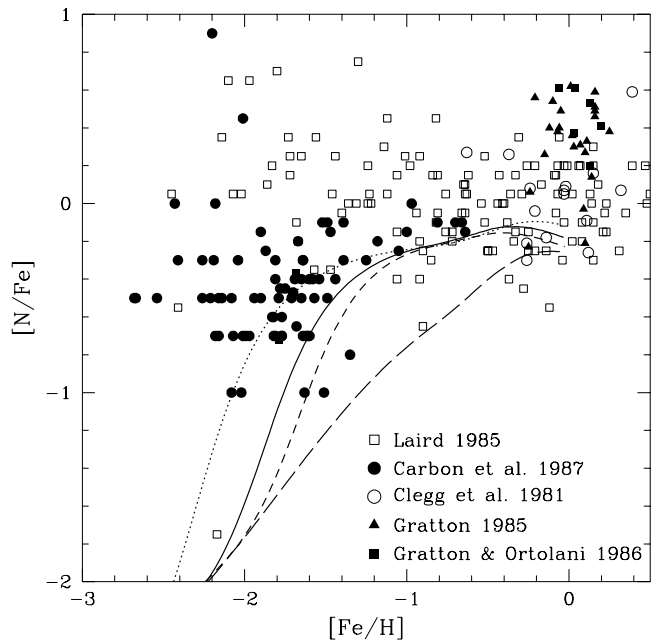,width=9truecm}
\caption{Observational data on the [N/Fe] ratio vs.\ [Fe/H]. Model predictions
are plotted as in Fig.~16. The long--dashed line shows separately the
contribution of the sole secondary component for Model B (see text)}
\label{NsuFefig}
\end{figure}

\begin{figure}[ht]
\psfig{file=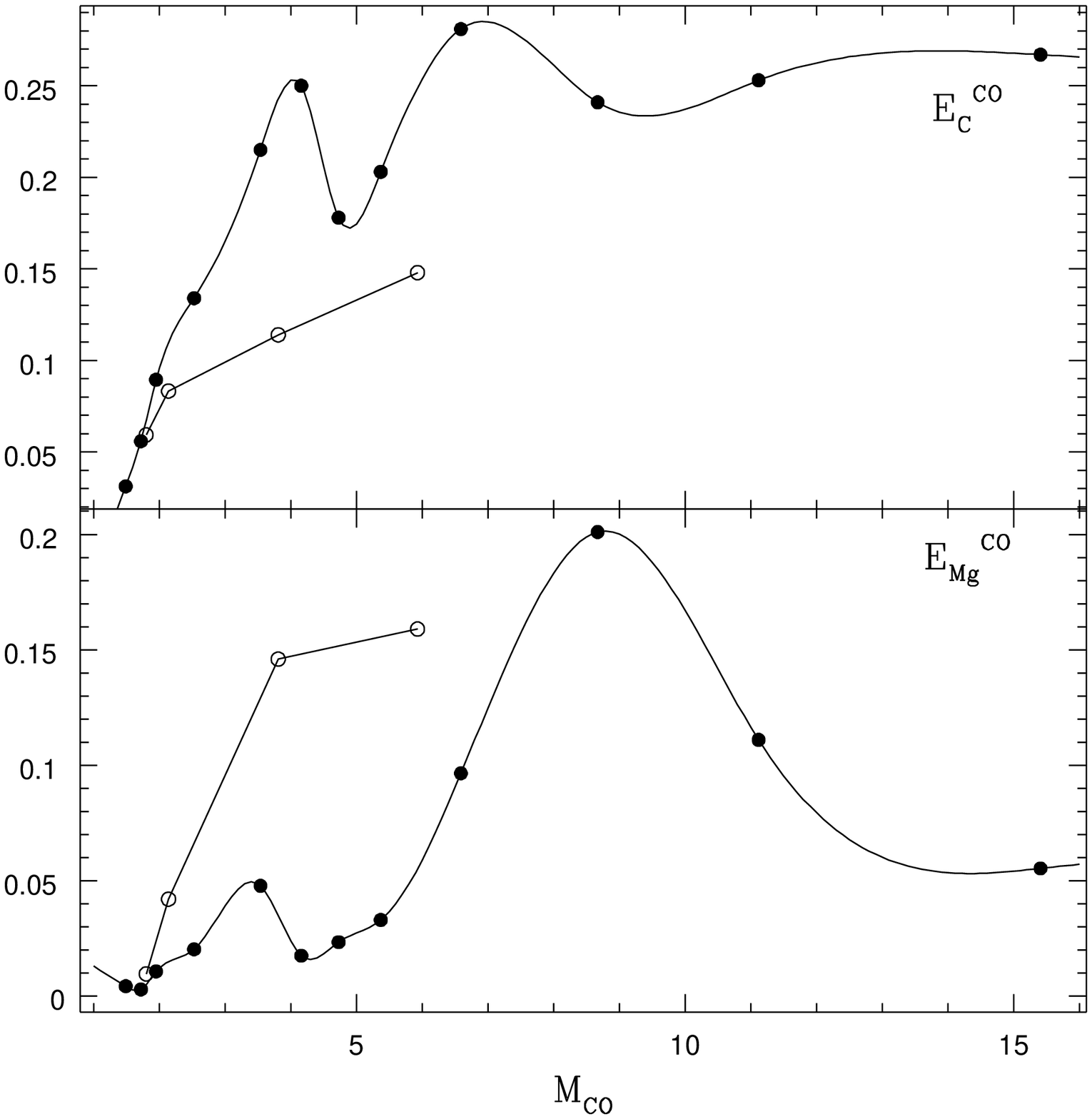,width=9truecm}
\caption{Ejecta of \Carbon\ (upper panel) and \Magnesium\ (lower panel)
of the CO--core vs.\ \MCO\ as deduced from Thielemann \etal 1996 (open circles)
compared with those from WW95 (case $Z =$\Zsol, filled circles)}
\label{ThMCOfig}
\end{figure}

\noindent
Figures~\ref{OsuFefig} to \ref{NsuFefig} display observational data on 
the abundance ratios of various elements with respect to iron, together
with the predictions of our models A, B and C. The data refer to spectroscopic
abundances in the atmospheres of nearby stars; detailed data sources are
indicated in the legends of the figures.

Predictions for the $\alpha$-elements (\Oxygen, \Magnesium, \Silicon,
\Sulfur, \Calcium) seem in most cases to well reproduce the trend of the data.
A remarkable exception is the
[Mg/Fe] ratio. Our yields for the SN explosion are derived from those of WW95,
and therefore display the same problem of \Magnesium\ underproduction 
(Timmes \etal 1995).
As discussed by Thomas \etal (1997), uncertainties in SN calculations still
allow for rather different yields among different authors, especially in the
case of \Magnesium\ production. For instance, \Magnesium\ yields by Thielemann 
\etal (1996) are sensitively higher and better suited at explaining the
[Mg/Fe] ratio observed in low--metallicity nearby stars. See, in this respect,
the lower panel of Fig.~\ref{ThMCOfig}, where we compare the \Magnesium\ 
ejecta from the CO--cores by WW95 with those by Thielemann \etal (1996)
(these latter obtained in a similar fashion as for WW95, see App.~A.
As in Timmes \etal (1995), there is a \Sulfur\ underproduction which could not
be solved by adopting Thielemann \etal's yields, which in the case of
sulfur are even lower than those by WW95.

\noindent
Our model predicts an overproduction of carbon with respect to the observational
data. The mismatch with the data starts at low metallicities, indicating that
the contribution of short-lived, massive stars to C-production is
overestimated. Timmes \etal (1995) do not display a similar problem when
using directly WW95 SN ejecta. At least two reasons might be tracked for this.
(1) Stellar models with overshooting give a lower limiting mass for SN explosion
($M_{up} \sim$6 \Msol\ rather than 8~\Msol) and therefore C-producing massive
stars have a larger global weight in the IMF is taken into account. (2) Our link
between mass-losing pre-SN structures and SN models requires some assumptions,
especially relevant to the edge of the He-burning region, where carbon is
produced; this might make \Carbon\ yields particularly uncertain in our grid
(see App.~B). With respect to the data, our carbon 
overproduction in massive stars amounts to a factor of $\sim$2, which is
still acceptable when taking into account the uncertainties in SN\ae\ II
yields. Again, if for instance we compare the results by WW95 and by
Thielemann \etal (1996) for a given \MCO, the latter give
\Carbon\ yields up to a factor of 2 lower (see Fig.~\ref{ThMCOfig}, upper
panel).

Unlike for the $\alpha$ elements, 
the predicted [N/Fe] ratio at low metallicities depends sensitively
on the past SF history. This effect is mainly due to the contribution of
primary $\Nitrogen$ from intermediate mass stars, which in the range 4--5~\Msol
undergo envelope CNO--burning during the TP-AGB phase (Marigo \etal 1997).
These stars release their contribution of primary \Nitrogen\ on timescales
of $\sim 1-2 \, 10^8$~yrs, independently of the initial metallicity.
In Fig.~\ref{NsuFefig} we display the relevant data versus the 
the predictions of our models A, B, C. As $\kappa$ decreases (from C to A),
the SF gets less efficient in the early phases, and the metallicity [Fe/H]
evolves more slowly. Therefore, the [N/Fe] ratio increases more rapidly in the
early phases if $\kappa$ is lower.

Abundance ratios of $\alpha$ elements do not show a similar dependence on the
SF efficiency in the early phases, since in the first Gyr or so iron and
$\alpha$ elements are produced by the same source (massive stars), and their
relative abundance is determined by the adopted yields and IMF for massive
stars, irrespectively of the SF history. On the contrary, in the case of
\Nitrogen\ the effect is related to the competition between different sources
for the two elements: if $\kappa$ is higher, more gas is recycled through
successive generations of massive stars in the first $\sim 10^8$~yrs, and 
intermediate mass stars contribute to the rise of the [N/Fe] ratio when
the metallicity [Fe/H] is already relatively high. In Fig.~\ref{NsuFefig} 
we also display the contribution of secondary \Nitrogen\ alone (long dashed
line); unlikely the primary component, the [N/Fe] ratio due to the bare
secondary component is basically independent of the SF history, therefore
we just show Model B for the secondary component. Evidently, 
primary \Nitrogen\ from intermediate
mass stars is a major component in the overall \Nitrogen\ production, and
must be properly taken into account. Still, even including this component,
models seem to underproduce \Nitrogen\ with respect to \Fe, especially
in the low-metallicity range. A classical
suggestion to overcome this mismatch is to assume some primary \Nitrogen\
contribution by massive stars (Matteucci 1986). This might be possible
in very low-metallicity massive stars if some peculiar mixing event is
allowed to occur (Timmes \etal 1995). Fig.~\ref{NsuFefig} might also suggest,
as an alternative, that a slow SF history ($\kappa=$1) combined with a more
efficient production of primary \Nitrogen\ in intermediate mass stars
might solve the problem. A more extreme alternative might be 
an initial IMF skewed to massive
stars, which quickly provide the ISM with a minimum amount of CNO isotopes
so that secondary production of \Nitrogen\ becomes effective already at very
low metallicities. But this alternative should need to be tested also with 
respect to the other abundance ratios and the other model constraints.


\section{Summary and conclusions}

We have presented a detailed study of the ejecta and chemical yields
from stars of different mass and initial chemical composition, taking into
account modern results of hydrostatic and explosive nucleosynthesis.
The analysis is made for stars that avoid the AGB phase. i.e. from 6
\Msol up 1000~\Msol. 

In this range of mass we distinguish three main 
groups: (i) the 6 to 9~\Msol\ stars,  which are not 
significantly affected by mass loss by stellar wind and
terminate their evolution  by the electron
capture instability and consequent supernova explosion; (ii) the 
9 to 120~\Msol\ stars,  which suffer from mass
loss by stellar wind starting from the zero age
main sequence and terminate as iron-core collapse or PC supernov\ae; 
and   finally (iii) the rather hypothetical range of very
massive
objects (120 to 1000 ~\Msol), which are characterized by pulsational
instability and violent mass loss during the core H-burning phase
and may terminate the evolution either as iron-core collapse
or ``PC'' supernov\ae or black holes, depending on the size of their
CO--core 

We start from the sets of hydrostatic stellar models  by
the Padua group that are calculated in presence of mild convective overshoot. 
This explains
why the upper mass limit for the occurrence of the AGB phase is set
at about 5~\Msol\ depending on the initial chemical composition. 
All these evolutionary sequences  are carried out from the ignition of
the 
core H-burning till the start of central C-burning, taking into
account  mass loss by stellar wind wherever appropriate. These sets of
models span a wide range of initial masses and chemical 
composition and are fully homogeneous as far as the input physics,
accuracy and numerical methodology are concerned. So they are the ideal
tool  to construct  equally homogeneous sets of ejecta and stellar
yields. 

From the models we  derive the basic relationships between
the initial and final (namely, at the onset of C-burning in the core)
mass of the stars, and between the initial mass and the final mass of the
helium and carbon--oxygen core. The subsequent evolutionary stages are so
short-lived that the above relationships will not change at all. 
If mass loss occurs by stellar wind during the core H- and He-burning stages,
the amounts of mass in the various species lost in the
wind are calculated.

Adopting the CO core mass as the driving parameter we  link 
the hydrostatic structure at early core C-burning to the final explosive stage.
Using literature data on  supernova models,  we calculate the
amount of mass ejected by each CO core  at the time of explosion for
each  species under consideration, to 
which the contribution from the overlying layers still existing in
the star at this stage, and the contribution from stellar wind in
the  earlier stages are added.

The results are summarized  in a series of tables, and  all details relative to
construction of the ejecta and yields for massive and very massive
stars  are given in the main text and  Appendices A and B.

The corresponding ejecta and yields for  low and intermediate mass stars (those
passing though the AGB phase) can be found in
Marigo et al. (1996, 1997). They are based on the same stellar models
in usage here up to the end of the E-AGB, and are derived from an
original semi-analytical method suitably tailored to follow the
TP-AGB 
phase in presence of envelope burning in the most massive stars of this
group. So the ejecta and yields for massive and very massive objects 
presented here together with those by  Marigo et
al. (1996, 1997)  constitute a dataset of unprecedented homogeneity.

Our results about stellar nucleosynthesis and ejecta have been applied
to study the chemical enrichment of the Solar Neighbourhood by means
of the infall model of Chiosi (1980), in which the matrix formalism
developed by Talbot \& Arnett (1973) to calculate  the fraction of a
star of initial mass $M$ that is ejected in form of chemical
species $i$ has been incorporated. This matrix formalism is particularly
useful to understand what fraction of a star originally in form of species
$j$ is eventually ejected as species $i$. 
The numerical method to solve the set of model equations governing
the chemical evolution of the ISM is the same as in Talbot \& Arnett (1971),
however adapted to the infall scheme. 

The chemical model contains a number of parameters, some of which are 
directly fixed by the observations or estimated independently
(e.g.\ the distance of the Sun from the Galactic center, the local
surface
mass density and the age of the galactic disk). Other parameters 
are let vary so that some
basic observational constraints are met. All our results refer to the
Scalo IMF with fixed slope and upper mass limit $M_u$=100~\Msol.

Constraints of the models are 
(i) the current gas fraction;
(ii) the rate of type~I and type~II SN\ae; 
(iii) the age metallicity relation;
(iv) the past and present estimated star formation rate; 
(v) the distribution of long-lived stars in metallicity (G-dwarf problem). 

Free parameters of the models are:
(i) the efficiency and exponent of the SFR ($\nu$ and $\kappa$, respectively);
(ii) the infall time-scale $\tau$; 
(iii) the mass fraction $\zeta$ of the IMF in stars with $M \geq$1~\Msol;
(iv) the amplitude factor $A$ and minimum mass of the binary $M_{B,l}$
     for Type~Ia SN\ae.

A careful comparison of model results with the  observational constraints
allows us to disentangle the effects of different parameters on the model
predictions and put successive limits on them. We are finally left with
the following ``good ranges'' for the various parameters:
1$\leq \kappa \leq$2, 0.3$\leq \nu \leq$3 (in suitable combination with
$\kappa$), 0.3$\leq \zeta \leq$0.4, 7$\leq \tau \leq$9~Gyrs, 
0.05$\leq A \leq$0.08 (for $M_{B,l}$=3~\Msol).  

Predictions for most elemental ratios as a function of [Fe/H]
seem to reproduce the trend of the data,
but for [Mg/Fe] and to a less extent for [C/Fe] and [N/Fe].
Possible reasons for this disagreement are briefly discussed. In the case of
\Magnesium\ and \Carbon, they might be reconduced to uncertainties
in the ejecta of SN\ae. In the case of \Nitrogen, the predicted [N/Fe] is
crucially dependent on the past SF history due to the contribution of primary 
\Nitrogen\ from intermediate mass stars.
Possible ways out to the low predicted \Nitrogen\ abundance are briefly
recalled: either primary \Nitrogen\ from massive stars of low metallicity 
(Matteucci 1986) or an IMF more skewed toward  massive stars than
predicted by a Salpeter--like law during the very early stages of galaxy
formation and evolution (cf. Chiosi et al. 1997). Both
hypotheses need to be thoroughly investigated. The problem is open.


Finally, we like to remark our effort to derive metallicity
and hence time dependent stellar yields, since limiting to consider
constant yields is not compatible with our current understanding
of stellar evolution.
But we also need to underline that for some elements many uncertainties still
exist on the nucleosynthetic yields of stars in the different mass ranges.
These uncertainties display already when considering model predictions
for the Solar Neighbourhood, and should be borne in mind when applying
chemical models to external galaxies, where detailed tests of the predictions
are more difficult.



\medskip

\acknowledgements{L.P.\ warmly thanks Paola Marigo for stimulating and helpful
discussions on stellar evolution and nucleosynthesis, and acknowledges kind
hospitality from the Nordita Institute of Copenhagen.
C.C.\ is pleased to acknowledge the hospitality and
stimulating environment provided by ESO in Garching where this paper
was finished during sabbatical leave from the Astronomy Department of
the Padua University.
This study has been financed by the Italian Ministry of
University, Scientific Research and Technology (MURST), the Italian Space 
Agency (ASI), and the European Community (TMR grant ERBFMRX-CT-96-0086).  }




\appendix

\section*{Appendix A: the link with SN models}
\label{link}
\setcounter{equation}{0}
\renewcommand{\theequation}{A\arabic{equation}}

Here we describe in detail the method we adopt to
establish the relations between \MCO\ and the amount of different elements
expelled by the core, so that we can ``complete'' the total
ejecta of our models with the outcome of the final iron--core collapse SN.
To this aim, we need to deduce the contribution of the sole
CO--core  $E_{i}^{CO}$ to the total ejecta of the SN models by WW95.
In the following, we adopt this notation: 

\begin{tabular}{l p{7truecm}}
$M$ & total mass of the model by WW95\\
\MCO\ & mass of the CO--core of the model\\
$Z$ & metallicity of the model\\
$E_{i}$ & total ejected amount of species $i$, taken from WW95 tables\\
$E_{i}^{new}$ & newly synthesized and ejected amount of species $i$\\
$X_{i}^{0}$ & initial abundance of species $i$ in the model\\
$E_{i}^{ext}$ & amount of species $i$ contained in the layers external to 
\MCO\\
$E_{i}^{CO}$ & amount of species $i$ ejected by the CO--core, i.e.\
contribution of the CO--core to the global ejecta of $i$\\
$M_{env}$ & mass of the region unaffected by CNO burning (roughly 
corresponding to the mass of the envelope, apart from dredge-up episodes)\\ 
\end{tabular}

First of all, for each tabulated total mass we need to deduce the relevant
value of \MCO. We assume that the expelled amounts of H and He originate solely
in the layers overlying \MCO, that is to say (1) in the H--rich envelope,
possibly enriched in \Helium\ and in \Nitrogen\ by convective dredge-up
episodes, and (2) in the fraction of the He--core that was not involved in the
He $\rightarrow$ CO burning, \MHe -- \MCO. (With these assumptions we neglect a
possible contribution in He from an $\alpha$ rich freeze--out, which anyway
should contribute only a small fraction of the total He ejecta.) The global
metallicity in the layers external to \MCO\ is assumed to correspond to the
initial one, since in these layers no nuclear burning takes place but
H--burning and the CNO cycle, though altering the relative abundances of CNO
isotopes, leaves the overall metallicity substantially unchanged. Therefore, on
the base of the expelled amount of H and He, and correcting for the initial
fraction of metals, we can derive the mass of the layers over the CO--core;
then we derive \MCO\ as: 

\begin{equation}
\label{MCOeq}
M_{CO} = M-\frac{E_{H}+E_{He}}{1-Z}
\end{equation}

\noindent
In the 15 and 25~\Msol, $Z$=\Zsol\ cases (Tab.~S in WW95), we get \MCO~=~2.53
and 6.59~\Msol\ respectively; these values look reasonable when compared
to Figs.~10a,b of WW95, that illustrate the chemical profiles
of the corresponding pre-SN structures. This gives us confidence in the method
we adopt (but see \S~\ref{linkproblems}). Once we get \MCO we consider the
production sites of the various elements, so that we can distinguish the ejecta
originating in the CO--core from the contribution of the overlying layers. 
\begin{description}
\item[\bf $^{14}$N---$^{13}$C]
are produced in the CNO cycle and destroyed by $\alpha$ capture during
He--burning. No \Nitrogen\ and \Ctredici\ are left inside the CO--core, and they
are not produced by explosive nucleosynthesis (see also Tab.~7 in WW95,
comparing the chemical composition of a 25~\Msol\ model before and after
explosion). These species can come only from layers outside \MCO, which doesn't
contribute to their ejecta; see also Fig.~13a,b of WW95.

\[ E_{C13}^{CO} = 0 \]
\[ E_{N}^{CO} = 0 \]

\item[\bf $^{12}$C---$^{16}$O]
are synthesized during He--burning inside \MCO. In order to remove the
contribution of the outer layers, we need to estimate carbon and oxygen
abundances outside \MCO. In layers experiencing the CNO cycle, their original
abundances are altered in favour of  \Nitrogen\ and \Ctredici. Consequently,
we estimate first the
amount of newly synthesized \Nitrogen\ and \Ctredici. Since these are located
only outside \MCO, by subtracting the initial amounts within M--\MCO\ to their
total ejecta we directly get the newly synthesized amounts: 

\[E_{C13}^{new}=E_{C13}-X_{C13}^{0}(M-M_{CO})\]
\[E_{N}^{new}=E_{N}-X_{N}^{0}(M-M_{CO})\]

\Ctredici\ is synthesized in the CNO cycle starting from \Carbon, while
\Nitrogen\ is synthesized starting from \Carbon\ and \Oxygen. Assuming that the
new \Nitrogen\ originates from \Carbon\ and \Oxygen\ proportionally to their
initial abundances, the amount of \Carbon\ and \Oxygen\ lying outside \MCO\ can
be expressed as: 

\[ E_{C}^{ext}=X_{C}^{0}(M-M_{CO})-E_{C13}^{new}-E_{N}^{new}
\frac{X_{C}^{0}}{X_{C}^{0}+X_{O}^{0}} \]
\[ E_{O}^{ext}=X_{O}^{0}(M-M_{CO})-E_{N}^{new}
\frac{X_{O}^{0}}{X_{C}^{0}+X_{O}^{0}} \]

Here, the first term is the amount of \Carbon\ and \Oxygen\ present outside 
\MCO\
since the beginning, from which we remove the amount that has been converted
into \Nitrogen\ and \Ctredici. The contribution of the CO--core to carbon and
oxygen ejecta is: 

\[ E_{C}^{CO} = E_{C} - E_{C}^{ext} \]
\[ E_{O}^{CO} = E_{O} - E_{O}^{ext} \]

\item[\bf $^{15}$N]
is quickly destroyed by the CNO cycle, which reduces its abundance by an
order of magnitude. The SN explosion can produce \Nquindici\ by neutrino
nucleosynthesis on \Oxygen\ ($^{16}$O($\nu_{x},\nu_{x}^{'}\,p$)$^{15}$N). We
assume that the abundance of \Nquindici\ is negligible where the CNO cycle takes
place, unaltered with respect to the initial one elsewhere. We estimate
the mass not involved in H--burning as:

\[ M_{env}=\frac{E_{H}}{X^{0}} \]

Starting with the total ejecta of \Nquindici, we subtract the contribution of
the ``envelope'' (mass unaffected by the CNO cycle) with the initial abundance
of \Nquindici. Whatever is left comes from neutrino nucleosynthesis on \Oxygen,
both on the \Oxygen\ in the CO--core and on the \Oxygen\ in the outer layers;
we further scale the resulting amount of \Nquindici\ with respect to the
fraction of \Oxygen\ within \MCO. The resulting contribution of \MCO\ to the
ejecta of \Nquindici\ is expressed as: 

\[ E_{N15}^{CO}=(E_{N15}-M_{env}\,X_{N15}^{0})\,\frac{E_{O}^{CO}}{E_{O}} \]

\item[\bf $^{17}$O]
abundance within \MCO\ is negligible because \Odiciassette\ is destroyed by
$\alpha$ capture during He--burning --- see also Figs.~13a,b of WW95.

\[ E_{O17}^{CO}=0 \]

\item[\bf $^{18}$O]
is destroyed in the CNO cycle and later produced, but also rapidly destroyed,
during He--burning; no \Odiciotto\ is left inside the CO--core.

\[ E_{O18}^{CO}=0 \]

\item[\bf $^{20}$Ne---$^{24}$Mg---$^{28}$Si---$^{32}$S---$^{40}$Ca]
are produced during He-- and C--burning; we assume that out of the CO--core
their abundances are unaltered with respect to the initial ones. To get
the contribution of the CO--core, we simply remove from the total ejecta the 
contribution of the overlying layers, where the initial abundances hold:

\[ E_{Ne}^{CO}=E_{Ne}-(M-M_{CO})\,X_{Ne}^{0} \]
\[ E_{Mg}^{CO}=E_{Mg}-(M-M_{CO})\,X_{Mg}^{0} \]
\[ E_{Si}^{CO}=E_{Si}-(M-M_{CO})\,X_{Si}^{0} \]
\[ E_{S}^{CO}=E_{S}-(M-M_{CO})\,X_{S}^{0} \]
\[ E_{Ca}^{CO}=E_{Ca}-(M-M_{CO})\,X_{Ca}^{0} \]

\item[\bf $^{56}$Fe]
is produced in the very last hydrostatic burning stage (Si--burning), but most
of the ejected amount of iron originates in the radioactive decay of \Nickel.
The tables from WW95 list the ejecta at a time immediately after
explosion, when the decay hasn't taken place yet; since all the released
\Nickel\ should
later decay in \Fe, we add the ejecta of \Nickel\ to those of \Fe: 

\[ E_{Fe}^{CO}= [E_{Fe}-(M-M_{CO})\,X_{Fe}^{0}]+E_{Ni} \]

\end{description}

\noindent
In the case of 30, 35 and 40~\Msol\ WW95 calculate 
several models (A, B, C) differing in the energy of the ejected material at 
infinity. This results mainly in 
different ejecta of \Fe, since this element is sensitive to the location of 
the {\it mass cut}, which changes with the explosion energy. In 
our link we referred to the A models of WW95; the effect of different
assumptions about the explosion energy can be included in the
uncertainty of a factor of two upon the amount of ejected iron, mentioned
by WW95.



\section*{Appendix B: some warnings about the link}
\label{linkproblems}
\setcounter{equation}{0}
\renewcommand{\theequation}{B\arabic{equation}}

Here we discuss possible drawbacks or incoherences of the method we adopted
to link our stellar models with WW95 SN models. 

(1) We have implicitly assumed that explosive nucleosynthesis only 
involves the CO--core leaving the overlying layers unaffected, as they are 
added from our pre-SN models. This sounds a reasonable approximation, but we 
neglect a possible production of \Nquindici\ from neutrino nucleosynthesis 
over the \Oxygen\ lying out of \MCO\ and a possible contribution of an $\alpha$
rich freeze--out in the ejecta of \Helium.

(2) One of the basic parameters in the adopted SN models is the kinetic energy
of the ejecta at infinity (WW95): the shockwave generating the explosion is
simulated by
means of a piston, regulated so that the final energy assumes a typical value,
$\sim$ 1.2 $\times$ 10$^{51}$ erg. Outside \MCO, the layers in our pre-SN
models have a different mass and structure from those of WW95, both because we
include mass loss and because of the different physical treatment of details of
stellar evolution (especially the inclusion of convective overshooting).
Therefore, for a given \MCO, the kinetic energy of our ejecta at infinity can
be different from that imposed in WW95 models. This affects mainly the remnant
mass, which depends on the explosion energy, on the pre-SN structure, on the
stellar mass and on metallicity (WW95). Most of all, the extent of a possible
{\it reverse shock} as found in WW95 models may not be directly transferred to
our models, because it depends on the density structure of the layers over the
core. A reverse shock can induce a fall back of material toward the collapsed
core, which increases the mass of the final remnant. Since we do not know 
how the remnant mass in WW95 models depends on this physical effect in detail,
we can't correct for it. The choice of adopting the same remnant mass for the
same \MCO\ as in WW95 models can influence our ejecta of
\Silicon, \Sulfur\ and \Fe, that are the elements produced next to the mass cut
of the SN. 

(3) An inconsistency between the two sets of models is the different
cross-section adopted for the fundamental reaction
\Carbon($\alpha$,$\gamma$)\Oxygen. In our tracks the adopted value is from
Caughlan \& Fowler (1988), while WW95 adopt 1.7 times that value. Both values
are within current uncertainties, but they result in a sensitively different
composition of the CO--core: in our models the resulting core consists of a
30\% of \Carbon\ and a 70\% of \Oxygen, while in WW95 models the carbon 
abundance in the core
is as low as $\sim$10\% (see Figs.~5ab, 10ab, 12ab, 14ab and 16ab of WW95). In
a way, by ``replacing'' our CO--cores with the cores of WW95, we adopt their
value of the cross-section, since that reaction is important only from
He--burning on, i.e.\ inside the CO--core. However, in cases where mass loss is
so efficient as to reveal He--burning processed material on the surface and
disperse it in the wind, the carbon/oxygen abundance ratio in the expelled
material
is due to a cross-section value which is different from that used inside the
core; this gives some inconsistency in case of extreme mass-loss. 

(4) Most of all, our method neglects the finite extension of the
He--burning shell: when we separate the CO--core contribution from the
contribution of the overlying layers, 
we somehow assume a sudden jump in the chemical
composition between the CO--core and the outer layers,
as if He--burning defined \MCO\ as a sharp edge.
Actually, there is a
gradient of chemical profile between the regions where helium is still abundant
and those where it has been completely processed. In cases where the
He--burning shell is
not thin and/or a shallow gradient of chemical composition has been
established, our assumption about a clean separation between the CO--core and
the rest of the star may lead to inconsistency. This would mainly affect the
early products of He--burning, such as carbon, whose abundance can change
sensitively even in the outer regions of He--burning, where the process is only
partial. This brings some scatter in the C--production vs.\ \MCO\ relation,
since at the time of explosion most of the carbon is located at the edge of the
CO--core or even outside it, and its distribution is sensitive to past history
especially in stars which suffered from efficient mass loss (M92). Therefore,
the linking method can be less reliable in the case of carbon, and this might
explain our problems in reproducing the observed [C/Fe] ratio (see
\S~\ref{abundanceratios}). To check this effect, we apply the method outlined
in App.~A to derive the ejecta of the CO--cores of the mass--losing
helium stars calculated by
Woosley \etal (1995). These stars build up a shallow gradient of carbon and
helium abundance at the outer edge of the CO--core, and for these stars the
assumption of a sharp--edged core is not properly correct. By applying
Eq.~(\ref{MCOeq}), we tend to overestimate the size of the CO--core for these
stars, as shown in Tab.~\ref{MCOchecktab} where the estimated core masses are
compared with the real ones as tabulated in Woosley \etal (1995, Tab.~6). In
Fig.~24 we plot the ejecta versus \MCO\ for these helium stars
and we compare these relations to those obtained for WW95 models of $Z$=\Zsol\
(App.~A and Fig.~\ref{WMCOfig}). While carbon evidently behaves in
quite a different way in the two cases, the ejecta of other elements and the
remnant mass seem to follow a better defined relation with \MCO. 


\begin{table}
\caption{1$^{st}$ column: model masses of mass--losing helium stars of Woosley
\etal 1995; 2$^{nd}$ column: corresponding CO--core masses as listed in Tab.~6
of Woosley \etal 1995; 3$^{rd}$ column: CO--core masses as deduced according to
the outlined method (see text)} 
\label{MCOchecktab}
\begin{center}
\begin{tabular}{|l|c|c|}
\hline
 $M$ & \MCO (1) & \MCO (2) \\
\hline
  4~ &   1.53   &   1.60   \\
  5~ &   1.87   &   2.05   \\
  7A &   2.30   &   2.73   \\
  7B &   2.30   &   2.80   \\
  7C &   2.30   &   2.80   \\
 10A &   2.50   &   3.26   \\
 10S &   2.50   &   3.26   \\
 20~ &   2.53   &   3.31   \\
\hline
\end{tabular}
\end{center}
\end{table}

\begin{figure*}
\psfig{file=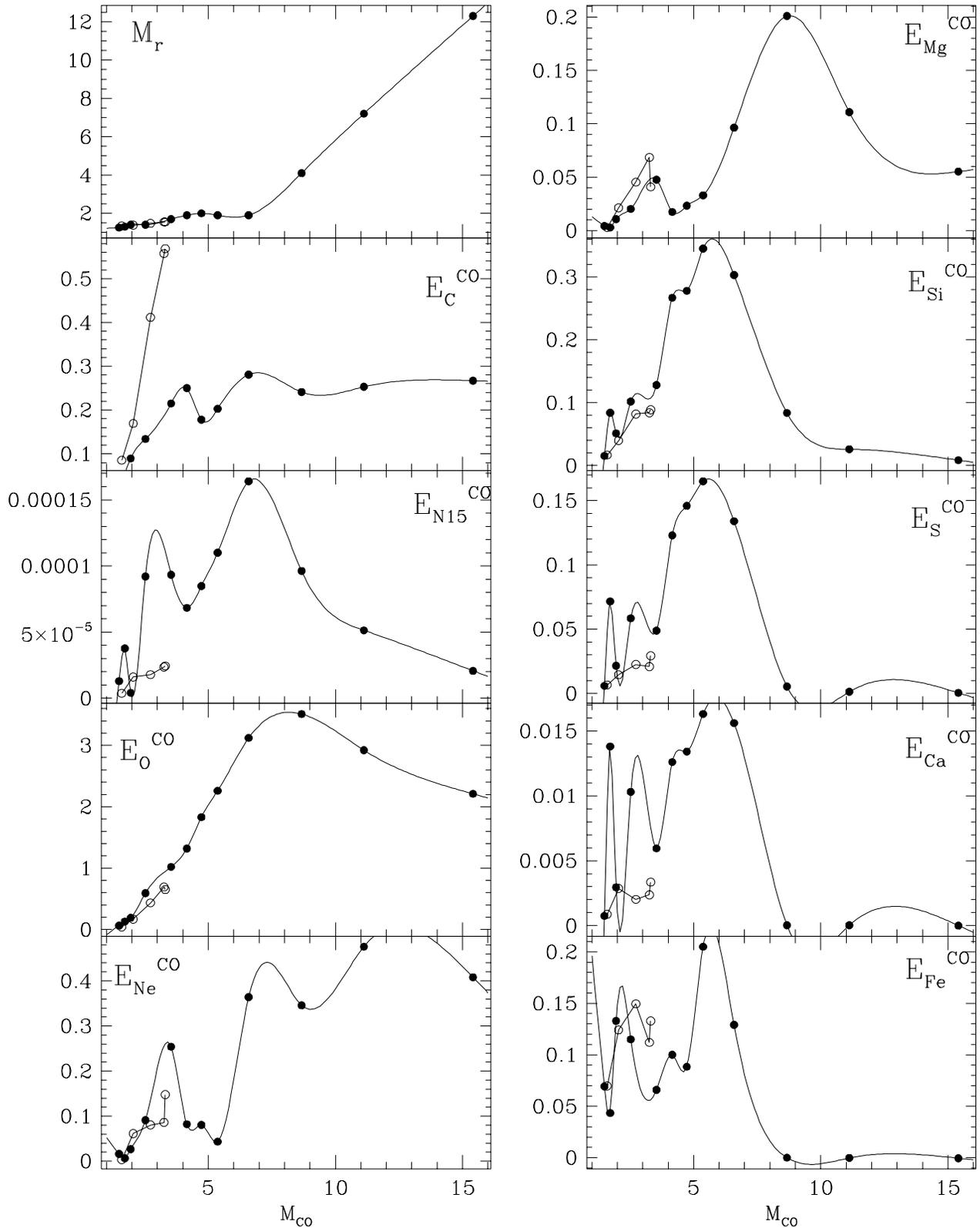,width=18truecm}
\begin{minipage}{18truecm}
\caption{Remnant mass and ejecta of the CO--core vs.\ \MCO as deduced for
mass-losing helium stars of Woosley \etal 1995 (open circles) compared with the
analogous relations deduced for the case $Z$=\Zsol\ of WW95 (filled circles).}
\end{minipage}
\label{MCOcheckfig}
\end{figure*}


\section*{Appendix C: calculating the $Q_{ij}$ matrix}
\label{Qijcomp}
\setcounter{equation}{0}
\renewcommand{\theequation}{C\arabic{equation}}

Here we explain the definition and the meaning of the non--zero elements of the 
$Q_{ij}$ matrix (Tab.~\ref{Qijtab}) and show how the fundamental quantities
entering the matrix are calculated.

\begin{description}
\item[\bf $^{1}$H]
For any stellar mass,
all the hydrogen which remains unburnt in the interiors is eventually
ejected. The only non--zero matrix element for \Hydrogen\
is $Q_{1,1}$, corresponding to the initial \Hydrogen\ which hasn't been
processed. As $q_{4}$ is the mass fraction where \Hydrogen\ is turned to
\Helium, 

\[ Q_{1,1}=1-q_{4} \]

\item[\bf $^{2}$He]
Helium is synthesized on \Hydrogen\ within $q_{4}$ and later turned to \Carbon,
\Oxygen\ and heavier species within $q_{C}$. All the surviving \Helium\ (at
least for stars of some interest, $M >$0.5 \Msol) is eventually ejected. The
term corresponding to the newly synthesised and ejected \Helium\ is 

\[ Q_{2,1}=q_{4}-q_{C} \]

while the term corresponding to the original unprocessed and ejected \Helium\ is

\[ Q_{2,2}=1-q_{C} \]

\item[\bf $^{12}$C]
Carbon is turned to \Ctredici\ and later to \Nitrogen\ within $q_{C13s}$, by
the CNO cycle; later, it is synthesised on \Helium\ within $q_{C}$. The original
\Carbon\ which is ejected is represented by 

\[ Q_{3,3}=1-q_{C13s} \]

Also the newly synthesized \Carbon\ within $w_{C}=q_{C}-d$ is not locked in the 
remnant and can be ejected:

\[ Q_{3,1} = Q_{3,2}=\chi_{C}w_{C} \]

The matrix element $Q_{3,1}$ is non--zero because the $3\alpha$ chain turns 
to \Carbon\ both the original \Helium\ and the \Helium\ which had been 
previously synthesized on the original \Hydrogen.

\item[\bf $^{13}$C]
The ejected original \Ctredici\ is the one outside $q_{Ns}$, where it is turned 
to \Nitrogen\ by the CNO cycle:

\[ Q_{4,4}=1-q_{Ns} \]

Some new \Ctredici\ can also be synthesized on \Carbon\ by the CN cycle within 
$q_{C13s}$, and the corresponding ejected fraction is represented by:

\[ Q_{4,3}=q_{C13s}-q_{Ns} \]

In intermediate mass stars some primary \Ctredici\ can also be produced, due to 
the combined effect of III dredge-up and CNO envelope burning, so for these 
stars further non--zero matrix elements can be:

\[ Q_{4,1}=Q_{4,2}=\chi_{C13}w_{C} \]

\item[\bf $^{14}$N]
Only the original \Nitrogen\ outside the He--burning region can be eventually 
expelled, otherwise it is turned to neutron--rich isotopes:

\[ Q_{5,5}=1-q_{C} \]

Secondary \Nitrogen\ is synthesized on \Carbon, \Oxygen\ and \Ctredici\ during
the CNO cycle; the fraction which remains outside $q_{C}$ is eventually ejected
for any stellar mass: 

\[ Q_{5,3}=Q_{5,4}=Q_{5,6}=q_{Ns}-q_{C} \]

We point out that, by taking $Q_{5,3}=Q_{5,4}=Q_{5,6}$, we implicitly assume
that in the CNO cycle secondary \Nitrogen\ comes from original \Carbon, \Oxygen\
and \Ctredici\ in the same proportions as their initial abundances.

In intermediate mass stars some primary nitrogen is produced by CNO envelope 
burning over dredged-up \Carbon, so for these stars we also have:

\[ Q_{5,1}=Q_{5,2}=\chi_{N}w_{C} \]

\item[\bf $^{16}$O]
The ejected original oxygen is that outside $q_{Ns}$, where it is turned to 
\Nitrogen\ by the ON cycle:

\[ Q_{6,6}=1-q_{Ns} \]

New oxygen is produced within $q_{C}$; the fraction which isn't locked in 
the remnant $d$ is expelled:

\[ Q_{6,1}=Q_{6,2}=\chi_{O}w_{C} \]

\item[\bf {\it nr}]
We indicate with {\it nr} the body of neutron rich isotopes synthesized on
\Nitrogen\ during He--burning (\Odiciotto, \Neventidue, \Mgventicinque) within 
$q_{C}$. All the original and newly synthesized nr isotopes are eventually 
ejected; the corresponding matrix elements are:

\[ Q_{7,7}=1-d \]

\[ Q_{7,3}=Q_{7,4}=Q_{7,5}=Q_{7,6}=w_{C} \]

\item[\bf $^{20}$Ne---$^{24}$Mg---$^{28}$Si---$^{32}$S---$^{40}$Ca---$^{56}$Fe]
All these species are produced in the He--burning stage or in later stages, 
within $q_{C}$. The fraction of the original amount of each of these species 
which is eventually ejected is represented by:

\[ Q_{i,i}=1-d~~~~~~~~~~~~~~~~i=8\div 13 \]

while the ejected fraction of the newly synthesized amount is:

\[ Q_{i,1}=Q_{i,2}=\chi_{i}w_{C}~~~~~~~~~~~~~~i=8\div 13 \]

\end{description}

\noindent
The relevant quantities entering the $Q_{ij}$ matrix (\S~\ref{Qij}) are
calculated on the base of the stellar ejecta derived from the tracks of the
Padua library and discussed in this paper. We impose, species
after species, Eq.~(\ref{Qijnorm}), so that by re-inserting the proper abundance
set $\left\{ X_{j}, \, j=1,...13 \right\}$ the original values of stellar ejecta
are preserved. 

\begin{description}
\item{$d~~~$}
The remnant mass fraction $d$ is, by definition:

\begin{equation}
\label{d}
d=\frac{M_{r}}{M}
\end{equation}

\item[$q_{4}~~~$]
By definition, $q_{4}$ is the mass fraction involved in the \Hydrogen
$\rightarrow$ \Helium\ burning. If we simply considered the mass fraction
of the
He--core, $q_{4}=M_{He}/M$, we would neglect that the core may have receded or
varied in size in the course of evolution and that some material has been
dredged-up in the envelope and/or lost in stellar winds. On the contrary, by
applying Eq.~(\ref{Qijnorm}) to the case of hydrogen ($i$=1): 

\[ (1-q_{4})\,X=\frac{E_{H}}{M} \]

we get an ``effective'' $q_{4}$ which automatically takes into account the
whole mass where hydrogen has been burnt, both the mass lying in the core and
the mass dredged-up to the surface. So we assume:

\begin{equation}
\label{q4}
q_{4}=1-\frac{E_{H}}{XM}
\end{equation}

\item[$q_{C}$~~]
By definition, $q_{C}$ is the mass fraction where \Helium\ is turned to \Carbon,
\Oxygen\ and heavier species. Again, if we simply took the mass fraction of
the CO--core, $q_{C}=M_{CO}/M$, we would neglect the material dredge-up in the
envelope and/or ejected with mass loss. In the case of low and intermediate
mass stars, we would lose completely the contribution of the III~dredge-up,
since
we'd find $q_{C}=d$ and $w_{C}=0$. On the contrary, by applying
Eq.~(\ref{Qijnorm}) to the case of helium ($i$=2), we get an ``effective''
$q_{C}$ which automatically takes into account the role of dredge-up episodes: 

\[ (q_{4}-q_{C})\,X+(1-q_{C})\,Y=\frac{E_{He}}{M} \]

\begin{equation}
\label{qC}
q_{C} = \left[ q_{4}X+Y-\frac{E_{He}}{M} \right] \frac{1}{X+Y}
\end{equation}

where $q_{4}$ is known from Eq.~(\ref{q4}). Now we can also determine:

\begin{equation}
\label{wC}
w_{C}=q_{C}-d
\end{equation}

\item[$q_{Ns}~$]
We obtain the mass fraction $q_{Ns}$ where secondary nitrogen is produced by 
imposing Eq.~(\ref{Qijnorm}) to the case of {\it secondary} \Nitrogen. This 
corresponds to 
all the ejected \Nitrogen\ in the case of massive stars, while in low and 
intermediate mass stars we keep the ejecta of primary and secondary \Nitrogen\
apart, since they involve distinct matrix elements.

\[ (q_{Ns}-q_{C})(X_{C}+X_{C13}+X_{O})+(1-q_{C})X_{N}=\frac{E_{Ns}}{M} \]

\begin{eqnarray}
\label{qNs}
q_{Ns} & = & \frac{E_{Ns}}{(X_{C}+X_{C13}+X_{O})M} \,-\, \nonumber \\
       &   & (1-q_{C})\frac{X_{N}}{X_{C}+X_{C13}+X_{O}} \,+\, q_{C}
\end{eqnarray}

where $q_{C}$ is known from Eq.~(\ref{qC}).

\item[$q_{C13s}$]
We obtain the mass fraction $q_{C13s}$ where the CN cycle turns \Carbon\ to 
\Ctredici\ in the same way as $q_{Ns}$, i.e.\ by 
imposing Eq.~(\ref{Qijnorm}) to the case of {\it secondary} \Ctredici:

\[ (q_{C13s}-q_{Ns})\,X_{C}+(1-q_{Ns})\,X_{C13}=\frac{E_{C13s}}{M} \]

\begin{equation}
\label{qC13s}
q_{C13s}=\frac{E_{C13s}}{X_{C}M}-(1-q_{Ns})\frac{X_{C13}}{X_{C}}+q_{Ns}
\end{equation}

where $q_{Ns}$ is known from Eq.~(\ref{qNs}).

\item[$\chi_{N}~$]
The abundance $\chi_{N}$ of the \Nitrogen\ synthesized within $w_{C}$ by
envelope burning in intermediate mass stars is defined by Eq.~(\ref{Qijnorm}),
applied to {\it primary} \Nitrogen: 

\[ (\chi_{N}w_{C})(X+Y)=\frac{E_{Np}}{M} \]

\begin{equation}
\label{chiN}
\chi_{N}=\frac{E_{Np}}{(X+Y) M w_C}
\end{equation}

where $w_{C}$ is known from Eq.~(\ref{wC}).

\item[$\chi_{C13}$]
Similarly, by imposing Eq.~(\ref{Qijnorm}) to {\it primary} \Ctredici\ we get:

\begin{equation}
\label{chi^{13}C}
\chi_{C13}=\frac{E_{C13p}}{(X+Y) M w_C}
\end{equation}

\item[$\chi_{C}~~$]
The abundance $\chi_{C}$ of new \Carbon\ synthesized within $w_{C}$ is obtained
by applying Eq.~(\ref{Qijnorm}) to carbon ($i$=3) 

\[ (\chi_{C}w_{C})(X+Y)+(1-q_{C13s})\,X_{C}=\frac{E_{C}}{M} \]

\begin{equation}
\label{chiC}
\chi_{C}=\frac{E_{C}}{(X+Y)M w_{C}}-\frac{1-q_{C13s}}{w_{C}} \frac{X_{C}}{X+Y}
\end{equation}

where $q_{C13s}$ is known from Eq.~(\ref{qC13s}).

\item[$\chi_{O}~$]
Similarly, by applying Eq.~(\ref{Qijnorm}) to \Oxygen\ ($i$=6) we get:

\begin{equation}
\label{chiO}
\chi_{O}=\frac{E_{O}}{(X+Y)M w_{C}}-\frac{1-q_{Ns}}{w_{C}} \frac{X_{O}}{X+Y}
\end{equation}

where $q_{Ns}$ is known from Eq.~(\ref{qNs}).

\item[$\chi_{Ne}$---$\chi_{Mg}$---$\chi_{Si}$---$\chi_{S}$---$\chi_{Fe}$]
The abundances of the newly synthesized component of these elements within 
$w_{C}$ is obtained by imposing Eq.~(\ref{Qijnorm}) to the cases $i$=8$\div$13.

\[ (\chi_{i}w_{C})(X+Y)+(1-d)\,X_{i}=\frac{E_{i}}{M} \]

\begin{equation}
\label{chii}
\chi_{i}=\frac{E_{i}}{(X+Y)M w_{C}}-\frac{1-d}{w_{C}} \frac{X_{i}}{X+Y}
\end{equation}

\end {description}

\noindent
We can now derive a
$Q_{ij}$ matrix for each star whose ejecta have been determined;
then, using Eq.~(\ref{RMi}), we get the 
``restitution fractions'' $R_{Mi}$ entering Eq.~(\ref{dGi/dt}). We have derived 
stellar ejecta and $Q_{ij}$ matrices for 5 different sets of 
metallicities; therefore in Eq.~(\ref{RMi}) we have an explicit dependence on
the initial composition of the star through the $X_{j}$'s, and also an indirect 
dependence through $Q_{ij}=Q_{ij}(M,Z)$. That's why the restitution fractions
$R_{Mi}$'s in Eq.~(\ref{dGi/dt}) are to be evaluated as a function of the 
birth-time of the star, to take the effect of initial composition into account:

\[ R_{Mi}(t-\tau_{M})= \sum_{j} Q_{ij}(M,Z(t-\tau_{M})) X_{j}(t-\tau_{M}) \]

\noindent
It is worth here underlining the advantages of this ``$Q_{ij}$ matrix''
formalism. It was originally introduced in order to compensate for the lack of
stellar model of different metallicities, by assuming that the influence of the
initial chemical composition over the $R_{Mi}$'s was included in the linear
dependence on the $X_{j}$'s, while the dependence of the $Q_{ij}$'s on $Z$ was
negligible (Talbot \& Arnett 1973). Now that stellar models and corresponding
ejecta for different metallicities are available, using the $Q_{ij}$ matrix
allows to take into account also possible differences of chemical composition
{\it within} a given $Z$. Indeed, stellar models with different metallicities
generally assume solar {\it relative} abundances of the various species within
a given $Z$; but abundance ratios are not constant in the course of galactic
evolution, nor they are in the evolution of a chemical model. The $Q_{ij}$
matrix links any ejected species to all its different nucleosynthetic sources,
allowing the model to scale the ejecta with respect to the detailed initial
composition of the star through the $X_{j}$'s.


{}

\end{document}